\documentclass[sn-basic]{sn-jnl}

\usepackage{graphicx}
\usepackage{multirow}
\usepackage{amsmath,amssymb,amsfonts}%
\usepackage{amsthm}%
\usepackage{mathrsfs}%
\usepackage[title]{appendix}%
\usepackage{xcolor}%
\usepackage{textcomp}%
\usepackage{manyfoot}%
\usepackage{booktabs}%
\usepackage{algorithm}%
\usepackage{algorithmicx}%
\usepackage{algpseudocode}%
\usepackage{listings}%
\usepackage{longtable}
\usepackage{caption}
\usepackage{subcaption}
\usepackage{hyperref}
\usepackage{xspace}
\usepackage{aas_macros}

\begin{document}

\title[Article Title]{Long-term X-ray variability of the NLS1 Ark~564}


\author[1,2]{\fnm{Neha } \sur{P R}}\email{nehapr.pr@gmail.com}

\author[3]{\fnm{Nikita} \sur{ S Arun}}\email{nikiarun2002@gmail.com}

\author[4]{ \sur{Savithri H. Ezhikode}}\email{savithrih@sfscollege.in}
\author[5]{ \sur{Ranjeev Misra}}\email{rmisra@iucaa.in}

\author[6]{ \sur{Bari Maqbool}}\email{barispn1@gmail.com}

\author[3]{ \sur{Jeena K}}\email{jeenakarunakaran@gmail.com}

\author[1]{ \sur{Joe Jacob}}\email{drjoephysics@gmail.com}

\affil[1]{\orgdiv{Department of Physics}, \orgname{Newman College,Thodupuzha}, \orgaddress \city{Idukki}, \postcode{685585}, \state{Kerala}, \country{India}}

\affil[2]{{\orgdiv{Department of Physics}, \orgname{Baselius College, Kottayam (Autonomous)}, \postcode{686001}, \state{Kerala}, \country{India}}}

\affil[3]{\orgdiv{Department of Physics}, \orgname{Providence Women's  College (Autonomous)}, \city{Kozhikode}, \postcode{673009}, \state{Kerala},\country{India}}

\affil[4]{\orgdiv{St. Francis de Sales College (Autonomous)}, \orgaddress \city{Electronics City, Benguluru}, \postcode{560100}, \state{Karnataka}, \country{India}}

\affil[5]{\orgdiv{Inter-University Centre for Astronomy \& Astrophysics}, \orgname{Post Bag 4}, \city{Ganeshkhind}, \state{Pune}, \country{India}}

\affil[6]{\orgdiv{Department of Physics}, \orgname{Islamic University of Science and Technology}, \city{Awantipora}, \postcode{192122}, \state{Kashmir},\country{India}}


\abstract{We present a systematic spectral study of the X-ray emission from the Narrow-Line Seyfert 1 (NLS1) galaxy Ark 564 using \textit{Swift}/XRT observations spanning sixteen years. The source exhibits significant variability in both the soft and hard X-ray bands, along with a strong positive correlation between them, indicating a closely coupled emission mechanism within the accretion flow-corona system. A simple absorbed power-law model is statistically insufficient to describe the spectra, confirming the presence of an additional soft component. To investigate the origin of the soft excess, we performed multi-model spectral fitting. Incorporating a warm Comptonization component yields a statistically improved fit, supporting the scenario in which the soft excess arises from Comptonization in a warm, optically thick corona. We also applied a relativistic reflection model to representative observations; however, given the moderate spectral resolution and limited high-energy coverage of \textit{Swift}/XRT, the current data do not allow us to robustly determine whether reflection provides a superior or physically preferred description. Our results indicate that warm Comptonization offers a consistent explanation for the observed soft excess in Ark 564 within the limitations of the available data, while deeper broadband observations are required to definitively distinguish between competing models and comprehensively characterise the long-term X-ray spectral variability of the source.}




\maketitle

\section{Introduction}\label{sec1}

An Active Galactic Nucleus (AGN) is a very bright and compact region found at the centre of a galaxy, often emitting more light than all the stars in the galaxy combined \citep{1993ARA&A..31..473A, 2015ARA&A..53..365N}. AGN derive their energy from accretion onto a central supermassive black hole (SMBH) \citep{1973A&A....24..337S,1997iagn.book.....P}, 
wherein the accretion disk emits thermal optical/UV photons that are subsequently Compton up-scattered to X-ray energies by hot electrons in a corona located close to the SMBH \citep{1993ApJ...413..507H,2001ApJ...561..131T}.

Seyfert galaxies are a class of active galaxies whose host systems typically exhibit spiral morphology \citep{2006agna.book.....O, 1997iagn.book.....P, 1974ApJ...192..581K}. 
They are known for pronounced X-ray variability over a broad range of timescales \citep{1997ARA&A..35..445U,1996A&A...305...53B, 1993ApJ...414L..85L}. This variability can occur on timescales from hours to days, indicating rapid changes in the inner accretion flow near the SMBH. Long-term variations over months to years are also common \citep[e.g.,][]{1981MNRAS.194..987M,2004ApJ...617..939M} and are likely associated with changes in the accretion rate or coronal structure. Observations further show that the variability amplitude tends to increase at softer X-ray energies and in lower-luminosity Seyfert 1 galaxies, consistent with models invoking smaller black holes or higher accretion rates. 
These variability characteristics provide important constraints on the geometry and physical processes of the accretion disk–corona system in AGN \citep{2015MNRAS.448.1541S}.

Narrow-line Seyfert 1 (NLS1) galaxies are a peculiar group of Seyfert 1 galaxies characterized by their unique spectral properties like strong Fe II emission, and relatively weak [O III] lines \citep{1985ApJ...297..166O,1989ApJ...342..224G}. X-ray studies of NLS1 galaxies reveal rapid X-ray variability on short timescales \citep{1996A&A...305...53B,1999ApJS..125..297L,2007A&A...465..107B}, which can be interpreted as evidence for relatively small black hole masses in these objects \citep{1996A&A...305...53B,2000MNRAS.314L..17M}. Early variability and multiwavelength studies of NLS1s highlighted the strong coupling between optical/UV and X-ray emissions, emphasizing the importance of disk-corona interactions in shaping the observed spectrum \citep{1991ApJ...380L..51H,2001ApJ...561..131T}. Previous studies suggest that NLS1 galaxies are characterized by accretion at rates close to the Eddington limit onto relatively low- to moderate-mass SMBHs \citep{1995MNRAS.277L...5P,2002ApJ...568..610E}. 

The X-ray spectra of NLS1 galaxies often show a pronounced soft X-ray excess below $\sim 2$ keV, appearing as an enhancement above the extrapolated hard power-law continuum. This primary continuum is generally attributed to Comptonization processes \citep{2006MNRAS.365.1067C}. Within the framework of a standard optically thick, geometrically thin accretion disk, the maximum disk temperature scales approximately as $T_{\max} \propto M_{\rm BH}^{-1/4}$ \citep{1973A&A....24..337S,2012MNRAS.420.1848D}. For SMBHs, this predicts characteristic temperatures in the extreme ultraviolet regime ($\sim 10$-$50$ eV), significantly lower than the observed soft excess temperatures. However, observations consistently find that the soft excess temperature remains clustered around $\sim 0.1$-$0.2$ keV, largely independent of black hole mass \citep{2006MNRAS.365.1067C,2004MNRAS.349L...7G}. This discrepancy indicates that the soft excess cannot be explained by direct thermal disk emission alone and instead requires additional physical processes operating in the inner accretion flow. Currently, two primary models are invoked to explain the soft X-ray excess in AGN. The first is the warm corona model, in which the excess arises from Compton up-scattering of seed photons from the accretion disk within a warm ($kT \approx 0.1$-$1$ keV), optically thick corona \citep[e.g.][]{2012MNRAS.420.1848D,2012MNRAS.420.1825J,2020A&A...634A..85P,2017MNRAS.472.3492E}. This interpretation is supported by the observed correlated variability between the optical/UV and soft X-ray bands \citep[e.g.,][]{2020A&A...634A..92U, 2016RAA....16..108E,1993A&A...274..105W}. The alternative explanation is the relativistically blurred reflection model, which attributes the soft excess to a blend of low-energy emission lines produced when hard X-ray radiation irradiates the ionized inner accretion disk \citep{2002MNRAS.335L...1F}. Strong relativistic effects in the vicinity of the central black hole broaden and smear these features, resulting in the observed smooth continuum-like soft excess \citep{2014MNRAS.439.2307F,2006MNRAS.365.1067C,2020MNRAS.498.3888J}. Both scenarios have been successfully applied to individual sources, highlighting the ongoing debate regarding the dominant physical mechanism.

In this study, we investigate the X-ray variability of the nearby ($z = 0.0247$) NLS1 galaxy \citep{1994MNRAS.271..958B} Arakelian 564 (Ark 564) using \textit{Swift}/XRT observations spanning from 2005 to 2021. 
Various estimates of the black hole mass of Ark~564 from variability and scaling relations constrain the mass to lie within $(0.6$--$5.9) \times 10^{6}\,M_{\odot}$ \citep{2007A&A...465..107B, 2004ApJ...613..682P, 2005MNRAS.356..789B, 2017MNRAS.468.3489K, 2013MNRAS.431.2441D}. Its relatively low black hole mass and high bolometric luminosity of about $10^{45}\,\rm erg\,s^{-1}$ \citep{2004ApJ...602..635R} imply that the source is accreting at or near the Eddington limit. Ark 564 is one of the brightest and most extensively studied NLS1 galaxies in the X-ray band. It is well known for exhibiting rapid and large-amplitude X-ray variability across multiple timescales, from hours to days \citep[e.g.,][]{2002A&A...391..875G,2002ApJ...568..610E,2007MNRAS.375.1479S,2007ApJ...661...38P,2016RAA....16..108E}. Its power spectral density (PSD) shows strong variability power at high frequencies, consistent with a low black hole mass and high accretion rate system \citep{2006Natur.444..730M}. The source also displays a characteristically steep hard X-ray continuum, with photon indices typically $\Gamma \gtrsim 2.3$–2.7 \citep[e.g.,][]{2007A&A...465..107B, 2020MNRAS.495.3373E}, 
indicative of efficient Compton cooling in a high Eddington ratio accretion flow. Broadband spectral studies using XMM-Newton and Suzaku have consistently reported a pronounced soft X-ray excess below $\sim$2 keV \citep{2006MNRAS.365.1067C, 2017MNRAS.468.3489K}. 
This soft excess has been successfully modeled using warm Comptonization scenarios 
\citep{2007ApJ...671.1284D, 2016RAA....16..108E, 2011A&A...534A..39M, 2015MNRAS.448.1541S}. Alternatively, relativistic reflection models have also been observed to contribute to soft excess in Ark 564 \citep{2024A&A...689A.116L,2007A&A...461..931P}. 
However, other studies show that a blurred reflection alone is not sufficient to model the soft excess \citep{2017MNRAS.468.3489K, 2023AN....34430042E}. In addition to its spectral properties, Ark 564 exhibits flux–spectral variability correlations, including spectral steepening with increasing flux and changes in the soft excess strength \citep{2002ApJ...568..610E}. These properties make it an ideal source for probing disk–corona coupling, origin of the soft excess and the nature of long-term X-ray variability in high-accretion-rate AGN.

Here, we make use of two decades of \textit{Swift}/XRT monitoring of Ark~564 to probe the 
long-term evolution of the coronal emissions through a systematic spectral and variability analysis in the 0.3--10\,keV band. The structure of the paper is as follows. \S\ref{data} presents the observation and data reduction methods. The analysis of the X-ray data is detailed in \S\ref{3}. \S\ref{4} and \S\ref{sum} describe the results and discussion of the present work.

\section{ Data Reduction Methods}
\label{data}

Ark 564 has been monitored by \textit{Swift} \citep{2004ApJ...611.1005G} since 2005. We retrieved 99 observations from the X-ray Telescope \citep[XRT;][]{2005SSRv..120..165B} onboard \textit{Swift} covering the 0.3–10 keV energy range between 2005 and 2021. Out of these, 85 observations were analysed, while 14 were excluded due to missing Level-1 event files. Data reduction and analysis were performed using HEASOFT v6.31.1. Level-2 data products for the XRT observations were derived from Level-1 data using the standard processing pipeline \textit{XRTPipeline} version 0.13.7, and calibration was performed using CALDB version 20240522. This pipeline generated a clean event list for each observation. These cleaned event lists in PC mode from each observation were then used to create source and background spectra using \textit{XSELECT}. The source and background regions were extracted by visual analysis with the DS9 software. The details of \textit{Swift}-XRT observations used in this analysis is presented in Table~\ref{tab:swift_obs}, including exposure times and net count rates of the source region in the 0.3-10 keV band. All 85 PC mode observations showed a count rate above 0.5 count $s^{-1}$, indicating pile-up. Pile-up correction was performed 
following the standard Swift XRT PC mode procedure\footnote{\url{https://www.swift.ac.uk/analysis/xrt/pileup.php}}. To correct for pile-up in PC mode observations, the radial surface brightness (PSF) profile was fitted with a King function combined with a Gaussian component. The radius at which the model deviated from the observed profile was identified from the residuals and adopted as the inner exclusion radius. Source spectra were extracted from annular regions centred on the source, with inner radii of 3-5.5 pixels (depending on the observation) and a fixed outer radius of 50 pixels. Background spectra were obtained from nearby source-free circular regions with radii of 40–50 pixels. 

\begin{table}
\centering
\footnotesize
\caption{List of Swift-XRT observations of Ark 564. Column~4 denotes X-ray count rate ($\rm{counts\,s^{-1}}$) in the 0.3-10 keV band. Exposure time is given in seconds (s).}
\label{tab:swift_obs}

\begin{tabular}{cccc}
\toprule
Obs ID & Exposure Time & Start Date & Count Rate\\
\midrule

00035062003 & 4721.00 & 2005-12-09 & 2.03\\
00035062001 & 4637.40 & 2005-04-19 & 1.80\\
00035062002 & 4515.60 & 2005-05-19 & 1.74\\
00081687001 & 2455.23 & 2015-05-22 & 1.18\\
00033282002 & 2198.94 & 2014-05-17 & 1.00\\
00096113002 & 1707.53 & 2021-04-29 & 0.96\\
00096113003 & 1659.81 & 2021-04-30 & 1.28\\
00096113001 & 1200.99 & 2021-04-24 & 1.18\\
00095000016 & 1110.73 & 2019-12-22 & 0.89\\
00093158012 & 1108.03 & 2017-11-21 & 1.78\\
00095653007 & 1098.78 & 2020-06-12 & 2.02\\
00092237005 & 1098.00 & 2016-06-06 & 1.39\\
00094000018 & 1090.67 & 2018-12-22 & 0.53\\
00093158011 & 1073.12 & 2017-11-14 & 1.99\\
00095000001 & 1070.61 & 2019-05-08 & 1.35\\
00094000017 & 1065.60 & 2018-12-15 & 1.97\\
00094000016 & 1065.60 & 2018-12-08 & 2.38\\
00093158014 & 1063.08 & 2017-12-05 & 1.29\\
00095000013 & 1058.08 & 2019-12-01 & 1.29\\
00093158010 & 1057.89 & 2017-11-10 & 1.48\\
00095000015 & 1040.53 & 2019-12-15 & 1.07\\
00095000003 & 1038.01 & 2019-05-22 & 1.71\\
00094000013 & 1027.99 & 2018-11-24 & 1.17\\
00092237001 & 1025.49 & 2016-05-05 & 1.52\\
00093158016 & 1025.48 & 2017-12-15 & 1.23\\
00094000011 & 1022.98 & 2018-11-14 & 0.99\\
00094000005 & 1020.46 & 2018-05-29 & 1.15\\
00092237003 & 1020.44 & 2016-05-21 & 1.53\\
00095653010 & 1017.96 & 2020-11-10 & 1.01\\
00092237016 & 1017.96 & 2016-12-27 & 0.94\\
00093158004 & 1017.95 & 2017-05-25 & 2.06\\
00092237006 & 1015.45 & 2016-06-14 & 2.02\\
00095000005 & 1013.20 & 2019-06-05 & 1.78\\
00093158003 & 1012.94 & 2017-05-18 & 0.93\\
00092237009 & 1010.44 & 2016-11-08 & 2.53\\
00093158008 & 1010.43 & 2017-06-22 & 1.95\\
00092237015 & 1007.97 & 2016-12-19 & 1.76\\
00092237013 & 1007.93 & 2016-12-03 & 1.63\\
00094000009 & 1007.93 & 2018-11-03 & 1.35\\
00093158006 & 1005.53 & 2017-06-08 & 1.42\\
00092237004 & 990.37 & 2016-05-29  & 1.80\\
00094000015 & 989.50 & 2018-12-05 & 1.72\\
00095000007 & 987.87 & 2019-06-16 & 0.98\\
00095000008 & 987.87 & 2019-06-19 & 1.34\\
00094000006 & 985.36 & 2018-06-05 & 1.71\\
00093158017 & 984.09 & 2017-12-19 & 0.55\\
00092237014 & 977.84 & 2016-12-11 & 1.89\\
00094000008 & 977.84 & 2018-06-19 & 0.46\\
00093158005 & 975.33 & 2017-06-01 & 1.13\\
00092237012 & 970.32 & 2016-11-25 & 1.85\\
00092237007 & 964.66 & 2016-06-22 & 1.93\\
00095000011 & 962.80 & 2019-11-17 & 2.19\\
00092237002 & 962.80 & 2016-05-13 & 1.27\\
00095653012 & 960.29 & 2020-11-24 & 1.88\\
00095653006 & 960.29 & 2020-06-05 & 1.77\\
00095653003 & 960.28 & 2020-05-20 & 1.60\\
00093158002 & 958.17 & 2017-05-11 & 1.96\\
00095000009 & 953.04 & 2019-11-03 & 1.61\\
\bottomrule
\end{tabular}
\end{table}

\begin{table}
\ContinuedFloat
\centering
\footnotesize
\caption{(continued)}

\begin{tabular}{cccc}
\toprule
Obs ID & Exposure Time & Start Date & Count Rate\\
\midrule

00095000004 & 952.83 & 2019-05-29 & 1.91\\
00095653013 & 952.77 & 2020-12-01 & 1.26\\
00095653016 & 950.27 & 2020-12-19 & 2.30\\
00093158013 & 950.25 & 2017-11-28 & 1.54\\
00092237008 & 937.76 & 2016-11-01 & 1.57\\
00095653014 & 927.70 & 2020-12-08 & 2.02\\
00095000014 & 922.68 & 2019-12-08 & 1.72\\
00093158007 & 907.61 & 2017-06-15 & 1.33\\
00093158001 & 905.28 & 2017-05-04 & 1.49\\
00095653009 & 902.62 & 2020-11-03 & 2.09\\
00095000010 & 898.01 & 2019-11-10 & 1.46\\
00094000012 & 883.03 & 2018-11-17 & 1.49\\
00095653001 & 880.05 & 2020-05-08 & 1.23\\
00095653017 & 860.00 & 2020-12-22 & 2.09\\
00094000007 & 860.00 & 2018-06-12 & 1.81\\
00094000003 & 858.03 & 2018-05-18 & 1.60\\
00095653011 & 852.48 & 2020-11-17 & 1.82\\
00095653005 & 838.13 & 2020-05-29 & 1.33\\
00095653008 & 814.86 & 2020-06-19 & 0.85\\
00096050002 & 755.22 & 2021-04-18 & 0.46\\
00094000001 & 571.65 & 2018-05-08 & 0.93\\
00092237010 & 468.18 & 2016-11-17 & 1.58\\
00033282001 & 463.05 & 2014-05-17 & 2.13\\
00096050001 & 407.28 & 2021-04-22 & 1.04\\
00092237011 & 377.99 & 2016-11-22 & 2.36\\
00095653004 & 288.31 & 2020-05-22 & 1.16\\
00094000002 & 198.03 & 2018-05-15 & 1.03\\
\bottomrule
\end{tabular}
\end{table}

\section{X-ray Spectral Analysis}
\label{3}

In this work, we used XSPEC \citep{1996ASPC..101...17A} version 12.13.0c to fit all the 85 spectra in the 0.3-10 keV energy range. The standard \textit{Swift}/XRT response matrix file (RMF), swxpc0to12s6-20130101v014.rmf, was adopted for all PC mode spectra. Ancillary response files (ARFs) were generated individually, using the standard \texttt{xrtmkarf} tool, for each observation using the corresponding annular extraction region to account for PSF losses and vignetting corrections\footnote{\url{https://www.swift.ac.uk/analysis/xrt/arfs.php}}. The source spectra were subsequently linked with the background, ARF, and RMF files using the \texttt{grppha} task. Spectra were regrouped to ensure a minimum of 20 counts per bin. All spectra were fitted in the 0.3–10 keV energy range using $\chi^{2}$ statistics, and uncertainties were estimated at the 90\% confidence level.

We first modeled the spectra with an absorbed \texttt{powerlaw}, using \texttt{tbabs} to account for the Galactic absorption.The absorbed power-law model is expressed as
\begin{equation}
  F_{1}(E) = exp[-N{_H} \sigma(E)] K_{p} E^{-\Gamma}  
\end{equation}
where $N_{H}$ is the equivalent hydrogen column density (cm$^{-2}$), $\sigma$(E) is the photoelectric absorption cross-section (cm$^{2}$), $\Gamma$ is the photon index or slope of the powerlaw (dimensionless), and $K_{p}$ is the power-law normalization in units photon/keV/cm$^{2}$/s at 1 keV. The hydrogen column density was fixed at the Galactic value, $N_{\rm H} = 4.98 \times 10^{20}$ cm$^{-2}$ \citep{2016A&A...594A.116H}\footnote{\url{https://heasarc.gsfc.nasa.gov/cgi-bin/Tools/w3nh/w3nh.pl}}, while the photon index ($\Gamma$) and normalisation were allowed to vary. The best-fit photon index was found in the range $\Gamma \sim 2.4$–2.9, providing a reasonable description of the overall continuum, given in Table~\ref{tab:tb_po}. However, systematic positive residuals in the 0.3–2 keV band indicated the presence of a soft X-ray excess. As a phenomenological approach, a blackbody component (\texttt{bbody}) was added to the absorbed power-law model. The absorbed blackbody plus power-law model is

\begin{equation}
    F_{2}(E) = exp[-N{_H} \sigma(E)] [\frac{K_{bb}\times8.0525E^{2}dE}{(kT)^{4}[exp(\frac{E}{kT})^{-1}]}  + K_{p} E^{-\Gamma}]  
\end{equation}

where \textit{E} represents the photon energy in keV, $K_{bb}$ is the blackbody normalization in units of photon/keV/cm$^{2}$/s at 1 keV, and kT in units of keV is the blackbody temperature. During the spectral fitting, the blackbody temperature was fixed at kT = 0.1 keV, consistent with the typical soft-excess temperatures reported for AGNs.
 Although this improved the fit in the soft band, the model remained physically non-self-consistent, as the power-law continuum and the soft component were treated independently. The best fit parameters of the model are given in Table~\ref{tab:tb_bbpo}. In addition, we tested two higher blackbody temperatures, kT = 0.2 and 0.3 keV, for representative high-flux (Obs ID 00096113002), low-flux (Obs ID 00092237011), and intermediate-flux (Obs ID 00094000003) observations. We found that most of the spectral parameters, including the photon index, power-law normalization, and fluxes in different energy bands, remain well constrained. However, the $\chi^2$ value increases slightly for kT = 0.3~keV, while the blackbody normalization is poorly constrained for both kT = 0.2~keV and 0.3~keV.

To obtain a more physically motivated description, we modelled the soft excess using the thermal Comptonization model \texttt{NthComp} \citep{1996MNRAS.283..193Z, 1999MNRAS.309..561Z}, assuming that disk photons serve as seed photons.
The redshift was fixed at $z = 0.0247$ \citep{1999ApJS..121..287H}, and the seed photon temperature was fixed at $kT_{\rm seed} = 0.01$ keV, representative of a standard accretion disk around a low-mass, high-accretion-rate black hole. In this framework, the soft excess arises from Compton up-scattering of disk photons in a warm, optically thick corona. To self-consistently link the soft and hard spectral components, the phenomenological power law was replaced by the convolution model \texttt{Simpl} \citep{2009PASP..121.1279S}, which represents the emission from an optically thin, hot corona that intercepts and scatters a fraction ($f_{\rm sc}$) of the warm corona output into a high-energy power-law tail. The up-scattering-only (UpScOnly) parameter was fixed at 1, so that only up-scattered photons contribute to the hard X-ray emission, and the input type parameter in \texttt{NthComp} was fixed at 0, corresponding to blackbody seed photons. 
The equation for \texttt{tbabs*Simpl*NthComp} may be represented as follows:
\begin{equation}
{F_{3}(E)=
e^{-N_{\rm H}\sigma(E)}
\left[
(1-f_{\rm sc})F_{\rm nthComp}(E)
+
f_{\rm sc}P(E)
\right]}
\end{equation}
where
    $F_{\rm obs}(E)$ - Observed X-ray spectrum,
    $f_{\rm sc}$ - Scattering fraction,
    $F_{\rm nthComp}(E)$ - Thermal Comptonization (warm-corona) spectrum,
    $P(E)$ - Hot Comptonization component computed by the convolution model \texttt{Simpl}, which takes the warm compontization spectrum (\texttt{NthComp}) as the
input seed photons.

\begin{figure}[t]
\centering

\begin{subfigure}{0.48\linewidth}
    \includegraphics[width=\linewidth]{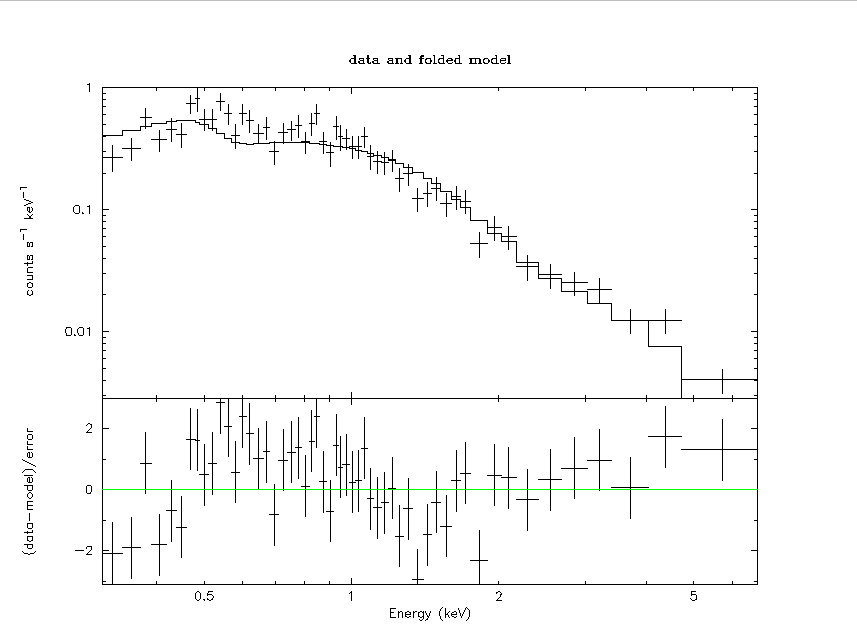}
    \caption{\texttt{tbabs*powerlaw}}
\end{subfigure}
\hfill
\begin{subfigure}{0.48\linewidth}
    \includegraphics[width=\linewidth]{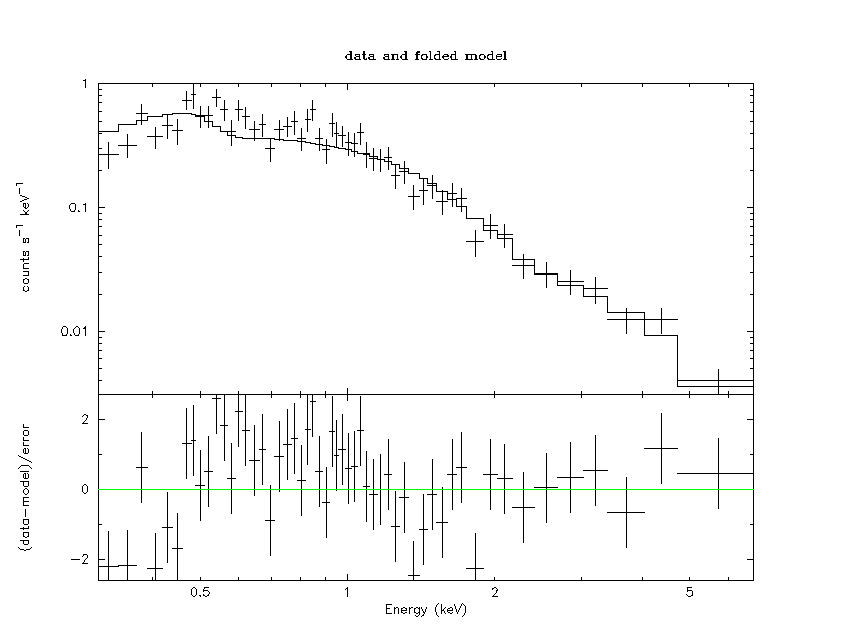}
    \caption{\texttt{tbabs*(powerlaw+blackbody)}}
\end{subfigure}

\vspace{0.4cm}

\begin{subfigure}{0.7\linewidth}
    \centering
    \includegraphics[width=\linewidth]{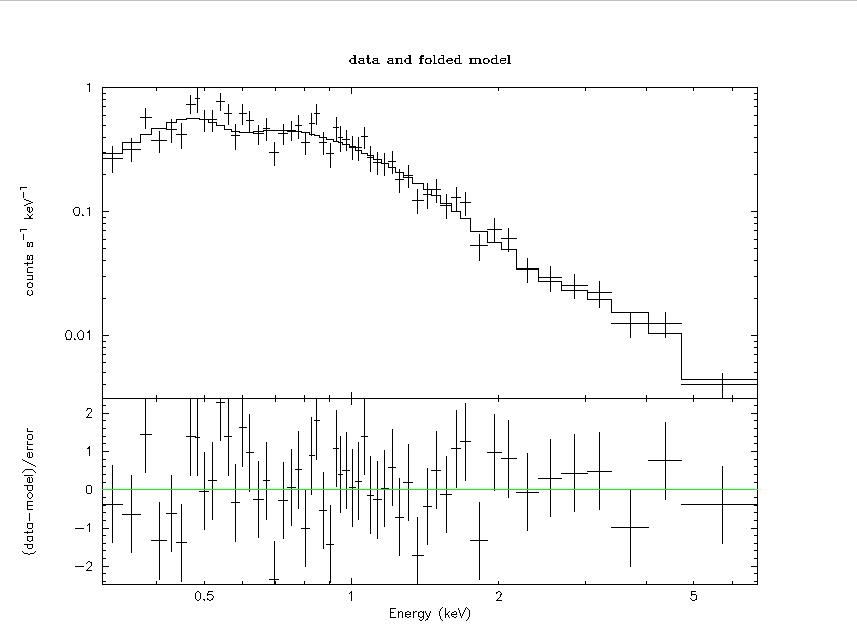}
    \caption{\texttt{tbabs*Simpl*NthComp}}
\end{subfigure}

\caption{
Spectral fits in the 0.3-10 keV band for the observation 00081687001 of Ark 564 using continuum-based models. In each figure, the top panel shows the unfolded spectrum and the bottom panel shows the data-to-model residuals. 
}
\label{fig:continuum_models}
\end{figure}

The photon index of the \texttt{Simpl} component was found to lie in the range of $\Gamma \sim 1.3-2.8$ across different observations. The best-fit parameters and the unabsorbed fluxes in the 0.3–10 keV band, calculated using the XSPEC convolution model cflux, are presented in 
Table~\ref{tab:tb_smnc}. The spectral fits in the 0.3-10 keV band is illustrated in Figure~\ref{fig:continuum_models}. We also examined the effect of varying the seed photon temperature $kT_{\rm seed}$ to 0.02 and 0.03 keV for representative high- (Obs ID 00096113002), low- (Obs ID 00092237011), and intermediate- (Obs ID 00094000003) flux observations. We find that the \texttt{NthComp} photon index, normalization, and the \texttt{Simpl} scattering fraction remain consistent with those obtained for $kT_{\rm seed}$ = 0.01 keV. However, the $\chi^2$ value increases slightly for $kT_{\rm seed} = 0.03$ keV, indicating that $kT_{\rm seed} = 0.01$ keV provides the statistically preferred fit.

The X-ray spectra of NLS1 galaxies can vary considerably across different flux states, and the broadband continuum and spectral components can become dominated by blurred or distant reflection components depending on the flux states, as has been observed in the case of Mrk~335 \citep{2015MNRAS.446..633G,2014MNRAS.443.1723P, 2019MNRAS.490..683P, 2021JApA...42...51E, 2025JHEAp..45..418A}. To investigate whether such behavior is present in Ark~564, we additionally fitted two representative observations corresponding to the highest (Obs~ID~00096113002) and lowest (Obs~ID~00092237011) flux states with the reflection models \texttt{relxillCp} and \texttt{xillverCp}. We compared the warm-corona model (\texttt{tbabs*Simpl*NthComp}) and the relativistic reflection model (\texttt{tbabs*relxillCp}) for the highest and lowest flux observations of Ark~564. The detailed comparison of the best-fit parameters for the two flux states is given in Appendix~\ref{tab:model_comparison}.

\begin{figure}[t]
\centering

\begin{subfigure}{0.48\linewidth}
    \includegraphics[width=\linewidth]{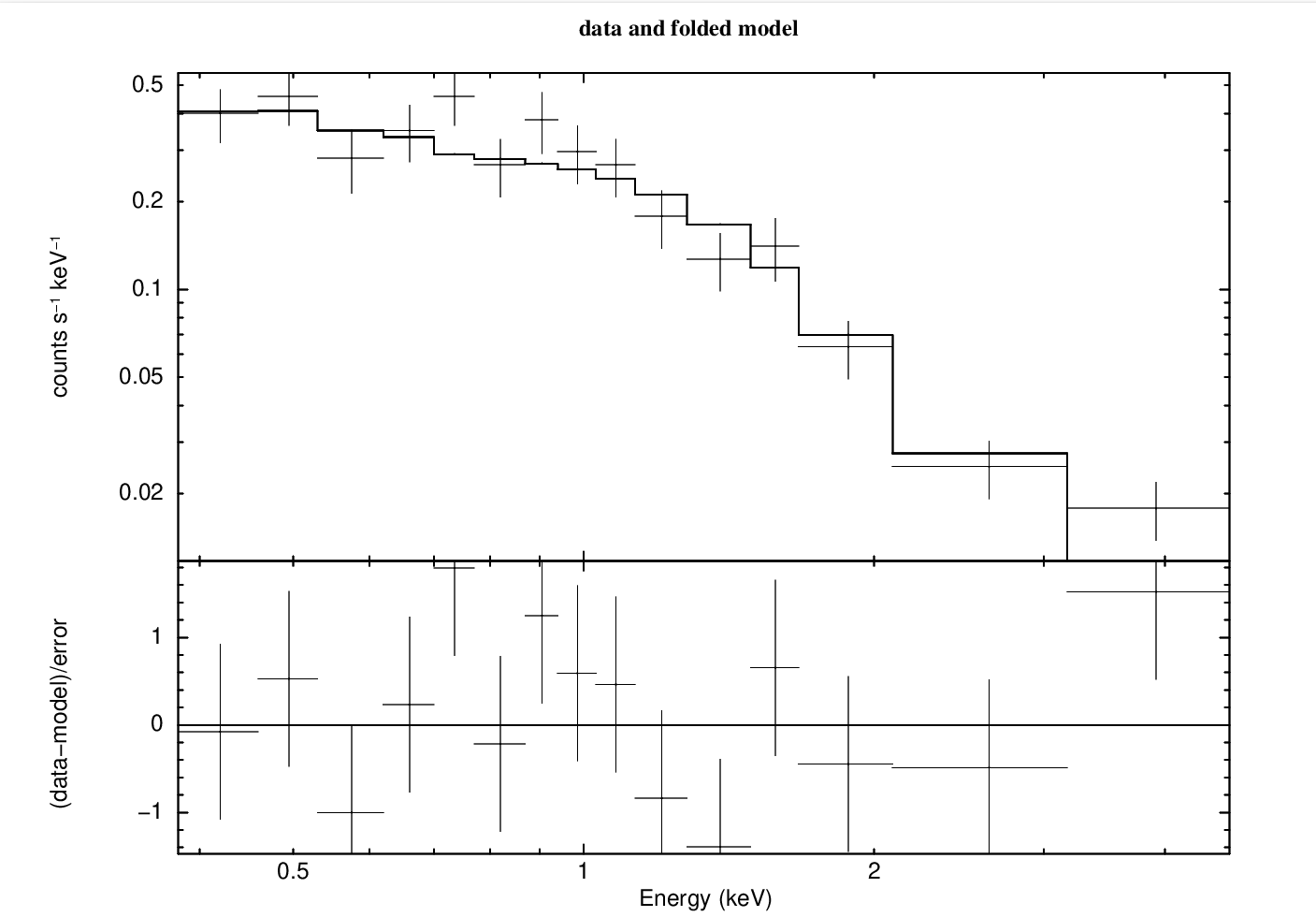}
    \caption{\texttt{tbabs*xillverCp}}
\end{subfigure}
\hfill
\begin{subfigure}{0.48\linewidth}
    \includegraphics[width=\linewidth]{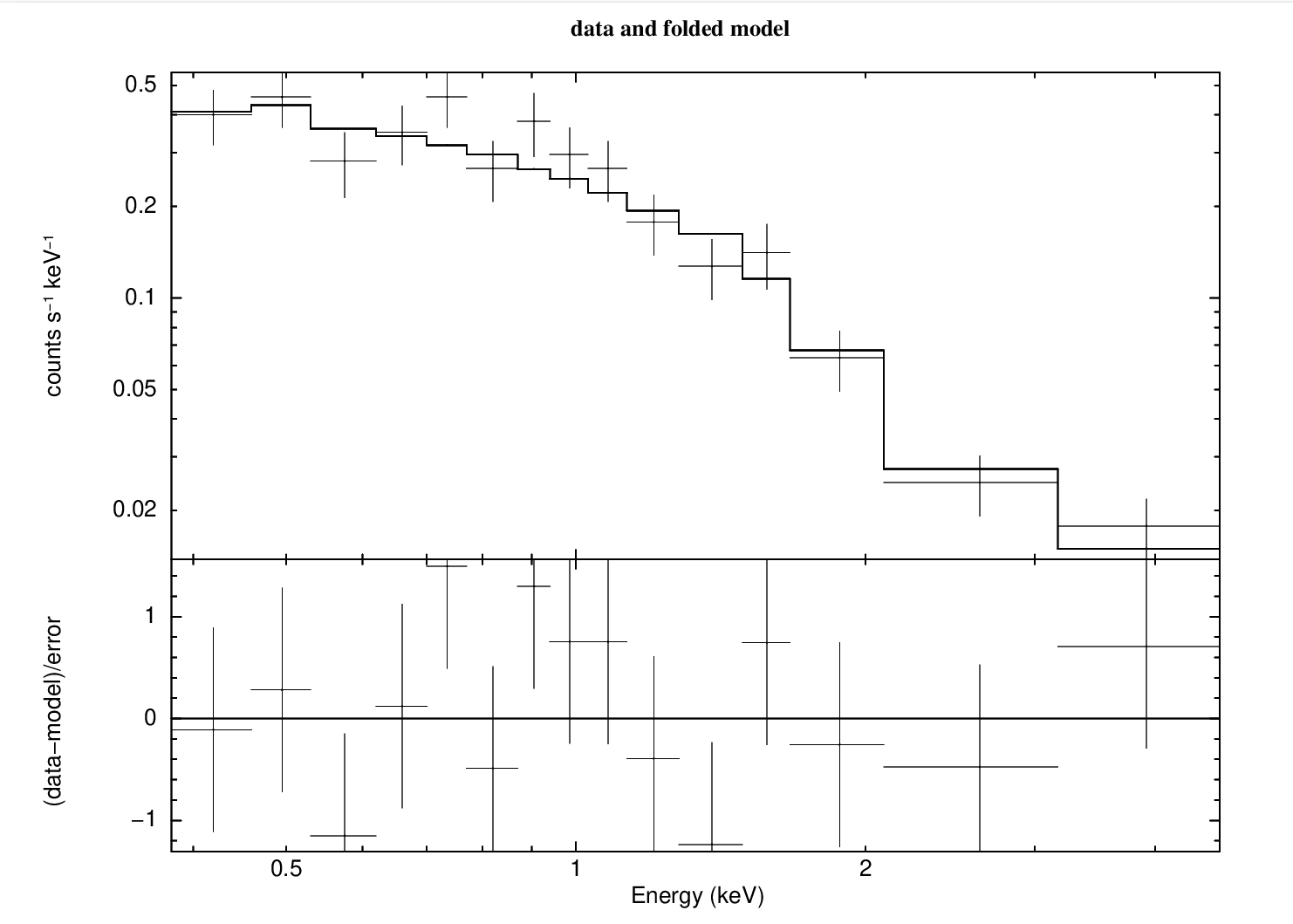}
    \caption{\texttt{tbabs*relxillCp}}
\end{subfigure}

\caption{
Spectral fits in the 0.3-10 keV band for the observation 00092237011 using reflection-based models. For the relativistic reflection model (\texttt{relxillCp}), the black hole spin was fixed at $a_* = 0.998$, the outer disk radius at $R_{\rm out} = 1000\,R_g$, the coronal electron temperature at $kT_e = 15$ keV, the iron abundance at solar, and the inclination angle at $30^\circ$.
}
\label{fig:reflection_models}
\end{figure}

The \texttt{relxillCp} model \citep{2014ApJ...782...76G,2014MNRAS.444L.100D, 2018ApJ...864...25G} describes relativistic reflection from an ionized accretion disk, incorporating angle-dependent reflection calculations 
convolved with relativistic blurring effects expected in the strong gravity regime near the black hole. The incident continuum in this model is assumed to arise from thermal Comptonisation (\texttt{Nthcomp}-like), characterised by the photon index ($\Gamma$) and coronal electron temperature ($kT_e$). The model self-consistently computes the reflected spectrum including fluorescent emission lines (e.g., Fe K$\alpha$), the Compton hump, and ionised disk features, with relativistic broadening governed by parameters such as black hole spin ($a_*$), disk inclination ($i$), and inner disk radius ($R_{\rm in}$). The ionization state of the reflector is governed by the ionization
parameter ($\xi = 4\pi F / n$), which determines the strength and shape of emission lines and continuum features. During fitting, several parameters were fixed due to the limited spectral resolution and bandpass of \textit{Swift} XRT. The black hole spin was fixed at $a_* = 0.998$ \citep[e.g.,][]{2022ApJ...939..109L}, and the outer disk radius was set to $R_{\rm out} = 1000\,R_g$, effectively approximating an extended disk and ensuring that the reflection spectrum is dominated by emission from the relativistically blurred inner regions. The coronal electron temperature was fixed at $kT_e = 15$ keV, the iron abundance was assumed to be solar, and the disk inclination angle was fixed at $i = 30^\circ$. The ionisation parameter, photon index, and reflection fraction were allowed to vary, and the resulting values are given in Table~\ref{tab:model_comparison}. We checked the dependence on black hole spin by repeating the spectral fitting with $a_* = 0$ and $a_* = 0.5$. We find that the best-fit spectral parameters are consistent within their uncertainties for both spin values, indicating that our results are not sensitive to the adopted black hole spin. We also tested the \texttt{xillverCp} model \citep{2010ApJ...718..695G}, which describes non-relativistic reflection from distant or weakly ionized material illuminated by the same thermal Comptonization continuum. It self-consistently computes the reflected spectrum, including narrow fluorescent lines and the Compton reflection hump, arising from material located at larger distances such as the outer disc and torus. The ionization parameter (log$\xi$) was fixed at zero during the fitting. The spectral fits for both models in the 0.3-10 keV band are shown in Figure~\ref{fig:reflection_models}.

The \texttt{relxillCp} model provides statistically acceptable fits, but the ionization and reflection fraction were poorly constrained. The \texttt{tbabs*xillverCp} model resulted in poor spectral fits with reduced $\chi^2$ greater than 2. Consequently, reliable flux estimates using \texttt{cflux} could not be obtained, and this model was excluded from further flux and luminosity analysis. The limited energy range and spectral resolution of \textit{Swift} XRT (0.3–10 keV) prevent robust detection of the Compton hump above 10 keV and do not allow precise measurement of broadened Fe K features. Therefore, while reflection cannot be ruled out, the current data do not allow a definitive confirmation of a relativistic reflection origin for the observed soft excess.

\begin{figure}
    \centering
     \includegraphics[width=0.6\linewidth]{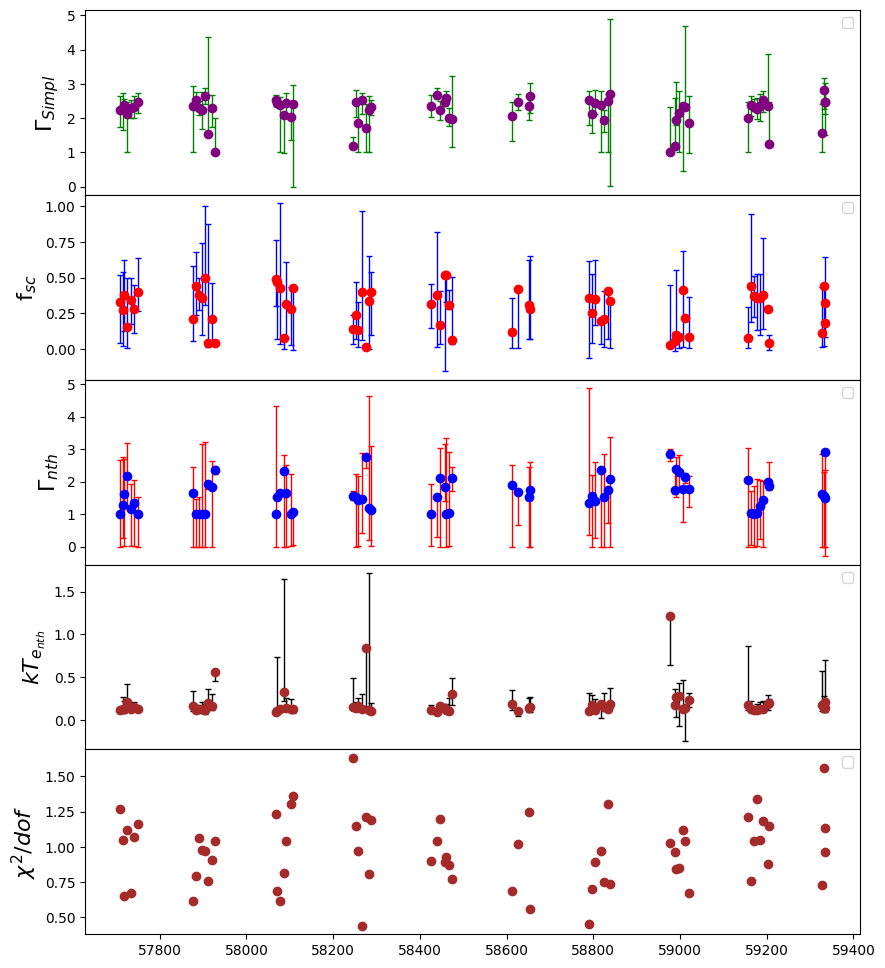}
    \caption{Temporal evolution of spectral parameters as a function of MJD obtained from the model \texttt{tbabs*Simpl*NthComp.} }
    \label{fig:evolution}
\end{figure}

\begin{figure}

     \centering
	\begin{subfigure}{\linewidth}
  
       \includegraphics[width=0.5\linewidth]{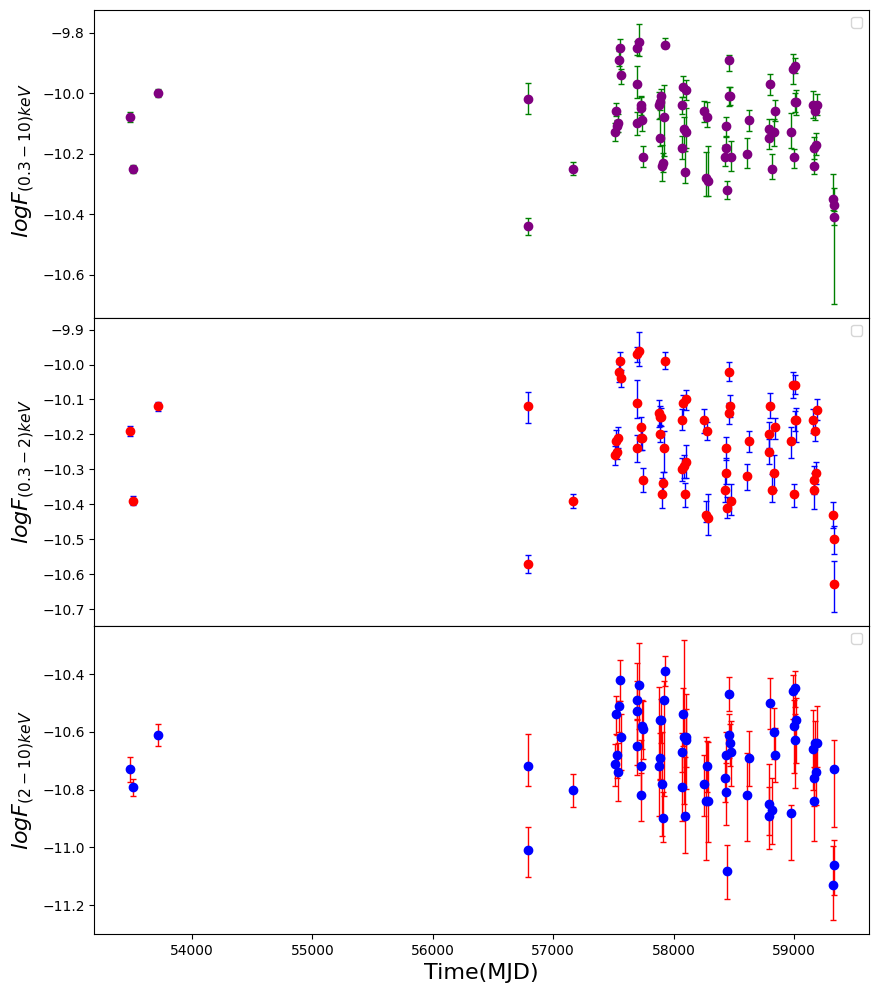}
             \includegraphics[width=0.5\linewidth]{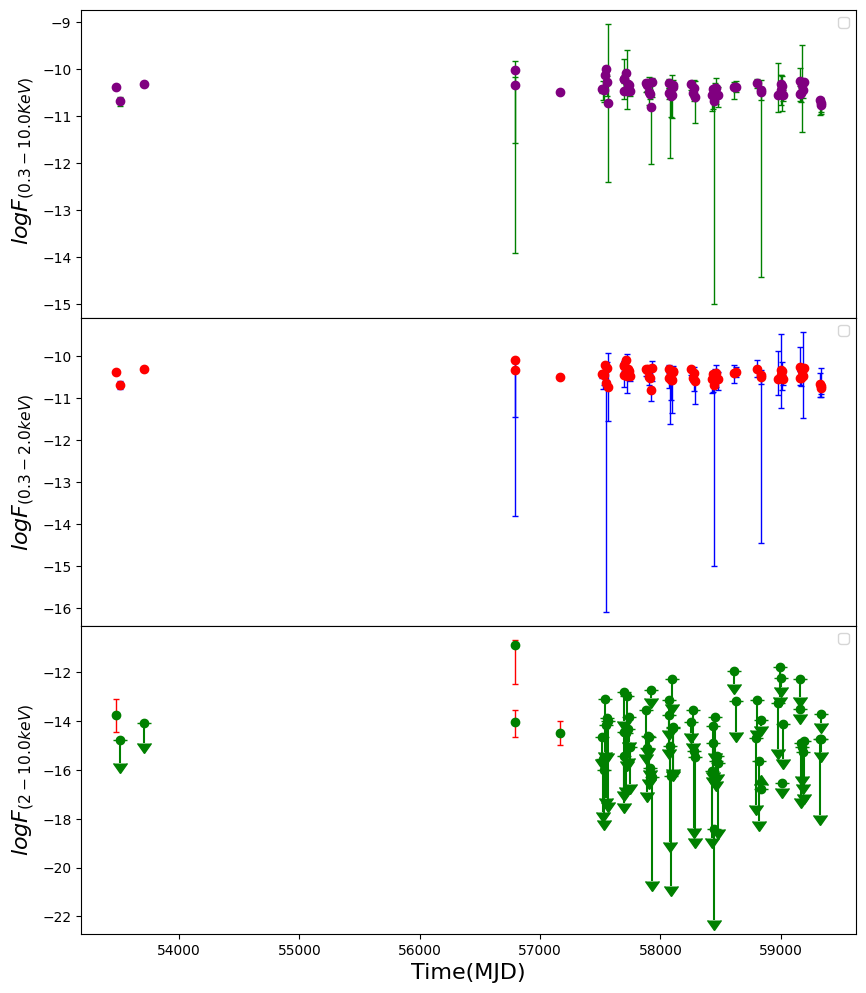}\\
	    \end{subfigure}
	   
	\caption{ Variation of logarithmic flux values across MJD. The left panel shows the logarithmic flux values of \texttt{tbabs*Simpl*NthComp}, and the right panel shows the logarithmic flux values of the \texttt{NthComp} model.}
	\label{fig:flux-mjd}

\end{figure}

{
\footnotesize
\centering
\begin{longtable}{l l l l l l l l  }

\caption{Best-fit spectral parameters obtained with the \texttt{tbabs*powerlaw} model. The normalization (Norm) is expressed in $\mathrm{keV^{-1}\,cm^{-2}\,s^{-1}}$, while the flux is given in $\mathrm{erg\,cm^{-2}\,s^{-1}}$.} \\

\label{tab:tb_po} \\
\toprule                 
 & & & & & &  &  \\ 

Obs Id & MJD & $\Gamma$  & Norm($10^{-2})$ & $\log F_{0.3-10} $ & $\log F_{0.3-2}$ & $\log F_{2-10} $ & $\chi^2/dof$\\ 
\midrule
\endfirsthead

\caption*{Table \thetable\ (continued)} \\
\toprule                 
 &  &  &    & &   \\  
Obs Id & MJD & $\Gamma$  & Norm($10^{-2})$ & $\log F_{0.3-10} $ & $\log F_{0.3-2}$  & $\log F_{2-10} $ & $\chi^2/dof$\\ 
\midrule
\endhead

\midrule
\multicolumn{8}{r}{\textit{Continued on next page}} \\
\midrule
\endfoot

\bottomrule
\endlastfoot

33282001&56794&$2.59\pm0.11$&$2.18\pm0.14$& $-9.97\pm0.03$   &$-10.08_{-0.03}^{+0.02} $  & $-10.61_{-0.08}^{+0.07}$  & 23.30/21          \\  
33282002&56794&$2.60\pm0.08$&$0.75\pm0.04$&$-10.43\pm0.02$&$-10.55\pm0.02$&$-11.08_{-0.06}^{+0.05}$&81.70/46         \\ 
35062001&53479&$2.78\pm0.04$&$1.62\pm0.04$&$-10.09\pm0.01$&$-10.18\pm0.01$&$-10.86\pm0.03$&215.19/144         \\ 
35062002&53509&$2.50\pm0.04$&$1.13\pm0.03$&$-10.26\pm0.01$&$-10.39\pm0.01$&$-10.84\pm0.02$&205.48/170          \\ 
35062003&53713&$2.68\pm0.04$&$2.04\pm0.04$&$-10.01\pm0.01$&$-10.10\pm0.01$&$-10.70_{-0.03}^{+0.02}$&201.26/167         \\ 
81687001&57164&$2.58\pm0.07$&$1.12\pm0.40$&$-10.26\pm0.02$&$-10.37\pm0.01$&$-10.90\pm0.04$& 140.98/75          \\ 
92237001&57513 &$2.52\pm0.08$&$1.58\pm0.09$& $ -10.12\pm0.02$ &  $-10.24\pm0.02$  & $-10.71\pm0.05$&64.47/46\\                                                                
92237002&57521 &$2.36\pm0.08$&$1.82\pm0.09$& $ -10.04_{-0.02}^{+0.02}$ &  $-10.20\pm0.02$  & $-10.55_{-0.05}^{+0.05}$&44.32/48                                      \\
92237003&57529&$2.55\pm0.08$ &$1.56\pm0.06$&$ -10.12\pm0.02$ &  $-10.24\pm0.02$  & $-10.73_{-0.06}^{+0.05}$ &48.89/51       \\
92237004&57537&$2.60\pm0.09$&$1.61\pm0.80$&$-10.10\pm0.02$&$-10.21\pm0.02$&$-10.75\pm0.06$&48.77/48 \\
92237005&57545&$2.54\pm0.07$&$2.59\pm0.13$&$-9.90\pm0.02$&$-10.02\pm0.01$&$-10.51\pm0.02$&56.52/59
\\
92237006&57553&$2.51_{-0.07}^{+0.08}$&$2.92\pm0.13$&$-9.84\pm0.02$&$-9.97_{-0.02}^{+0.01}$&$-10.44\pm0.05$&38.59/65\\
92237007&57561&$2.63_{-0.07}^{+0.08}$&$2.37\pm0.10$&$-9.93\pm0.02$&$-10.04\pm0.01$&$-10.60\pm0.05$&55.95/58 \\
92237008&57693&$2.55\pm0.10$&$1.59\pm0.09$&$-10.11_{-0.03}^{+0.02}$&$-10.23\pm0.02$&$-10.73_{-0.07}^{+0.06}$&46.87/37       \\
92237009&57700&$2.58\pm0.07$&$3.03\pm0.13$&$-9.83\pm0.02$&$-9.94\pm0.01$&$-10.47\pm0.04$&82.07/64             \\
92237010&57709&$2.67\pm0.02$&$2.19\pm0.16$&$-9.97\pm0.03$&$-10.07\pm0.03$&$-10.66_{-0.09}^{+0.08}$&30.73/23        \\
92237011&57714&$2.70\pm0.01$&$3.03\pm0.20$&$-9.83\pm0.03$&$-9.92_{-0.03}^{+0.02}$&$-10.54_{-0.09}^{+0.08}$&26.69/23                         \\ 92237012&57717&$2.54\pm0.08$&$2.42\pm0.10$&$-9.93\pm0.02$&$-10.05_{-0.02}^{+0.01}$&$-10.54\pm0.05$&38.07/56     \\
92237013&57725&$2.50\pm0.09$&$1.74\pm0.09$&$-10.07\pm0.02$&$-10.20\pm0.02$&$-10.66\pm0.06$&51.01/43            \\
92237014&57733&$2.55\pm0.09$&$1.90\pm0.10$&$-10.03\pm0.02$&$-10.02\pm0.02$&$-10.65_{-0.06}^{+0.05}$&39.55/44         \\ 92237015&57741&$2.67\pm0.09$&$1.70\pm0.09$&$-10.08\pm0.02$&$-10.18\pm0.02$&$-10.77_{-0.06}^{+0.05}$&51.88/40         \\ 92237016&57749&$2.56\pm0.10$&$1.32\pm0.08$&$-10.19\pm0.03$&$-10.31\pm0.02$&$-10.82_{-0.07}^{+0.06}$&47.69/32         \\
93158001&57877&$2.79\pm0.11$&$1.89\pm0.11$&$-10.03\pm0.03$&$-10.11\pm0.02$&$-10.80\pm0.07$&26.73/34          \\ 
93158002&57884&$2.62\pm0.08$&$1.98\pm0.09$&$-10.02_{-0.02}^{+0.02}$&$-10.12\pm0.02$&$-10.68\pm0.05$&51.56/53\\

93158003&57891 &$2.54\pm0.09$&$1.74\pm0.10$&$-10.07_{-0.03}^{+0.02}$&$-10.19\pm0.02$&$-10.68\pm0.06$&38.20/38        \\

93158004 & 57898   &$ 2.51\pm0.08$&$2.02\pm0.10$ & $-10.01\pm0.02$&$-10.13\pm0.02$&$-10.60\pm0.05$&67.71/52              \\
93158005 & 57905   & $2.57_{-0.01}^{+0.02}$ &$1.23\pm0.10$  &$ -10.22\pm0.03 $  &$-10.33\pm0.03$&$-10.85_{-0.10}^{+0.09}$& 24.88/22        \\
93158007 &57919     &$2.89\pm0.16$ &$1.40\pm0.12$&  $-10.15\pm0.04$&$-10.21\pm0.03$&$-10.98_{-0.11}^{+0.10}$  & 16.03/18     \\

93158008 &57926 &$2.45_{-0.06}^{+0.07}$&$2.89\pm0.12$&$-9.85\pm0.02$  &$-9.99\pm0.01$& $ -10.41\pm0.04$&69.92/69           \\
93158010 &58067   & $ 2.67\pm0.10$&$1.33\pm0.07$  & $-10.19\pm0.02 $&$-10.28\pm0.02$   &$-10.88_{-0.07}^{+0.06}$& 40.87/42       \\
93158011 &58071     & $ 2.62_{-0.08}^{+0.09}$ &$2.90\pm0.09$& $-10.03\pm0.02$   &$-10.14\pm0.02$&$-10.69_{-0.06}^{+0.05}$& 58.87/48      \\
93158012 &58078     & $ 2.53\pm0.07$ & $2.10\pm0.09$&$-9.99\pm0.02$&$-10.11_{-0.02}^{+0.01}$   &$-10.59\pm0.05$& 39.76/58      \\
93158013 &58085     & $2.48\pm0.10$&$1.44\pm0.09$ &$-10.15\pm0.03$&$-10.28\pm0.02$&$-10.73_{-0.07}^{+0.06}$   & 20.32/33   \\
93158014 & 58092    & $2.60\pm0.10$ &$1.16\pm0.07$& $ -10.25_{-0.03}^{+0.02}$&$-10.36\pm0.02$&$-10.90_{-0.07}^{+0.06}$ & 29.93/36                \\
93158016 & 58102    & $2.66_{-0.08}^{+0.09} $ &$2.09\pm0.10$     & $-9.99\pm0.02 $&$-10.09\pm0.02$& $-10.68\pm0.06$ & 53.64/49                                          \\
93158017 & 58106    & $2.61_{-0.14}^{+0.15}$ &$1.49\pm0.12$     & $-10.14\pm0.03 $ &$-10.25\pm0.03$   & $-10.79_{-0.10}^{+0.09}$&31.90/21           \\
94000001 & 58246      &$2.51\pm0.19$ &$0.85\pm0.09$& $ -10.37_{-0.05}^{+0.04}$  & $-10.50\pm0.04$&$-10.97_{-0.12}^{+0.11}$&14.60/12                  \\
94000002 & 58253    &$2.64\pm0.03$ &$1.68\pm0.20$&$-10.08_{-0.06}^{+0.05}$&$-10.19_{-0.05}^{+0.04}$& $ -10.76_{-0.21}^{+0.20}$  & 7.54/6          \\
94000003 & 58256   & $2.81_{-0.09}^{+0.10}$ &$1.76\pm0.10$&$-10.06_{-0.03}^{+0.02}$   &$-10.13\pm0.02$&$-10.84_{-0.07}^{+0.08}$& 50.53/37      \\
94000005 &58267   & $2.74_{-0.12}^{+0.13}$ &$0.99\pm0.70$ & $-10.31\pm0.03$&$-10.39\pm0.02$   &$-11.05\pm0.09$ &33.55/25                                            \\
94000006 & 58274    & $2.67_{-0.09}^{+0.10}$ &$1.81\pm0.10$ & $-10.05_{-0.03}^{+0.02}$   &$-10.15\pm0.10$ &$-10.74\pm0.06$&19.93/40             \\
94000007 & 58281 &$2.58\pm0.10$&$1.83\pm0.11$&$-10.05_{-0.03}^{+0.02}$  &$-10.16\pm0.02$ &$-10.68_{-0.07}^{+0.06}$&43.15/37             \\
94000008 & 58288 & $ 2.58\pm0.02$&$1.03\pm0.08$&$-10.30_{-0.04}^{+0.03}$& $-10.41\pm0.03$&$-10.94_{-0.10}^{+0.09}$  & 15.41/18          \\
94000009 & 58425    & $2.54_{-0.10}^{+0.11}$      &$1.24\pm0.07$ &$-10.22_{-0.03}^{+0.02}$& $ -10.34\pm0.02   $    &$-10.83_{-0.07}^{+0.06}$ &45.81/37         \\
94000011 & 58436   & $ 2.63\pm0.10$&$1.51\pm0.08$ &$-10.13\pm0.02$  &$-10.24\pm0.02$&$-10.80_{-0.07}^{+0.06}$&52.54/41             \\
94000012 & 58439    & $2.70\pm0.10$ &$1.34\pm0.08$& $ -10.18_{-0.03}^{+0.02}$ &$-10.28\pm0.10$&$-10.89_{-0.07}^{+0.06}$&43.07/37              \\

94000013 & 58446    & $2.85_{-0.11}^{+0.12}$ &$0.94\pm0.06$ &$-10.32\pm0.03$ &$-10.40\pm0.02$ &$-11.13\pm0.08$ &32.30/32        \\
94000015 & 58457    & $2.66\pm0.09$ &$1.88\pm0.09$ &$ -10.03\pm0.02$ &$-10.13\pm0.02$ &$-10.72\pm0.06$&65.19/50              \\
94000016 & 58460    & $2.49\pm0.07$ & $2.71\pm0.11$ &$ -9.88\pm0.02$ &$-10.01\pm0.01$ &$-10.46\pm0.04$  & 58.08/65         \\
94000017 & 58467    & $2.59_{-0.08}^{+0.09}$ &$2.16\pm0.12$ &$ -9.88\pm0.02$ &$-10.09\pm0.02$ &$-10.62_{-0.06}^{+0.05}$&41.74/41             \\
94000018 & 58474   & $2.50_{-0.14}^{+0.15}$ &$1.13\pm0.08$ &$ -10.26\pm0.03$ &$-10.39_{-0.03}^{+0.02}$ &$-10.85\pm0.09$&28.61/24             \\
95000001 & 58611    & $2.65_{-0.10}^{+0.01}$&$1.30\pm0.08$&$-10.19\pm0.02$&$-10.30\pm0.02     $ &$-10.88_{-0.07}^{+0.06}$ & 31.23/33          \\
95000003 &58625    & $ 2.69_{-0.08}^{+0.09}$ &$1.59\pm0.08$&  $ -10.12\pm0.02$  & $-10.20\pm0.02$&$-10.81_{-0.06}^{+0.05}$&45.52/49          \\
95000004 &58632    & $2.60\pm0.09$ &$2.11\pm0.90$&  $ -9.99\pm0.02$ & $-10.10\pm0.02$&$-10.64\pm0.06$&35.45/44                                                 \\
95000005 & 58639 & $2.67\pm0.07$ &$2.88\pm0.12$ &$ -9.85\pm0.02$ &$-9.95\pm0.01$ &$-10.54_{-0.05}^{+0.04}$&68.07/68                                             \\
95000007 &58650 & $2.59_{-0.13}^{+0.14}$ &$0.83\pm0.62$ &$-10.39\pm0.03$&$-10.51_{-0.03}^{+0.02}$& $-11.03\pm0.09$   & 20.06/22  \\
95000008 &58653     & $2.60\pm0.20$      &$1.26\pm0.08$ & $ -10.21\pm0.03$ &$-10.32\pm0.02$&$-10.86\pm0.07$& 38.61/30 \\     

95000009 &58790    & $2.87\pm0.10$  &$1.56\pm0.09$    &  $ -10.10_{-0.03}^{+0.02}$  &$-10.18\pm0.02$ &$-10.92_{-0.07}^{+0.06}$&24.18/36   \\ 

95000010 &58797     & $2.75\pm0.11$ &$1.45\pm0.09$ & $-10.15\pm0.03$   & $-10.23\pm0.02$&$-10.89_{-0.08}^{+0.07}$&15.85/30           \\ 

95000011 & 58804    & $2.49\pm0.08$ &$2.18\pm0.10$ & $-9.97\pm0.02$  &$-10.10\pm0.02$ &$-10.56\pm0.05$&47.09/51                 \\ 

95000013 & 58818   & $2.69\pm0.11$&$1.18\pm0.07$  & $-10.24\pm0.03$   &$-10.33\pm0.02$&$-10.94_{-0.08}^{+0.07}$&29.49/32                  \\
95000014 & 58825    & $2.66_{-0.10}^{+0.11}$  &$1.71\pm0.10$&$-10.08_{-0.03}^{+0.02}$& $-10.18\pm0.02$  & $-10.76\pm0.07$&34.30/37                 \\
95000015 & 58832   & $2.49\pm0.12$  &$1.4\pm0.09$& $-10.16\pm0.02$  &$-10.29_{-0.03}^{+0.02}$ &$-10.75_{-0.08}^{+0.07}$&47.09/51                 \\
95000016 & 58839   & $2.59\pm0.10$  &$1.82\pm0.10$& $-10.05\pm0.03$  &$-10.17\pm0.02$ &$-10.70_{-0.07}^{+0.06}$&41.90/34                 \\
95653001 & 58977    & $2.76\pm0.12$&$1.56\pm0.10$& $-10.11\pm0.03$    &$-10.20\pm0.02$&$-10.86_{-0.08}^{+0.07}$&20.33/27           \\
95653003 &58989    & $2.69\pm0.12$ &$2.17\pm0.13$ &$-9.98\pm0.03$& $-10.07\pm0.02$  &$-10.68\pm0.08$ &34.57/33            \\
95653004 & 58991    & $2.88_{-0.20}^{+0.21}   $    &$0.79\pm0.082$&$ -10.39_{-0.05}^{+0.04}  $    &$-10.46\pm0.03$ &$-11.22_{-0.14}^{+0.13}$&14.07/12           \\
95653005 & 58998    & $2.39 _{-0.11}^{+0.12}$ &$1.23\pm0.80$ &$-10.22\pm0.03$   &$-10.37\pm0.02$&$-10.75_{-0.08}^{+0.07}$& 24.46/29     \\
95653006 &59005    & $2.65\pm0.09$  &$1.80\pm0.09$ &$-10.05\pm0.02$&$-10.16\pm0.02$ &$-10.73\pm0.06$& 35.54/40        \\
95653007 &59012     & $2.46\pm0.07$  &$2.50\pm0.12$ &$-9.91\pm0.02$ &$-10.05\pm0.01$ &$-10.48\pm0.05$&70.73/64        \\

95653008 &59019     & $2.59\pm0.12$ &$1.76\pm0.12$ & $ -10.06\pm0.03$&$-10.18\pm0.02$ &$-10.70_{-0.08}^{+0.07}$  & 28.98/28                                          \\
95653009 & 59156    & $2.63\pm0.10$ & $1.87\pm0.12$&$-10.04\pm0.03$&$-10.14\pm0.02$&$-10.70_{-0.07}^{+0.06}$ & 35.54/31                                         \\
95653010 & 59163    & $2.78_{-0.14}^{+0.15}$&$1.09\pm0.08$  &$ -10.27_{-0.04}^{+0.03}$ &$-10.35\pm0.03$ &$-11.03_{-0.10}^{+0.09}$&29.86/21                                        \\
95653011 & 59170    & $2.46\pm0.10 $  &$1.39\pm0.94$    & $-10.16\pm0.03 $   &$-10.30\pm0.21$& $-10.73_{-0.07}^{+0.06}$&25.63/27                                         \\
95653012 & 59177    & $2.60\pm0.09$ &$1.80\pm0.96$ &$-10.05\pm0.02$ &$-10.17_\pm0.02$  &$-10.70\pm0.06$& 50.74/43              \\
95653013 &59184    & $2.58\pm0.01$&$1.36\pm0.08$ &$-10.18_{-0.03}^{+0.02}$  &$-10.29\pm0.02$&$-10.82\pm0.07$ &52.86/36             \\
95653014 & 59191    & $2.59\pm0.09$ &$2.08\pm0.10$     &$-9.99\pm0.02$ &$-10.10\pm0.02$ &$-10.64\pm0.06$      & 56.17/46                      \\
95653016 & 59202   & $2.67_{-0.08}^{+0.09}$ &$2.48\pm0.12$& $ -9.91\pm0.02$   &$-10.02\pm0.02$&$-10.60\pm0.04$& 59.24/47           \\
95653017 &59205    & $2.54\pm0.10$ &$2.76\pm0.16$& $ -9.87_{-0.03}^{+0.02}$   &$-9.99\pm0.02$&$-10.48\pm0.06$& 32.32/37             \\
96050001 &59326    & $2.76_{-0.18}^{+0.19}$ &$2.09\pm0.18$  &$ -9.98\pm0.04 $&$-10.07\pm0.03$ &$-10.73_{-0.12}^{+0.11}$& 20.67/16                  \\

96050002 &59332     & $2.55\pm0.19$ &$0.65\pm0.60$  &$ -10.49\pm0.04$ &$-10.61_{-0.04}^{+0.03}$&$-11.11_{-0.13}^{+0.11}$& 13.24/13                       \\

96113001 & 59328    & $2.78\pm0.10$&$0.97\pm0.60$ & $-10.32_{-0.03}^{+0.02}$  &$-10.40\pm0.02$&$-11.08_{-0.07}^{+0.06}$& 62.42/38    \\        
96113002 & 59333   & $2.70\pm0.09$&$0.81\pm0.04$&  $-10.40\pm0.02$  & $-10.50\pm0.02$&$-11.11\pm0.06$&46.17/43      \\
96113003 & 59334   & $2.65_{-0.08}^{+0.09}$&$1.29\pm0.06$  &$-10.20\pm0.02$  &$-10.30\pm0.02$& $-10.88_{-0.06}^{+0.05}$&57.89/53       \\

\end{longtable}
}

{
\footnotesize
\centering
\begin{longtable}{l l l l l l l l l l }

\caption{Best fit parameters for the model \texttt{{tbabs*(bbody+powerlaw)}}. The normalisation (Norm) is given in units of $\mathrm{keV^{-1}cm^{-2}s^{-1}}$, while the flux is given in unit of erg\,cm$^{-2}$\,s$^{-1}$.} \\ 
\label{tab:tb_bbpo} \\
\toprule                 
 & & & Bbody & Powerlaw &  &  \\ 

Obs Id & MJD & $\Gamma$ & Norm $(10^{-4})$ & Norm ($10^{-2})$ & $\log F_{0.3-10} $ & $\log F_{0.3-2}$ & $\log F_{2-10} $ & $\chi^2/dof$\\ 
\midrule
\endfirsthead

\caption*{Table \thetable\ (continued)} \\
\toprule                 
 &  &  & Bbody & Powerlaw & &   \\  
Obs Id & MJD & $\Gamma$ & Norm $(10^{-4})$ & Norm $(10^{-2})$ & $\log F_{0.3-10} $ & $\log F_{0.3-2}$  & $\log F_{2-10} $ & $\chi^2/dof$\\ 
\midrule
\endhead

\midrule
\multicolumn{9}{r}{\textit{Continued on next page}} \\
\midrule
\endfoot

\bottomrule
\endlastfoot

33282001&56794&$2.59\pm0.07$&$<0.41$&$2.18\pm0.10$& $-9.97\pm0.02$   &$-10.08\pm0.02 $  & $-10.61\pm0.04$  & 23.33/20          \\  
33282002&56794&$2.50\pm0.09$&$0.54\pm0.50$&$0.7\pm0.05$&$-10.43_{-0.02}^{+0.01}$&$-10.55\pm0.02$&$-11.05\pm0.04$&80.56/45         \\ 
35062001&53479&$2.53\pm0.06$&$2.98\pm0.50$&$1.38\pm0.05$&$-10.08\pm0.01$&$-10.18\pm0.01$&$-10.77\pm0.02$&193.29/143         \\ 
35062002&53509&$2.35\pm0.04$&$1.13\pm0.30$&$1.03\pm0.03$&$-10.24\pm0.01$&$-10.38\pm0.01$&$-10.80\pm0.02$&190.89/169          \\ 
35062003&53713&$2.48\pm0.05$&$2.89\pm0.50$&$1.77\pm0.06$&$-9.99\pm0.01$&$-10.10\pm0.01$&$-10.64\pm0.02$&179.65/166         \\ 
81687001&57164&$2.34\pm0.08$&$1.95\pm0.40$&$0.95\pm0.05$&$-10.24\pm0.01$&$-10.37\pm0.01$&$-10.82\pm0.03$& 127.04/74          \\ 
92237001&57513 &$2.51_{-0.09}^{+0.06}$&$0.16\pm0.90$&$1.57_{-0.10}^{+0.07}$ &$ -10.11\pm0.02$ &  $-10.24\pm0.02$  & $-10.71_{-0.02}^{+0.04}$&64.44/45 \\                                                                
92237002&57521 &$2.33\pm0.08$&$<0.43$&$1.78_{-0.10}^{+0.09}$ &$ -10.04\pm0.02$ &  $-10.20\pm0.02$  & $-10.54\pm0.04$&44.12/47                                     \\
92237003&57529&$2.36_{-0.09}^{+0.10}$ &$1.97\pm0.90$&$1.39\pm0.09$&$ -10.10\pm0.02$ &  $-10.23 _{-0.02}^{+0.01}$  & $-10.67\pm0.04$ &44.58/50       \\
92237004&57537&$2.60\pm0.07$&$<0.49$&$1.61_{-0.07}^{+0.05}$&$-10.10\pm0.01$&$-10.21\pm0.02$&$-10.75\pm0.04$&48.78/47 \\
92237005&57545&$2.47\pm0.09$&$1.26\pm0.14$&$2.48\pm0.15$&$-9.89\pm0.01$&$-10.02\pm0.01$&$-10.49\pm0.04$&55.87/58
\\
92237006&57553&$2.48_{-0.09}^{+0.06}$&$<0.57$&$2.87_{-0.17}^{+0.13}$&$-9.84\pm0.01$&$-9.97_{-0.02}^{+0.01}$&$-10.43_{-0.03}^{+0.04}$&38.49/64\\
92237007&57561&$2.63_{-0.08}^{+0.04}$&$<0.12$&$2.37_{-0.13}^{+0.06}$&$-9.93\pm0.01$&$-10.04\pm0.01$&$-10.60_{-0.03}^{+0.04}$&55.95/57 \\
92237008&57693&$2.49\pm0.12$&$<0.75$&$1.52_{-0.12}^{+0.11}$&$-10.10\pm0.02$&$-10.23\pm0.02$&$-10.71\pm0.05$&46.54/36       \\
92237009&57700&$2.58_{-0.06}^{+0.04}$&$<0.06$&$3.04_{-0.12}^{+0.08}$&$-9.83\pm0.01$&$-9.94\pm0.01$&$-10.47\pm0.03$&82.07/63             \\
92237010&57709&$2.40\pm0.02$&$3.74\pm0.20$&$1.89\pm0.21$&$-9.95\pm0.02$&$-10.07\pm0.02$&$-10.56\pm0.08$&28.24/22        \\
92237011&57714&$2.70\pm0.02$&$4.23\pm0.30$&$2.67_{-0.28}^{+0.30}$&$-9.82\pm0.02$&$-9.93\pm0.02$&$-10.47_{-0.08}^{+0.07}$&25.15/22                         \\ 92237012&57717&$2.44\pm0.09$&$1.62\pm0.13$&$2.27\pm0.14$&$-9.92\pm0.01$&$-10.05_{-0.02}^{+0.01}$&$-10.51\pm0.04$&36.76/55     \\
92237013&57725&$2.42\pm0.01$&$<0.91$&$1.65_{-0.12}^{+0.11}$&$-10.06\pm0.02$&$-10.20\pm0.02$&$-10.63\pm0.05$&50.49/42            \\
92237014&57733&$2.45\pm0.10$&$1.32\pm0.10$&$1.79_{-0.12}^{+0.13}$&$-10.02\pm0.02$&$-10.15\pm0.02$&$-10.62\pm0.04$&38.47/43         \\ 92237015&57741&$2.62_{-0.01}^{+0.09}$&$<0.67$&$1.64_{-0.12}^{+0.09}$&$-10.08\pm0.02$&$-10.18\pm0.02$&$-10.75\pm0.05$&51.64/39         \\ 92237016&57749&$2.56\pm0.10$&$<0.18$&$1.30_{-0.10}^{+0.06}$&$-10.19\pm0.02$&$-10.31\pm0.02$&$-10.81\pm0.05$&47.66/31         \\
93158001&57877&$2.76_{-0.15}^{+0.13}$&$<0.46$&$1.84_{-0.20}^{+0.10}$&$-10.03\pm0.02$&$-10.11\pm0.02$&$-10.79_{-0.05}^{+0.06}$&26.67/33          \\ 
93158002&57884 &$2.60_{-0.09}^{+0.08}$ &$<0.34$&$1.95_{-0.12}^{+0.08}$ &$-10.02\pm0.01$ &$-10.12\pm0.02$&$-10.67\pm0.04$&51.49/52\\

93158003&57891 &$2.54_{-0.09}^{+0.06}$ &$<0.10$&$1.73_{-0.10}^{+0.05}$ &$-10.07\pm0.02$&$-10.19\pm0.02$&$-10.68\pm0.04$&38.20/37        \\

93158004 & 57898   &$ 2.41\pm0.09$&$1.50\pm0.11$ &$1.89\pm0.11$& $-10.00\pm0.02$&$-10.13\pm0.02$&$-10.57\pm0.04$&67.71/51              \\
93158005 & 57905   & $2.48_{-0.02}^{+0.02}$ &$<0.70$  &$1.18_{-0.12}^{+0.10}$&$ -10.21_{-0.02}^{+0.03} $  &$-10.33\pm0.03$&$-10.82\pm0.08$& 24.63/21        \\
93158007 &57919     &$2.81_{-0.29}^{+0.14}$ &$<0.79$&$1.36_{-0.19}^{+0.12}$ & $-10.14\pm0.02$&$-10.21\pm0.03$&$-10.95_{-0.09}^{+0.12}$  & 15.95/17     \\

93158008 &57926 &$2.45_{-0.07}^{+0.04}$ &$<4.05$&$2.89_{-0.14}^{+0.07}$ &$-9.85\pm0.01$  &$-9.99\pm0.01$& $ -10.41\pm0.03$&69.92/68           \\
93158010 &58067   & $ 2.60_{-0.14}^{+0.11}$&$<0.65$ &$1.28_{-0.10}^{+0.08}$ &$-10.18\pm0.02 $&$-10.29\pm0.02$   &$-10.86\pm0.06$& 40.59/41      \\
93158011 &58071     & $ 2.53_{-0.09}^{+0.10}$ &$1.30\pm0.12$&$1.79_{-0.11}^{+0.12}$& $-10.03_{-0.02}^{+0.01}$   &$-10.14\pm0.02$&$-10.66\pm0.04$& 57.84/47      \\
93158012 &58078     & $ 2.53_{-0.07}^{+0.07}$ & $2.10\pm0.86$&$2.01\pm0.12$&$-9.99\pm0.02$&$-10.11_{-0.02}^{+0.01}$   &$-10.59\pm0.05$& 39.76/58      \\
93158013 &58085     & $2.48_{-0.12}^{+0.10}$&$<0.29$ &$1.41_{-0.10}^{+0.07}$&$-10.15\pm0.02$&$-10.28\pm0.02$&$-10.72_{-0.04}^{+0.05}$   & 20.25/32   \\
93158014 & 58092    & $2.61_{-0.07}^{+0.06}$ &$<0.2$&$1.15_{-0.05}^{+0.04}$ &$ -10.25_{-0.02}^{+0.01}$&$-10.36\pm0.02$&$-10.90\pm0.04$ & 29.94/35               \\
93158016 & 58102    & $2.56_{-0.11}^{+0.12} $ &$<1.48$&$1.97\pm0.14$& $-9.98\pm0.02 $&$-10.09\pm0.02$& $-10.64\pm0.05$ &52.75/48                                          \\
93158017 & 58106    & $2.38_{-0.18}^{+0.19}$ &$2.16\pm0.20$     &$1.31_{-0.14}^{+0.15}$& $-10.12\pm0.03 $ &$-10.25\pm0.02$   & $-10.71\pm0.08$&30.15/20           \\
94000001 & 58246      &$2.52_{-0.16}^{+0.11}$ &$<0.2$&$0.86_{-0.08}^{+0.05}$ &$ -10.37\pm0.03$  & $-10.50\pm0.04$&$-10.97\pm0.08$&14.60/11                  \\
94000002& 58253 & $2.17\pm0.60$ &$<3.64$ &$1.39_{-0.32}^{+0.36}$ &$-10.04_{-0.07}^{+0.11}$ &$-10.20\pm0.04$ & $ -10.55\pm0.30$  & 6.89/5          \\
94000003 & 58256   & $2.66_{-0.13}^{+0.12}$ &$2.10\pm0.20$&$1.60\pm0.13$&$-10.05\pm0.02$   &$-10.14\pm0.02$&$-10.79\pm0.05$& 48.83/36      \\
94000005 &58267    & $2.67_{-0.20}^{+0.13}$ &$<5.12$ &$0.96_{-0.10}^{+0.07}$& $-10.31\pm0.02$&$-10.39\pm0.02$   &$-11.02_{-0.19}^{+0.13}$ &33.38/24                                            \\
94000006 & 58274    & $2.63_{-0.11}^{+0.08}$ &$<5.12$ &$1.77_{-0.13}^{+0.09}$ &$-10.05\pm0.02$   &$-10.15\pm0.02$ &$-10.73_{-0.02}^{+0.05}$&19.81/39             \\
94000007 & 58281 &$2.44_{-0.12}^{+0.13}$&$1.82\pm0.14$&$1.67_{-0.13}^{+0.14}$&$-10.04\pm0.02$  &$-10.17\pm0.02$ &$-10.64\pm0.05$&41.59/36             \\
94000008 & 58288 & $ 2.35\pm0.20$&$1.59\pm0.10$&$0.89\pm0.10$&$-10.28\pm0.03$& $-10.41\pm0.03$&$-10.86_{-0.08}^{+0.07}$  & 13.40/17          \\
94000009 & 58425    & $2.34_{-0.11}^{+0.10}$      &$1.70\pm0.70$ &$1.09\pm0.08$&$-10.20\pm0.02$& $ -10.30\pm0.02   $    &$-10.77\pm0.05$ &41.92/36         \\
94000011 & 58436   & $ 2.37_{-0.11}^{+0.12}$&$2.58\pm0.10$ &$1.29\pm0.10$&$-10.12\pm0.02$  &$-10.24\pm0.02$&$-10.70\pm0.05$&47.32/40             \\
94000012 & 58439   & $2.49\pm0.13$ &$1.96\pm0.10$&$1.18_{-0.10}^{+0.09}$& $ -10.17\pm0.02$ &$-10.28\pm0.02$&$-10.82\pm0.05$&39.91/36              \\

94000013 & 58446    & $2.66_{-0.15}^{+0.16}$ &$1.37\pm0.10$ &$0.84_{-0.08}^{+0.09}$&$-10.32\pm0.02$ &$-10.40\pm0.02$ &$-11.07\pm0.06$ &30.61/31        \\
94000015 & 58457    & $2.41_{-0.10}^{+0.11}$ &$3.29\pm0.10$ &$1.61\pm0.11$&$ -10.02_{-0.02}^{+0.01}$ &$-10.14\pm0.02$ &$-10.64_{-0.05}^{+0.04}$&58.78/49              \\
94000016 & 58460    & $2.48_{-0.04}^{+0.05}$ & $<0.80$ &$2.71_{-0.09}^{+0.06}$&$ -9.88\pm0.02$ &$-10.01\pm0.01$ &$-10.46\pm0.02$  & 58.08/64         \\
94000017 & 58467    & $2.59_{-0.07}^{+0.05}$ &$<0.06$ &$2.16_{-0.10}^{+0.07}$&$ -9.88\pm0.01$ &$-10.09\pm0.02$ &$-10.62\pm0.03$&41.75/40             \\
94000018 & 58474  & $2.06_{-0.14}^{+0.15}$ &$2.81\pm0.70$ &$0.87\pm0.09$&$ -10.21\pm0.03$ &$-10.38\pm0.03$ &$-10.69\pm0.07$&19.21/23             \\
95000001 & 58611    & $2.66\pm0.06$&$<1.44$&$1.30\pm0.05$&$-10.19\pm0.02$&$-10.30\pm0.02     $ &$-10.88\pm0.04$ & 31.24/32          \\
95000003 &58625    & $ 2.51\pm0.11$ &$1.96\pm0.10$&$1.43\pm0.10$&  $ -10.10_{-0.02}^{+0.01}$  & $-10.21\pm0.02$&$-10.75\pm0.05$&42.75/48          \\
95000004 &58632    & $2.57_{-0.12}^{+0.09}$ &$<0.42$& $2.07_{-0.15}^{+0.10}$& $ -9.99\pm0.02$ & $-10.10\pm0.02$&$-10.63\pm0.05$&35.39/43                                                 \\
95000005 & 58639    & $2.61\pm0.08$ &$<1.44$ &$2.77\pm0.15$&$ -9.85\pm0.01$ &$-9.95\pm0.01$ &$-10.52_{-0.04}^{+0.03}$&67.37/67                                             \\
95000007 &58650 & $2.47_{-0.15}^{+0.16}$ &$<0.65$ &$0.7\pm0.08$&$-10.38\pm0.02$&$-10.51\pm0.02$& $-10.99\pm0.07$   & 19.47/21  \\
95000008 &58653     & $2.54_{-0.11}^{+0.13}$      &$<0.58$ &$1.21_{-0.10}^{+0.09}$ &$ -10.21\pm0.02$ &$-10.32\pm0.02$&$-10.84\pm0.05$& 38.34/29 \\      
95000009 &58790    & $2.80\pm0.13$  &$<0.85$    & $1.28_{-0.12}^{+0.13}$& $ -10.10\pm0.02$  &$-10.18\pm0.02$ &$-10.90\pm0.05$&23.89/35   \\

95000010 &58797     & $2.54\pm0.20$ &$2.19\pm0.90$ &$1.28_{-0.12}^{+0.13}$& $-10.13\pm0.02$   & $-10.23\pm0.02$&$-10.81_{-0.07}^{+0.06}$&13.68/29           \\ 

95000011 & 58804    & $2.46\pm0.02$ &$<0.59$ &$2.12_{-0.13}^{+0.10}$&$-9.97_{-0.02}^{+0.01}$  &$-10.10\pm0.02$ &$-10.55\pm0.04$&46.9/50                 \\

95000013 & 58818   & $2.53_{-0.14}^{+0.15}$&$1.28\pm0.90$  &$1.07_{-0.10}^{+0.09}$& $-10.23\pm0.02$   &$-10.33\pm0.02$&$-10.89\pm0.06$&28.18/31                  \\
95000014 & 58825    & $2.59_{-0.16}^{+0.12}$  &$<0.84$&$1.64_{-0.14}^{+0.10}$&$-10.07\pm0.02$& $-10.18\pm0.02$  & $-10.73\pm0.02$&34.06/36                 \\
95000015 & 58832   & $2.19_{-0.13}^{+0.12}$  &$2.61\pm0.90$&$1.15\pm0.10$& $-10.13\pm0.02$  &$-10.29\pm0.02$ &$-10.65\pm0.05$&26.34/30                 \\
95000016 & 58839    & $2.53\pm0.11$  &$<0.92$&$1.73_{-0.14}^{+0.13}$& $-10.05\pm0.02$  &$-10.16\pm0.02$ &$-10.68\pm0.05$&41.53/33                 \\
95653001 & 58977    & $2.76\pm0.10$&$<0.42$&$1.56_{-0.09}^{+0.06}$& $-10.11\pm0.02$    &$-10.20\pm0.02$&$-10.86\pm0.04$&20.33/26           \\
95653003 &58989    & $2.41\pm0.20$ &$3.70\pm0.20$ &$1.87_{-0.18}^{+0.19}$&$-9.95\pm0.02$& $-10.07\pm0.02$  &$-10.57\pm0.07$ &31.77/32            \\
95653004 & 58991    & $2.52\pm0.41   $    &$1.67\pm0.17$&$0.67_{-0.12}^{+0.14}$&$ -10.38_{-0.03}^{+0.04}  $    &$-10.46\pm0.03$ &$-11.08_{-0.17}^{+0.16}$&13.25/11           \\
95653005 & 58998    & $2.31 _{-0.14}^{+0.13}$ &$<0.67$&$1.16\pm0.10$ &$-10.21\pm0.02$   &$-10.37\pm0.02$&$-10.72\pm0.06$& 24.01/28     \\
95653006 &59005     & $2.65_{-0.08}^{+0.05}$  &$<0.25$ &$1.79_{-0.08}^{+0.06}$&$-10.05_{-0.02}^{+0.01}$&$-10.16\pm0.02$ &$-10.73\pm0.04$& 35.54/39        \\
95653007 &59012     & $2.37_{-0.08}^{+0.09}$  &$1.44\pm0.12$ &$2.37\pm0.13$&$-9.90\pm0.01$ &$-10.05\pm0.01$ &$-10.45\pm0.04$&69.56/63        \\

95653008 &59019     & $2.36\pm0.14$ &$2.68\pm0.15$ &$1.53_{-0.14}^{+0.15}$& $ -10.05\pm0.02$&$-10.18\pm0.02$ &$-10.63_{-0.06}^{+0.05}$  & 26.03/27                                          \\
95653009 & 59156   & $2.63\pm0.06$ & $<0.60$&$1.87_{-0.08}^{+0.07}$&$-10.04\pm0.02$&$-10.14\pm0.02$&$-10.70\pm0.04$ & 35.54/30                                         \\
95653010 & 59163    & $2.43\pm0.20$&$2.47\pm0.10$  &$0.89_{-0.10}^{+0.11}$&$ -10.25\pm0.02$ &$-10.35\pm0.02$ &$-10.91\pm0.09$&26.53/20                                        \\
95653011 & 59170    & $2.46_{-0.11}^{+0.06} $  &$<0.80$    &$1.39_{-0.10}^{+0.11}$& $-10.16\pm0.02 $   &$-10.30\pm0.21$& $-10.73\pm0.04$&25.63/26                                         \\
95653012 & 59177    & $2.43\pm0.11$ &$2.14\pm0.10$ &$1.62\pm0.11$&$-10.04\pm0.02$ &$-10.16\pm0.02$  &$-10.65\pm0.04$& 47.82/42              \\
95653013 &59184    & $2.38\pm0.13$&$1.86\pm0.90$&$1.20\pm0.10$ &$-10.16\pm0.02$  &$-10.29\pm0.02$&$-10.75\pm0.05$ &49.92/35             \\
95653014 & 59191    & $2.52\pm0.02$ &$1.05\pm0.10$&$1.99_{-0.12}^{+0.13}$     &$-9.99\pm0.02$ &$-10.10\pm0.02$ &$-10.64\pm0.05$      & 55.73/45              &         \\
95653016 & 59202   & $2.65_{-0.07}^{+0.11}$ &$<0.45$& $2.44_{-0.16}^{+0.11}$&$ -9.91\pm0.01$   &$-10.02\pm0.02$&$-10.60\pm0.04$& 59.18/46           \\
95653017 &59205    & $2.54_{-0.12}^{+0.06}$ &$<0.79$& $2.77_{-0.19}^{+0.10}$&$ -9.87\pm0.02$   &$-9.99\pm0.02$&$-10.49_{-0.04}^{+0.06}$& 32.32/36             \\
96050001 &59326    & $2.68_{-0.17}^{+0.32}$ &$<1.14$  &$2.01_{-0.31}^{+0.19}$&$ -9.98\pm0.02 $&$-10.07\pm0.03$ &$-10.70_{-0.14}^{+0.10}$& 20.60/15                  \\

96050002 &59332     & $2.25\pm0.30$ &$1.15\pm0.1$  &$0.55\pm0.08$&$ -10.47\pm0.03$ &$-10.61\pm0.03$&$-11.00_{-0.11}^{+0.09}$& 11.48/12                       \\

96113001 & 59328   & $2.78\pm0.10$&$0.97\pm0.01$ & $0.97^{+0.03}_{-0.04}$ &$-10.32_{-0.03}^{+0.02}$  &$-10.40\pm0.02$&$-11.08_{-0.07}^{+0.06}$& 62.42/38    \\        
96113002 & 59333   & $2.61\pm0.12$&$0.81\pm0.40$&$0.76\pm0.06$&  $-10.40\pm0.02$  & $-10.50\pm0.02$&$-11.08\pm0.04$&45.47/42     \\
96113003 & 59334   & $2.58_{-0.12}^{+0.11}$&$<0.65$  &$1.24_{-0.09}^{+0.08}$&$-10.19\pm0.01$  &$-10.30\pm0.02$& $-10.86\pm0.05$&57.46/52       \\

\end{longtable}
}

{
\footnotesize
\centering
\begin{longtable}{l l l l l l l l l  }

\caption{
Best-fit parameters obtained from fitting with the model \texttt{tbabs*Simpl*NthComp}. The flux is expressed in $\mathrm{erg\,cm^{-2}\,s^{-1}}$ and the unit of $kT_{e}$ is keV.}\\

\label{tab:tb_smnc} \\
\toprule

Obs Id & MJD  & $\Gamma_{Simpl}$ & $f_{\mathrm{sc}}$ &  $\Gamma_{NthComp}$ & $kT_{e}$&$\chi^2/dof$ &  $\log F_{0.3-10}$  \\
\midrule
\endfirsthead

\caption{Table \thetable\ (continued)} \\
\toprule

Obs Id & MJD  &   $\Gamma_{Simpl}$ & $f_{\mathrm{sc}}$ &  $\Gamma_{NthComp}$ & $kT_{e}$& $\chi^2/dof$ &  $\log F_{0.3-10}$  \\ 
\midrule
\endhead

\midrule
\multicolumn{7}{r}{\textit{Continued on next page}} \\
\midrule
\endfoot

\bottomrule
\endlastfoot

00033282001 & 56794 & $2.59\pm0.10$ & $0.93_{-0.02}^{+0.04}$ & $1.89_{-0.23}^{+0.67}$ &$0.17_{-0.01}^{+0.02}$ &15.10/18 & $-10.02_{-0.04}^{+0.05}$
\\
00095653011 & 59170 & $1.99^{+0.48}_{-0.99}$ & $0.44_{-0.25}^{+0.50}$ & $1.04_{-1.0}^{+0.68}$ & $0.13_{-0.03}^{+0.09}$&21.90/18 & $-10.18\pm0.03$  
\\
00095653005 & 58998 & $1.95_{-0.15}^{+1.09}$ & $0.95_{-0.001}^{+0.46}$ & $2.40_{-0.86}^{+0.34}$ &$0.27_{-0.02}^{+0.03}$& 22.02/26 & $-10.21_{-0.03}^{+0.02}$ 
\\
00095653008 & 59019 & $2.33\pm0.64$ & $0.22_{-0.002}^{+0.003}$ & $2.13_{-0.16}^{+0.13}$ &$0.14_{-0.39}^{+0.42}$ &26.22/25 & $-10.03_{-0.04}^{+0.03}$
\\
00092237010 & 57710 & $2.23_{-0.29}^{+0.31}$ & $0.32_{-0.27}^{+0.09}$ & $<1.37$ & $0.18_{-0.06}^{+0.69}$&25.35/20 & $-9.97_{-0.04}^{+0.05}$
\\
00092237011 & 57714 & $2.31_{-0.30}^{+0.32}$ & $0.31_{-0.24}^{+0.09}$ & $1.27_{-1.00}^{+1.49}$ &$ 0.13_{-0.03}^{+0.13}$&20.97/20 & $-9.83_{-0.04}^{+0.05}$
\\
00094000008 & 58288 & $2.24_{-1.24}^{+0.39}$ & $0.34_{-0.33}^{+0.31}$  & $1.19_{-1.00}^{+3.44}$ &$0.19_{-0.04}^{+1.60}$ &12.09/15 & $-10.29_{-0.04}^{+0.01}$
\\
00093158005 & 57905 & $2.24_{-0.54}^{+0.51}$ & $0.36_{-0.26}^{+0.38}$ & $<3.17$ &$0.13_{-0.05}^{+0.08}$ &18.72/19 & $-10.04\pm0.03$
\\
00092237007 & 57561 & $2.58^{+0.16}_{-1.57}$ & $>0.02$ & $2.59^{+0.10}_{-0.16}$ & $0.13_{-0.36}^{+0.41}$&55.46/55 & $-9.94_{-0.02}^{+0.18}$
\\
00095000011 & 58804 & $2.13_{-0.56}^{+0.37}$ & $0.24_{-0.20}^{+0.28}$ & $<2.18$ &$0.17_{-0.06}^{+0.12}$ &33.68/48 & $-9.97\pm0.03$
\\
00095653012 & 59177 & $2.33_{-0.13}^{+0.23}$ & $0.36_{-0.14}^{+0.13}$ & $1.00_{-1.00}^{+0.84}$ &$0.17_{-0.02}^{+0.04}$& 41.91/40 & $-10.06_{-0.03}^{+0.02}$
\\
00095653006 & 59005 & $2.14_{-1.14}^{+0.64}$ & $>0.79$ & $2.28_{-0.50}^{+0.54}$ &$0.28_{-0.35}^{+0.15}$ &31.42/37 & $-10.03\pm0.03$
\\
00093158002 & 57884 & $2.54_{-0.13}^{+0.21}$ & $0.44_{-0.19}^{+0.24}$ & $<1.45$ & $0.12_{-0.02}^{+0.06}$&39.29/50 & $-9.83_{-0.04}^{+0.05}$ 
\\
00095000009 & 58790 & $2.64_{-0.48}^{+0.37}$ & $0.28_{-0.20}^{+0.37}$ & $<2.60$ & $0.15_{-0.05}^{+0.12}$&18.65/33 & $-10.12_{-0.01}^{+0.03}$
\\
00095653013 & 59184 & $2.27_{-0.30}^{+0.31}$ & $0.35_{-0.22}^{+0.17}$ & $1.03_{-1.00}^{+1.04}$ &$0.12_{-0.04}^{+0.07}$ &44.50/33 & $-10.17\pm0.03$
\\
00092237008 & 57693 & $2.09^{+0.42}_{-0.67}$ & $0.13^{+0.31}_{-0.12}$ & $2.08^{+0.69}_{-0.88}$ &$0.21_{-0.09}^{+0.01}$ &42.12/34 & $-10.10\pm0.03$
\\
00095653014 & 59191 & $2.36_{-0.49}^{+0.34}$ & $0.36_{-0.26}^{+0.17}$ & $1.24_{-1.00}^{+0.80}$ & $0.13_{-0.02}^{+0.08}$&45.17/43 & $-10.04\pm0.03$
\\
00093158001 & 57877 & $2.36_{-1.36}^{+0.57}$ & $0.21_{-0.15}^{+0.37}$ & $1.66_{-1.66}^{+0.78}$  & $0.16_{-0.06}^{+0.18}$&19.18/31 & $-10.04\pm0.04$
\\
00095653009 & 59156 & $1.87_{-0.87}^{+0.76}$ & $0.08_{-0.07}^{+0.27}$ & $1.78_{-0.57}^{+0.39}$ & $0.23\pm0.08$&18.81/28 & $-10.04\pm0.04$
\\
00094000012 & 58439 & $2.35_{-0.32}^{+0.31}$ & $0.32_{-0.17}^{+0.13}$ & $<1.90$ & $0.12_{-0.18}^{+0.05}$&30.86/34 & $-10.18\pm0.03$
\\ 
00095653001 & 58977 & $2.69^{+0.95}_{-1.28}$ & $>0.01$ & $<3.36$ &$0.19_{-0.05}^{+0.19}$ &17.75/24 & $-10.13_{-0.05}^{+0.06}$
\\
00093158016 & 58102 & $2.43_{-0.40}^{+0.29}$ & $0.31_{-0.23}^{+0.29}$ & $<2.51$ &$0.14_{-0.04}^{+0.12}$& 48.01/46 & $-9.99\pm0.03$
\\
00094000011 & 58436 & $2.54_{-0.13}^{+0.21}$ & $0.44_{0.19}^{+0.24}$ & $<1.45$ &$0.15_{-0.13}^{+0.25}$& 39.29/50 & $-10.11\pm0.03$
\\
00094000005 & 58267 & $1.85_{-0.85}^{+0.65}$ & $0.14_{-0.11}^{+0.19}$ & $1.43_{-1.42}^{+0.74}$ &$0.16_{-0.04}^{+0.09}$ &21.44/22 & $-10.28_{-0.06}^{+0.08}$
\\
00092237003 & 57529 & $2.33_{-0.18}^{+0.19}$ & $0.40_{-0.20}^{+0.16}$ & $<2.39$ &$0.11_{-0.02}^{+0.06}$ &42.25/48 & $-10.11\pm0.02$
\\
00092237016 & 57749 & $2.46_{-0.30}^{+0.28}$ & $0.40_{-0.13}^{+0.24}$ & $<1.52$ &$0.13_{-0.14}^{+0.03}$ &33.77.29 & $-10.21\pm0.03$
\\
00093158004 & 57898 & $2.28_{-0.20}^{+0.22}$ & $0.38_{-0.10}^{+0.12}$ & $<1.53$ & $0.12_{-0.02}^{+0.03}$&52.03/49 & $-10.05\pm0.03$
\\
00092237006 & 57553 & $2.38^{+0.19}_{-1.26}$ & $0.40^{+0.59}_{-0.39}$ & $<2.27$ & $0.17_{-0.06}^{+0.29}$&34.91/62 & $-9.85\pm0.02$
\\
00092237009 & 57700 & $2.60_{-0.17}^{+0.15}$ & $>0.36$ & $1.89_{-0.23}^{+0.67}$ & $0.11_{-0.02}^{+0.03}$&69.18/61 & $-9.85\pm0.02$
\\
00092237015 & 57741 & $2.33_{-0.32}^{+0.30}$ & $0.28_{-0.16}^{+0.17}$ & $1.33_{-1.32}^{+0.72}$ & $0.15_{-0.03}^{+0.06}$&39.63/37 & $-10.09\pm0.03$
\\
00092237013 & 57725 & $2.12_{-1.11}^{+0.36}$ & $0.16_{-0.14}^{+0.34}$ & $2.17_{-0.89}^{+1.02}$ &$0.21_{-0.10}^{+0.21}$ &47.33/40 & $-10.05\pm0.03$
\\
00094000009 & 58425 & $2.31_{-0.21}^{+0.20}$ & $0.40_{-0.30}^{+0.14}$  & $<2.51$ &$0.11_{-0.03}^{+0.09}$ &40.79/34 & $-10.21_{-0.03}^{+0.04}$
\\
00092237004 & 57537 & $2.54\pm1.5$ & $>0.07$ & $>1.82$ & $0.17_{-0.02}^{+0.06}$&48.68/45 &  $-10.10\pm0.03$
\\
00094000006 & 58274 & $2.53_{-0.42}^{+0.19}$ & $0.40_{-0.33}^{+0.56}$ & $1.47_{-1.06}^{+1.40}$ &$0.13_{-0.01}^{+0.18}$ &16.59/37 & $-10.08_{-0.03}^{+0.04}$
\\
00093158017 & 58106 & $2.03_{-0.66}^{+0.47}$ & $0.28_{-0.25}^{+0.15}$ & $1.01_{-1.00}^{+1.21}$  &$0.13_{-0.03}^{+0.11}$ &23.55/18 & $-10.13_{-0.05}^{+0.07}$ 
\\
00092237014 & 57733 & $2.28_{-0.27}^{+0.25}$ & $0.34_{-0.16}^{+0.15}$ & $1.15_{-1.12}^{+0.76}$ &$0.13_{-0.02}^{+0.05}$ &25.51/41 & $-10.04\pm0.03$
\\
00035062003 & 53713 & $2.29\pm0.08$ & $0.23_{-0.06}^{+0.05}$ & $0.14_{-0.01}^{+0.02}$ & $0.14_{-0.02}^{+0.03}$&144.7/165 & $-10.00\pm0.01$
\\
00035062001 & 53479 & $2.25\pm0.10$ & $0.14_{-0.05}^{+0.06}$ & $0.16_{-0.01}^{+0.02}$ & $0.16_{-0.02}^{+0.03}$&167.4/142 & $-10.08\pm0.01$ 
\\
00035062002 & 53509 & $2.28\pm0.06$ & $0.33_{-0.09}^{+0.07}$ & $0.13_{-0.01}^{+0.33}$ &$0.13_{-0.02}^{+0.33}$ &180.4/168 &  $-10.25\pm0.01$
\\
00081687001 & 57164 & $2.15_{-0.13}^{+0.13}$ & $0.28_{-0.04}^{+0.03}$ & $<1.24$ & $0.13_{-0.02}^{+0.01}$&87.56/73 & $-10.25\pm0.02$
\\
00033282002 & 56794 & $2.51_{-1.99}^{+0.22}$ & $0.26_{-0.06}^{+0.04}$ & $<1.37$ & $0.14_{-0.01}^{+0.02}$&46.08/44 & $-10.44\pm0.02$
\\
00095000016 & 58839 & $2.49_{-1.48}^{+0.21}$ & $>0.07$ & $>0.71$ &$0.13_{-0.05}^{+0.08}$ &40.42/31 & $-10.06\pm0.03$
\\
00095653007 & 59012 & $2.34_{-1.88}^{+0.14}$ & $0.41_{-0.39}^{+0.27}$ & $>0.77$ &$0.13_{-0.01}^{+0.34}$& 68.54/61 & $-9.91\pm0.02$
\\
00092237005 & 57545 & $2.42^{+1.15}_{-1.27}$ & $>0.05$ & $2.34^{+0.15}_{-0.13}$ & $0.16_{-0.03}^{+0.02}$&55.48/56 & $-9.89^{+0.03}_{-0.02}$
\\
00094000018 & 58474 & $2.01_{-0.25}^{+0.28}$ & $0.31_{-0.02}^{+0.10}$ & $<2.87$ & $0.11_{-0.02}^{+0.15}$&18.36/21 & $-10.21_{-0.04}^{+0.05}$
\\
00095000001 & 58611 & $1.97_{-0.81}^{+1.25}$ & $0.06_{-0.03}^{+0.43}$ & $2.11_{-0.39}^{+0.34}$ &$0.30_{-0.13}^{+0.20}$ &23.19/30 & $-10.20_{-0.04}^{+0.05}$
\\
00094000017 & 58467 & $2.60_{-0.13}^{+0.18}$ & $<0.71$ & $<3.33$ &$0.11_{-0.02}^{+0.08}$ &35.36/38 & $-10.01_{-0.02}^{+0.03}$
\\
00093158014 & 58092 & $2.67^{+0.20}_{-0.25}$ & $>0.17$ & $<2.83$ & $0.33_{-0.10}^{+1.32}$&26.86/33 & $-10.26_{-0.03}^{+0.06}$
\\
00093158010 & 58067 & $2.28_{-0.54}^{+0.38}$ & $0.21_{-0.18}^{+0.25}$ & $<2.63$ &$0.16_{-0.05}^{+0.14}$& 35.81/39 & $-10.18\pm0.03$
\\
00095000015 & 58832 & $1.94_{-0.33}^{+0.28}$ & $0.21\pm0.19$ & $1.52_{-1.53}^{+1.31}$ & $0.15_{-0.04}^{+0.17}$&21.16/28 & $-10.13\pm0.04$
\\
00095000003 & 58625 & $2.06_{-0.72}^{+0.41}$ & $0.11_{-0.10}^{+0.23}$ & $1.90_{-1.91}^{+0.60}$ &$0.18_{-0.07}^{+0.17}$ &32.12/46 & $-10.09\pm0.03$
\\
00094000013 & 58446 & $2.68_{-0.23}^{+0.20}$ & $<0.87$  & $2.67^{+1.67}_{-1.32}$ & $0.09\pm0.01$&30.03/29 & $-10.32\pm0.03$
\\
00094000015 & 58457 & $2.25^{+0.25}_{-0.31}$ & $0.15^{+0.27}_{-0.12}$ & $<3.06$ &$0.16_{-0.18}^{+0.03}$& 56.50/47 & $-10.01\pm0.03$
\\
00096113001 & 59328 & $2.84^{+0.43}_{-1.49}$ & $>0.05$ & $<2.17$ &$0.14_{-0.04}^{+0.15}$ &54.77/35 & $-10.29\pm0.04$
\\
00095653010 & 59163 & $1.99_{-0.99}^{+0.48}$ & $<0.91$ & $<3.02$ &$0.18_{-0.06}^{+0.69}$ &21.90/18 & $-10.24_{-0.02}^{+0.04}$
\\
00096050002 & 59332 & $1.25_{-0.01}^{+1.23}$ & $0.40_{-0.05}^{+0.54}$ & $1.87_{-0.12}^{+0.73}$ &$0.17_{-0.06}^{+0.40}$ &15.03/13 & $-10.37_{-0.06}^{+0.05}$
\\
00093158006 & 57912 & $2.64\pm0.23$ & $0.50_{-0.19}^{+0.50}$ & $<3.22$ & $0.11_{-0.04}^{+0.07}$&33.06/34 & $-10.09\pm0.03$
\\
00092237001 & 57513 & $2.44\pm0.20$ & $0.43\pm0.08$ & $1.00_{-1.00}^{+0.49}$ & $0.12\pm0.02$&49.39/44 & $-10.13\pm0.30$
\\
00092237002 & 57521 & $2.28_{-0.13}^{+0.15}$ & $0.51_{-0.19}^{+0.09}$ & $1.00_{-1.00}^{+1.25}$ &$0.11_{-0.01}^{+0.09}$ &40.67/46 & $-10.05_{-0.01}^{+0.02}$
\\

00093158007 & 57919 & $1.54_{-0.02}^{+2.82}$ & $0.39_{-0.02}^{+0.84}$ & $1.91_{-0.14}^{+0.11}$ &$0.20_{-0.03}^{+0.17}$ &11.45/15 & $-10.08_{-0.02}^{+0.10}$
\\
00093158011 & 58071 & $2.52\pm0.17$ & $0.49_{-0.19}^{+0.28}$ & $<3.31$ &$0.09_{-0.03}^{+0.05}$ &55.45/45 & $-10.04\pm0.02$
\\
00093158012 & 58078 & $2.44_{-0.10}^{+0.13}$ & $>0.07$ & $>1.52$ & $0.11_{-0.04}^{+0.63}$&37.95/55 & $-9.98\pm0.04$
\\
00093158013 & 58085 & $2.39_{-1.39}^{+0.23}$ & $0.43_{-0.40}^{+0.59}$ & $>1.65$&$0.13_{-0.03}^{+0.01}$ & 18.87/30 & $-10.12_{-0.07}^{+0.04}$
\\
00094000016 & 58460 & $2.48_{-0.14}^{+0.13}$ & $>0.16$ & $1.83_{-0.42}^{+1.33}$ & $0.14_{-0.04}^{+0.02}$&55.55/62 & $-9.89_{-0.04}^{+0.02}$
\\
00094000017 & 58467 & $2.60_{-0.13}^{+0.18}$ & $>0.33$ & $1.00_{-1.0}^{+2.33}$ & $0.11_{-0.02}^{+0.08}$&35.56/38 & $-10.01\pm0.03$
\\
00095000003 & 58625 & $2.06_{-0.73}^{+0.41}$ & $0.12_{-0.11}^{+0.24}$ & $1.90_{-1.91}^{+0.60}$ & $0.18_{-0.07}^{+1.71}$&32.12/46 & $-10.09\pm0.04$
\\
00095000004 & 58632 & $2.37_{-1.41}^{+0.99}$ & $<0.24$ & $2.22_{-1.17}^{+6.42}$ &$0.15_{-0.04}^{+0.07}$ &33.85/41 & $-9.98_{-0.04}^{+0.05}$
\\

00095000007 & 58650 & $2.48_{-0.23}^{+0.24}$ & $>0.04$ & $>0.67$ &$0.10\pm0.06$ &19.41/19 & $-10.41\pm0.05$
\\
00095000013 & 58818 & $2.45_{-0.33}^{+0.36}$ & $0.35_{-0.19}^{+0.27}$ & $1.41_{-1.41}^{+1.19}$ &$0.12_{-0.04}^{+0.13}$ &26.09/29 & $-10.25_{-0.03}^{+0.05}$
\\
00095630013 & 59184 & $2.27\pm0.31$ & $0.36_{-0.22}^{+0.17}$ & $1.03_{-1.00}^{+1.04}$ & $0.12_{-0.04}^{+0.07}$&44.50/33 & $-10.17_{-0.03}^{+0.04}$
\\
00096113002& 59333 & $2.47_{-0.43}^{+0.31}$ & $0.32_{-0.24}^{+0.29}$ & $<2.35$ & $0.13_{-0.04}^{+0.11}$&38.80/40 & $-10.41_{-0.30}^{+0.02}$
\\
00092237012 & 57717 & $2.30_{-0.22}^{+0.19}$ & $0.38_{-0.26}^{+0.24}$ & $<2.55$ & $0.13_{-0.05}^{+0.09}$&34.22/53 & $-9.93\pm0.03$
\\
00093158008 & 57926 & $2.42\pm0.10$ & $0.56_{-0.19}^{+0.11}$ & $<2.82$ &$0.16_{-0.06}^{+0.09}$ &77.88/71 & $-9.84_{-0.01}^{+0.02}$
\\
\end{longtable}
}

\section{Results}
\label{4}

We carried out spectral analysis of 85 Swift/XRT observations of Ark 564 in the 0.3–10 keV band. All spectra show a clear soft X-ray excess, which was initially modelled using a blackbody component. To investigate the physical origin of the soft excess, we applied a warm corona model, which provided statistically improved fits across the full dataset. The best-fitting parameters from the warm corona modelling indicate electron temperatures in the range $\sim 0.1$–0.3 keV, while the photon index of the primary continuum varies between $\Gamma \sim 1.3$–2.8. In addition, two representative observations corresponding to the highest and lowest flux states were fitted with relativistic reflection model 
to explore alternative explanations. Although reflection models also provided acceptable fits, the ionization parameter and reflection fraction remained poorly constrained. Overall, the warm corona scenario offers a consistent description of the soft excess across the observed flux states.

The X-ray fluxes of the individual spectral components were estimated in the 0.3--10 keV energy range using the XSPEC convolution model \texttt{cflux}. The total unabsorbed flux was obtained from the best-fitting model \texttt{tbabs*Simpl*NthComp}, while the flux corresponding to the \texttt{NthComp} component was calculated separately in the same energy range. Since \texttt{Simpl} is a convolution Comptonization model that scatters a fraction of the seed photons into a power-law component, the flux associated with the \texttt{Simpl} component was estimated using

\begin{equation}
F_{\rm Simpl} = F_{\rm tot} - (1-f_{\rm sc})F_{\rm nthcomp},
\end{equation}

where $F_{\rm tot}$ is the total unabsorbed flux of the model, $F_{\rm nthcomp}$ is the flux of the \texttt{NthComp} component, and $f_{\rm sc}$ is the scattering fraction parameter of \texttt{Simpl}.

The luminosity corresponding to each spectral component was then calculated using L = 4$\pi D^2F$, 
%
%
assuming a luminosity distance of $D = 98.5$ Mpc for Ark~564. 
%
%
%
From the estimated 0.3-10 keV luminosities, we computed the X-ray Eddington ratio defined as $\lambda_{\rm Edd,~0.3-10 keV}=L_{0.3-10 keV}/L_{\rm Edd}$, where the Eddington luminosity is given by
\begin{equation}
L_{\rm Edd} = 1.26 \times 10^{38}
\left(\frac{M_{\rm BH}}{M_\odot}\right)
~{\rm erg~s^{-1}}
\end{equation}
%
Averaging over all \textit{Swift} observations and adopting the black hole mass of $M_{\rm BH} = 2.6 \times 10^{6}\,M_{\odot}$ obtained from optical virial estimates \citep{2005MNRAS.356..789B} yields a mean X-ray Eddington ratio of $\langle \lambda_{\rm Edd,~0.3-10 keV} \rangle \approx 0.27$, indicating that Ark~564 accretes at a significant fraction of its Eddington limit. 

\subsection{X-ray variability and correlations}

We investigated the variability of various spectral parameters and the flux values by fitting a constant to the best-fit values obtained from different observations using $\chi^ 2 $ analysis. The reduced $\chi^ 2 $ values are listed in Table~\ref{tab:chi2}. The flux components in the 0.3-10 keV, 0.3-2 keV, and 2-10 keV bands exhibit large reduced $\chi^ 2 $ values ($\chi^ 2_{v} >1$), confirming strong flux variability across observations. The scattered fraction ($f_{sc}$) also shows significant variability. In contrast, the photon index $\Gamma_{Simpl}$ displays a reduced $\chi_\nu ^ 2 $ value below unity, indicating that its variability is comparatively modest within statistical uncertainties. Given the pronounced variability in the flux components, shown in Figures~\ref{fig:evolution} and ~\ref{fig:flux-mjd}, we focus on examining correlations between fluxes and spectral parameters.


\begin{table*}[!htbp]

    \caption{Reduced $\chi^2$ values for constant model fits applied to the spectral parameters.}
    \label{tab:chi2}
    \centering
    \begin{tabular}{c c c}
    \toprule
        Parameter  & $\chi^2_{v}$\\
        \midrule
        $\Gamma_{pl}$ & 4.27 \\
       $\Gamma_{Simpl}$ & 0.64 \\
       $\log F_{0.3-10}$  & 28.68 \\
       $\log F_{0.3-2}$  & 33.49 \\
       $\log F_{2-10}$  & 5.64 \\
       $f_{\mathrm{sc}}$  & 13.31 \\
       $\log F_{0.3-2}^{\mathrm{nthcomp}}$  & 2.81\\
       $\log F^{Simpl}_{0.3-10}$ & 22.08 \\

        \bottomrule
    \end{tabular}
    
\end{table*}

\begin{table*}
  \caption{Fractional Variability of luminosity in different energy bands.}
 \label{tab:fvar}
 \centering
 
 \begin{tabular}{ccc}
 \toprule
 Energy Band (keV) & Reduced $\chi^2$ & $F_{\mathrm{var}}$ \\
 \midrule
0.3-10  & 30.47 & $0.30 \pm 0.01$ \\
0.3-2   & 33.70 & $0.31 \pm 0.09$ \\
2-10    & 7.29  & $0.19 \pm 0.04$ \\
 \bottomrule
 \end{tabular}
\end{table*}

We also quantified the intrinsic variability using the fractional variability amplitude $F_{\mathrm{var}}$ \citep{2003MNRAS.345.1271V}, defined as

\begin{equation}
F_{\mathrm{var}} =
\sqrt{\frac{S^2 - \langle \sigma^2_{\mathrm{err}} \rangle}{\langle x \rangle^2}},
\end{equation}

where $S^2$ is the sample variance of the light curve, $\langle \sigma^2_{\mathrm{err}} \rangle$ is the mean squared measurement uncertainty, and $\langle x \rangle$ is the mean flux. The calculated fractional variability amplitudes are $F_{\mathrm{var}} = 0.30 \pm 0.01$ in the 0.3-10 keV band, $0.31 \pm 0.09$ in the 0.3-2 keV band, and $0.19 \pm 0.04$ in the 2-10 keV band, mentioned in Table~\ref{tab:fvar}.  The larger variability amplitude observed in the soft band compared to the hard band suggests that the soft X-ray component contributes significantly to the overall variability of the source.

We further evaluated the relationships among the variable quantities listed in Tables~\ref{tab:pc} and ~\ref{tab:sc} using Pearson and Spearman correlation analyses. A strong positive correlation is found between the soft ($\log F_{0.3-2}$) and hard ($\log F_{2-10}$) X-ray fluxes, with very small null-hypothesis probabilities ($p < 10^{-10}$). This indicates coherent broadband variability, suggesting that the soft and hard X-ray emissions are driven by the same underlying physical process. In comparison, the correlations involving $\Gamma_{Simpl}$ and $f_{sc}$ are weaker and less statistically significant. While mild trends are present in some cases, their null-hypothesis probabilities are substantially larger than those of the flux–flux correlations.

\begin{table}[!htbp]
\centering
    \caption{Pearson Correlation between different parameters. Column 3: pearson’s rank order correlation coefficient. Column 4: p-value.}
    \label{tab:pc}
    \centering
    \begin{tabular}{c c c c c}
    \toprule
      Model & Parameter 1  & Parameter 2 & r & p \\
       \midrule
    & $\Gamma_{Pl}$ & $\log F_{0.3-10}$ & 0.09 & 0.36 \\
\texttt     { tbabs*(bbody+powerlaw)} & $\Gamma_{Pl}$ & $\log F_{0.3-2}$ & 0.21 & 0.05 \\ 
       &  $\Gamma_{Pl}$ & $\log F_{2-10}$ & -0.25 & 0.01 \\
       &  $\log F_{0.3-2}$ & $\log F_{2-10}$ & 0.81 & $9.58\times10^{-21}$ \\
        
        \midrule

       & $f_{\mathrm{sc}}$ & $\log F_{0.3-10}$ & -0.06 & 0.66 \\
       & $f_{\mathrm{sc}}$ & $\log F_{0.3-2}$ & -0.04 & 0.75 \\
       & $f_{\mathrm{sc}}$ & $\log F_{2-10}$ & 0.003 & 0.98 \\

     \texttt{tbabs*Simpl*NthComp}  &  $\log F_{0.3-2}$ & $\log F_{2-10}$ & 0.75 & $1.84\times10^{-11}$ \\
        & $\log F^{nthcomp}_{0.3-2}$ & $f_{\mathrm{sc}}$  & 0.17 & $0.21$\\
        & $\log F^{Simpl}_{0.3-10}$ & $\log F^{nthcomp}_{0.3-2}$ & 0.20 & 0.16 \\
        
        \bottomrule
    \end{tabular}
    
\end{table}

\begin{table}[!htbp]
\centering
    \caption{Spearman’s correlation between different parameters. Column 3: Spearman’s rank order correlation coefficient. Column 4: p-value.}
    \label{tab:sc}
    \centering
    \begin{tabular}{c c c c c}
         \toprule
     Model &  Parameter 1  & Parameter 2 & r & p \\
       \midrule
       &  $\Gamma_{Pl}$ & $\log F_{0.3-10}$ & 0.08 & 0.44\\
       \texttt{tbabs*(bbody+powerlaw)} & $\Gamma_{Pl}$ & $\log F_{0.3-2}$ & 0.18 & 0.09 \\
       &  $\Gamma_{Pl}$ & $\log F_{2-10}$ & -0.28 & 0.007 \\
       &  $\log F_{0.3-10}$ & $\log F_{2-10}$ &  0.78 & $1.01\times10^{-18}$ \\
        \midrule
         
       &  $f_{\mathrm{sc}}$ & $\log F_{0.3-10}$ & -0.09 & 0.53 \\
       &  $f_{\mathrm{sc}}$ & $\log F_{0.3-2}$ & 0.05 & 0.71\\
       &  $f_{\mathrm{sc}}$ & $\log F_{2-10}$ & -0.04 & 0.78 \\
        
      \texttt{tbabs*Simpl*NthComp} &  $\log F_{0.3-2}$ & $\log F_{2-10}$ & 0.74 & $2.13\times10^{-11}$ \\
       &  $\log F^{nthcomp}_{0.3-2}$ & $f_{\mathrm{sc}}$  & 0.02 & $0.84$\\
       & $\log F^{Simpl}_{0.3-10}$ & $\log F^{nthcomp}_{0.3-2}$ & 0.38 & 0.007 \\
        
        \bottomrule
    \end{tabular}
   
\end{table}

\section{Discussion and Summary}
\label{sum}

We analysed the Swift/XRT spectra of Ark~564 obtained between 2005 and 2021 in the 0.3-10\,keV band using various phenomenological and physically motivated models. The simplest model, an absorbed power law (\texttt{tbabs}*\texttt{powerlaw}), describes the hard continuum above $\sim$2\,keV adequately but leaves systematic positive residuals below $\sim$1-2\,keV, confirming the presence of a soft excess component in Ark~564. The existence of a soft excess in this source has been well established from earlier observations with ASCA \citep{2001ApJ...561..131T}, XMM-Newton \citep[e.g.][]{2004MNRAS.347..854V, 2007ApJ...671.1284D, 2016RAA....16..108E}, and it is a common feature in NLS1 galaxies and a significant fraction of the broader Seyfert population \citep[e.g.][]{2004MNRAS.349L...7G, 2009A&A...495..421B, 2012MNRAS.420.1825J,2017MNRAS.472.3492E}. 
The addition of a blackbody component to the power law (\texttt{tbabs*(bbody+powerlaw)}) improves the fit significantly, but this parameterisation is purely phenomenological. The blackbody temperatures typically obtained from such fits ($\sim$0.1-0.2\,keV) show a remarkable uniformity across AGN spanning several orders of magnitude in black hole mass, which is difficult to reconcile with direct thermal emission from the accretion disc \citep{2004MNRAS.349L...7G}. For Ark~564, with a black hole mass of $\sim 2.6\times10^{6}\,M_{\odot}$, the inferred blackbody temperature exceeds what the inner disc is expected to produce even after allowing for colour correction effects. Hence, while the blackbody model serves as a useful diagnostic, it does not provide a self-consistent physical picture of the soft excess emission.

A two-corona framework, in which a warm, optically thick ($\tau \sim 10$-20) corona with $kT_{\rm e} \sim 0.1$-1\,keV comptonizes the thermal disc photons to produce the soft excess, while a hot, optically thin corona gives rise to the hard X-ray power law, has been developed and applied extensively to AGN spectra \citep{2012MNRAS.420.1848D, 2018A&A...611A..59P}. In this work, the warm Comptonization model (\texttt{tbabs*Simpl*NthComp}), following the approach of \citet{2007ApJ...671.1284D}, \citet{2015MNRAS.448.1541S}, \citet{2016RAA....16..108E}, and \citet{2024RAA....24f5025A}, gives a statistically improved and physically self-consistent description of the broadband spectrum. 
The photon indices and Comptonisation parameters recovered from our spectral fits fall within the typical ranges reported for NLS1 galaxies studied with \textit{XMM-Newton} and \textit{NuSTAR} \citep{2015MNRAS.448.1541S, 2013A&A...549A..73P, 2018A&A...611A..59P, 2020A&A...640A..99M}. 
The consistency between our \textit{Swift}/XRT results and these earlier \textit{XMM-Newton} findings supports the warm corona interpretation.

We also applied a relativistic reflection model (\texttt{tbabs*relxillCp}) and a distant reflection model (\texttt{tbabs*xillverCp}), to the highest and lowest flux observations to test whether a reflection-dominated scenario could account for the observed soft excess. Blurred reflection from the inner accretion disc, where the soft excess is attributed to a blend of relativistically smeared emission lines produced by irradiation of the disc by the X-ray 
coronal continuum \citep{2005MNRAS.358..211R,2014ApJ...782...76G}, has been widely 
employed in AGN spectroscopy. \citet{2013MNRAS.434.1129K} detected reverberation soft lags of $\sim$100\,s in Ark~564, which clearly point to some contribution of blurred reflection in the soft X-ray band. In our fits, both \texttt{relxillCp} and \texttt{xillverCp} return formally acceptable statistics; however, key parameters such as the ionization parameter and reflection fraction remain weakly constrained. This is not unexpected, as the \textit{Swift}/XRT bandpass (0.3--10\,keV) lacks the high-energy coverage above 10\,keV that is needed to disentangle the reflection fraction from the coronal continuum slope and the high-energy cut-off \citep{2013MNRAS.428.2901W}. The recent \textit{NuSTAR} observation by \citet{2020MNRAS.492.3041B} extending up to 50\,keV has revealed one of the coolest coronae ($kT_{\rm e} \sim 15$\,keV) in Ark~564. Other broadband studies combining \textit{NuSTAR} and \textit{XMM-Newton} data \citep{2022ApJ...939..109L, 2023AN....34430042E} have confirmed the presence of a strong, relativistically broadened iron line, but found that reflection alone cannot fully account for the prominent soft excess, requiring an additional Comptonisation component.
Therefore, while we cannot rule out a reflection origin for the soft excess from our data alone, the present \textit{Swift}/XRT observations do not demand it either. Disentangling the contributions of the warm corona and blurred reflection would require broadband coverage extending to 50--80\,keV, as provided by \textit{NuSTAR}, and ideally simultaneous observations with \textit{XMM-Newton}.

Ark~564 shows substantial X-ray variability over the sixteen-year monitoring baseline. The 0.3-10\,keV luminosity varies between $\sim 4.2\times10^{43}$ and $\sim 9.5\times10^{43}\,\rm erg\,s^{-1}$, and the constant-flux hypothesis is strongly rejected in all bands, 
with reduced $\chi^{2}$ values of 28.68, 33.49, and 5.64 for the 0.3-10\,keV, 0.3-2\,keV, and 2--10\,keV bands, respectively. This leaves no ambiguity that the variability is intrinsic to the source. The fractional variability amplitudes, $F_{\rm var} \approx 0.30$ in the full band, $0.31$ in the soft band, and $0.19$ in the hard band, are consistent with the strong X-ray variability that is a hallmark of NLS1 galaxies \citep{1999ApJ...524..667T, 1999ApJS..125..297L}. The larger variability in the soft band compared to the hard band is in agreement with the behavior reported for Ark~564 by \citet{2001ApJ...561..131T}, who found that the soft hump varied by a factor of $\sim$6 over a 35-day ASCA observation compared to a factor of $\sim$4 in the power law. 
NLS1 galaxies exhibit the largest variability amplitudes among Seyferts at a given luminosity. This is understood as a consequence of their relatively low black hole masses and high accretion rates, which produce compact emission regions with short dynamical and radiative timescales.
 
Within the model framework adopted in this work, the thermal disc photons are first comptonized by the warm, optically thick corona, giving rise to the observed soft excess. A fraction of this warm corona emission is subsequently intercepted and upscattered by the hot, optically thin corona, producing the hard X-ray power law continuum. The strong variability observed in the scattering fraction across the monitoring period implies that the overall geometry of the coronal components undergoes considerable changes over time. Such geometrical variability is consistent with the findings of \citet{2016RAA....16..108E}, who suggested that the variability in Ark~564 is associated with changes in the geometry of the inner X-ray emitting region rather than with the X-ray luminosity alone. A similar interpretation has been invoked for other NLS1 galaxies, where the coronal geometry is thought to respond dynamically to fluctuations in the accretion rate \citep{2015MNRAS.449..129W, 2004MNRAS.349.1435M}. 

The mean 0.3-10\,keV luminosity of $\sim 8.8\times10^{43}\,\rm erg\,s^{-1}$ gives a mean X-ray Eddington ratio of $\lambda_{\rm Edd,0.3-10 keV} \approx 0.27$, placing Ark~564 among the high-accretion-rate AGN. High Eddington ratios favor the development of radiatively efficient, geometrically thin accretion flows where the disc extends close to the innermost stable circular orbit, providing ample seed photons for Comptonization. The high accretion rate also supports the warm corona picture, since theoretical work has shown that a warm, optically thick coronal layer can be sustained in radiative equilibrium above the disc at accretion rates typical of NLS1 galaxies \citep{2015A&A...580A..77R, 2020A&A...634A..85P}. 
However, the bolometric luminosity is likely higher than the value inferred from the 0.3--10\,keV band alone, given the considerable ultraviolet and extreme-ultraviolet contribution in high-accretion-rate sources. Applying a typical bolometric correction factor of 10--30 for NLS1 galaxies \citep{2007MNRAS.381.1235V} would place Ark~564 near the Eddington limit, consistent with earlier multi-wavelength and X-ray estimates \citep[e.g.,][]{2015MNRAS.448.1541S}.

Within the warm corona framework, however, the \texttt{Simpl} photon index $\Gamma_{\rm simpl}$ does not show significant variability across these observations, most likely a consequence of the limited quality of the individual \textit{Swift}/XRT spectra. 
Higher-quality long-term monitoring observations with \textit{XMM-Newton} or simultaneous broadband overage with \textit{NuSTAR} would be needed to investigate whether the softer-when-brighter behavior widely observed in Seyfert galaxies \citep{2001ApJ...547..684M, 2009MNRAS.399.1597S} is also encoded in the warm corona Comptonization parameters of Ark~564 on longer timescales.

\bmhead{Acknowledgements}
We sincerely acknowledge the Editor and the Reviewer for their valuable comments and suggestions, which have greatly improved the quality of the manuscript.
We acknowledge the use of public data from the \textit{Swift} data archive. This work has made use of the NASA/IPAC Extragalactic Database, which is operated by the Jet Propulsion Laboratory, California Institute of Technology and data obtained through the High Energy Astrophysics Science
Archive Research Center Online Service, provided by NASA/GSFC. SHE acknowledges the support from Mane Kancor Ingredients Pvt. Ltd. through its CSR initiative.

\section*{Data Availability}
The Swift/XRT data used in the study are freely available from the HEASARC Archive (\url{https://heasarc.gsfc.nasa.gov/docs/archive.html}).

\section*{Funding Declaration}
This research received no specific grant from any funding agency in the public,  commercial,  or not-for-profit sectors.


\newpage
\begin{appendices}

\section{Figures}\label{secA1}

\begin{figure}
    \centering
     \includegraphics[width=0.90\linewidth]{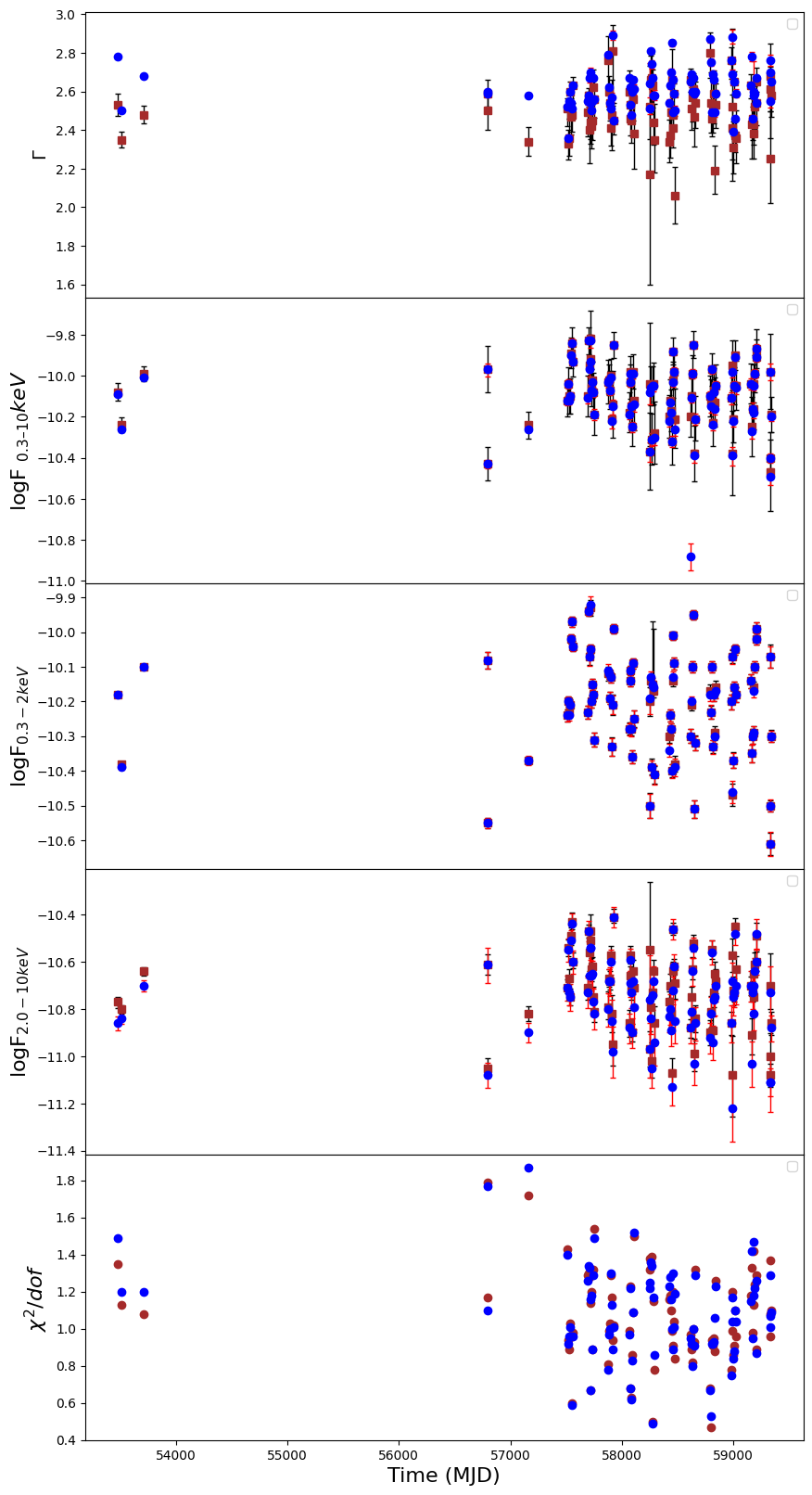}
 \caption{Comparison of the different parameters obtained from the \texttt{tbabs*(powerlaw+bbody)} model (brown) and the \texttt{tbabs*powerlaw} model  (blue).}
\end{figure}

{
\footnotesize  
\centering
\begin{longtable}{l l l l l l l }
\caption{Logarithmic flux values of the spectral components in the model \texttt{tbabs*Simpl*NthComp}. Flux is given in units of erg\,cm$^{-2}$\,s$^{-1}$.}    \\
\toprule
 & \multicolumn{3}{c}{tbabs*cflux*Simpl*NthComp} & \multicolumn{3}{c}{tbabs*Simpl*cflux*Nthcomp} \\
\cmidrule(lr){2-4} \cmidrule(lr){5-7}
Obs Id & $\log F_{0.3-10}$ & $\log F_{0.3-2.0}$ & $\log F_{2-10}$ 
& $\log F_{0.3-10}$ & $\log F_{0.3-2.0}$ & $\log F_{2-10}$ \\

\midrule
\endfirsthead

\caption{Table \thetable\ (continued)} \\
\toprule
 & \multicolumn{3}{c}{tbabs*cflux*Simpl*NthComp} & \multicolumn{3}{c}{tbabs*Simpl*cflux*Nthcomp} \\
\cmidrule(lr){2-4} \cmidrule(lr){5-7}
Obs Id & $\log F_{0.3-10}$ & $\log F_{0.3-2.0}$ & $\log F_{2-10}$ 
& $\log F_{0.3-10}$ & $\log F_{0.3-2.0}$ & $\log F_{2-10}$ \\

\midrule
\endhead

\midrule
\multicolumn{4}{r}{\textit{Continued on next page}} \\
\midrule
\endfoot

\bottomrule
\endlastfoot

00033282001 & $-10.02_{-0.04}^{+0.05}$ & $-10.12\pm0.04$ & $-10.72_{-0.06}^{+0.01}$ & $-10.02_{-1.55}^{+0.20}$ & $-10.09_{-1.35}^{+0.05}$ & $-10.86_{-1.63}^{+0.19}$
\\
00095653011 &   $-10.18\pm0.03$ & $-10.33_{-0.04}^{+0.03}$ & $-10.76_{-0.09}^{+0.10}$ & $-10.51_{-0.19}^{+0.10}$ & $-10.51_{-0.20}^{+0.10}$ & $<-12.60$ 
\\
00095653005 & $-10.21_{-0.03}^{+0.02}$ & $-10.37_{-0.03}^{+0.02}$ & $-10.58_{-0.16}^{+0.08}$ & $-10.44_{-0.01}^{+0.32}$ & $-10.45_{-0.78}^{+0.15}$ &  $<-10.52$
\\
00095653008 & $-10.03_{-0.04}^{+0.03}$ & $-10.16_{-0.06}^{+0.03}$ & $-10.60_{-0.04}^{+0.07}$ & $-10.54_{-0.13}^{+0.10}$ & $-10.54_{-0.13}^{+0.10}$ & $>-15.56$
\\
00092237010 & $-9.97_{-0.04}^{+0.05}$ & $-10.11_{-0.04}^{+0.06}$ & $-10.53_{-0.18}^{+0.16}$ & $-10.20_{-0.38}^{+0.41}$ & $-10.23_{-0.26}^{+0.12}$ & $<-12$
\\
00092237011 & $-9.83_{-0.04}^{+0.05}$ & $-9.96_{-0.04}^{+0.05}$ & $-10.44_{-0.15}^{+0.14}$ & $-10.08_{-0.22}^{+0.06}$ & $-10.08_{-0.22}^{+0.06}$ & $<-12.23$
\\
00094000008 & $-10.29_{-0.04}^{+0.01}$ & $-10.44_{-0.04}^{+0.07}$ & $-10.84_{-0.14}^{+0.20}$ & $-10.59_{-0.54}^{+0.34}$ & $-10.59_{-0.54}^{+0.05}$ & $<-12.13$ 
\\
00093158005 & $-10.04\pm0.03$ & $-10.18_{-0.02}^{+0.03}$ & $-10.59_{-0.08}^{+0.07}$ & $-10.48_{-0.28}^{+0.31}$ & $-10.48_{-0.21}^{+0.05}$ & $<-12.91$
\\
00092237007 & $-9.94_{-0.02}^{+0.18}$ & $-10.04\pm0.02$ & $-10.62_{-0.11}^{+0.08}$ & $<-9.03$ & $<-10.41$ & $<-10.70$
\\
00095000011 & $-9.97\pm0.03$ & $-10.12_{-0.03}^{+0.02}$ & $-10.50_{-0.09}^{+0.08}$ & $-10.29_{-0.11}^{+0.07}$ & $-10.29_{-0.17}^{+0.20}$ & $<-11.94$
\\
00095653012 &  $-10.06_{-0.03}^{+0.02}$ & $-10.19_{-0.02}^{+0.03}$ & $-10.64_{-0.08}^{+0.07}$ & $-10.34_{-1.00}^{+0.86}$ & $-10.34_{-0.15}^{+0.03}$ & $<-13.86$
\\
00095653006 &  $-10.03\pm0.03$ & $-10.16\pm0.02$ & $-10.63_{-0.16}^{+0.09}$ & $-10.31_{-0.46}^{+0.18}$ & $-10.33_{-0.07}^{+0.85}$ & $<-11.88$
\\
00093158002 &  $-9.83_{-0.04}^{+0.05}$ & $-9.96_{-0.04}^{+0.05}$ & $-10.44_{-0.15}^{+0.14}$ & $-10.33_{-0.14}^{+0.07}$ & $-10.33_{-0.14}^{+0.07}$ & $<-13.21$
\\
00095000009 & $-10.12_{-0.01}^{+0.03}$ & $-10.20\pm0.03$ & $-10.89_{-0.11}^{+0.10}$ & $-10.40_{-0.27}^{+0.08}$ & $-10.40_{-0.27}^{+0.08}$ & $<-11.98$
\\
00095653013 & $-10.17_\pm0.03$ & $-10.31\pm0.03$ & $-10.74_{-0.11}^{+0.10}$ & $-10.45_{-0.16}^{+0.03}$ & $-10.46_{-1.00}^{+1.05}$ & $<-14.00$
\\
00092237008 & $-10.10\pm0.03$ & $-10.24\pm0.03$ & $-10.65_{-0.10}^{+0.09}$ & $-10.46\pm0.17$ & $-10.45_{-0.28}^{+0.16}$ & $<-11.60$  
\\
00095653014 & $-10.04\pm0.03$ & $-10.13_{-0.02}^{+0.03}$ & $-10.64_{-0.08}^{+0.13}$ & $-10.27_{-0.15}^{+0.02}$ & $-10.28_{-0.14}^{+0.05}$ & $<-12.61$
\\
00093158001 & $-10.04\pm0.04$ &  $-10.14_{-0.04}^{+0.03}$ & $-10.72_{-0.17}^{+0.27}$ & $-10.29_{-0.18}^{+0.54}$ & $-10.29_{-0.18}^{+0.45}$ & $<-11.79$
\\
00095653009 & $-10.04\pm0.04$ & $-10.16\pm0.03$ & $-10.66_{-0.14}^{+0.13}$ & $-10.25_{-0.10}^{+0.28}$ & $-10.25_{-0.10}^{+0.48}$ & $<-11.54$
\\
00094000012 &  $-10.18\pm0.03$ & $-10.31_{-0.03}^{+0.04}$ & $-10.81_{-0.11}^{+0.10}$ & $-10.42_{-0.12}^{+0.02}$ & $-10.42_{-0.12}^{+0.03}$ & $<-13.51$
\\ 
00095653001 & $-10.13_{-0.05}^{+0.06}$ & $-10.22\pm0.04$ & $-10.88_{-0.16}^{+0.02}$ & $-10.54_{-0.37}^{+0.67}$ & $-10.54_{-0.38}^{+0.66}$ & $<-11.37$
\\
00093158016 & $-9.99\pm0.03$ & $-10.10_{-0.03}^{+0.02}$ & $-10.62_{-0.10}^{+0.09}$ & $-10.33_{-0.25}^{+0.09}$ & $-10.34_{-0.25}^{+0.10}$ & $<-12.52$
\\
00094000011 &  $-10.11\pm0.03$ &  $-10.24\pm0.03$ & $-10.68_{-0.09}^{+0.08}$ & $-10.54_{-0.29}^{+0.12}$ & $-10.54_{-0.30}^{+0.06}$ & $<-12.34$
\\
00094000005 & $-10.28_{-0.06}^{+0.08}$ & $-10.43_{-0.02}^{+0.04}$ & $-10.84_{-0.20}^{+0.22}$ & $-10.51_{-0.09}^{+0.06}$ & $-10.51_{-0.09}^{+0.06}$ & $<-12.22$
\\
00092237003 & $-10.11\pm0.02$ & $-10.25_{-0.02}^{+0.03}$ & $-10.68^{+0.07}_{-0.08}$ &$-10.45^{+0.04}_{-0.24}$ & $-10.45^{+0.04}_{-0.24}$ & $<-13.97$
\\
00092237016 & $-10.21\pm0.03$ & $-10.33\pm0.03$ & $-10.82_{-0.10}^{+0.09}$ & $-10.46_{-0.11}^{+0.05}$ & $-10.46_{-0.11}^{-0.05}$ & $<-13.51$
\\
00093158004 &  $-10.05\pm0.03$ & $-10.21\pm0.03$ & $-10.58_{-0.09}^{+0.08}$ & $-10.29_{-0.09}^{+0.02}$ & $-10.29_{-0.09}^{+0.03}$ & $<-14.10$
\\
00092237006 & $-9.85\pm0.02$ & $-9.99\pm0.02$ & $-10.42_{-0.07}^{+0.06}$ & $-10.27_{-0.30}^{+0.12}$ & $-10.27_{-0.31}^{+0.13}$ & $<-12.43$ 
\\
00092237009 & $-9.85\pm0.02$ &  $-9.97\pm0.02$ & $-10.49\pm0.06$ & $-10.20_{-0.21}^{+0.08}$ & $-10.20_{-0.15}^{+0.08}$ & $<-13.49$
\\
00092237015 & $-10.09\pm0.03$ & $-10.21\pm0.03$ & $-10.72_{-0.07}^{+0.09}$ & $-10.34_{-0.11}^{+0.04}$ & $-10.34_{-0.10}^{+0.04}$ & $<-12.81$
\\
00092237013 & $-10.05\pm0.03$ & $-10.21_{-0.03}^{+0.03}$ & $-10.58_{-0.09}^{+0.08}$ & $-10.47_{-0.37}^{+0.88}$ & $-10.47_{-0.41}^{+0.53}$ & $<-10.29$
\\
00094000009 &  $-10.21_{-0.03}^{+0.04}$ & $-10.36_{-0.03}^{+0.04}$ & $-10.76_{-0.08}^{+0.07}$ & $-10.55_{-0.33}^{+0.17}$ & $-10.55_{-0.33}^{+0.17}$ & $<-13.29$
\\
00092237004 & $-10.10\pm0.03$ & $-10.21\pm0.03$  & $-10.74\pm0.10$ & $-10.12^{+0.04}_{-0.02}$ & $-10.21^{+0.03}_{-0.02}$ &  $<-10.90$
\\
00094000006 & $-10.08_{-0.03}^{+0.04}$ & $-10.19_{-0.05}^{+0.04}$ & $-10.72\pm0.08$ & $-10.40_{-0.27}^{+0.16}$ & $-10.40_{-0.42}^{+0.15}$ & $<-12.10$
\\
00093158017 & $-10.13_{-0.05}^{+0.07}$ & $-10.28_{-0.04}^{+0.05}$ & $-10.63_{-0.16}^{+0.15}$ & $-10.37_{-0.15}^{+0.05}$ & $-10.37_{-0.15}^{+0.006}$ & $<-12.53$
\\
00092237014 & $-10.04\pm0.03$ & $-10.18_{-0.02}^{+0.03}$ & $-10.59_{-0.08}^{+0.07}$ & $-10.31_{-0.13}^{+0.04}$ & $-10.31_{-0.13}^{+0.05}$ & $<-13.16$
\\
00035062003 & $-10.00\pm0.01$ & $-10.12\pm0.01$ & $-10.61\pm0.03$ & $-10.31\pm0.04$ & $-10.31\pm0.04$ & $-10.68\pm0.06$
\\
00035062001 & $-10.08\pm0.01$ & $-10.19\pm0.01$ & $-10.73\pm0.04$ & $-10.38_{-0.04}^{+0.03}$ & $-10.38_{-0.04}^{+0.03}$ & $-13.73_{-0.72}^{+0.63}$ 
\\
00035062002 & $-10.25\pm0.01$ & $-10.39\pm0.01$ & $-10.79_{-0.03}^{+0.02}$  & $-10.68_{-0.10}^{+0.08}$ & $-10.68_{-0.09}^{+0.08}$ & $<-13.95$
\\
00081687001 & $-10.25\pm0.02$ & $-10.39_{-0.02}^{+0.01}$ & $-10.80_{-0.06}^{+0.05}$ & $-10.48_{-0.03}^{+0.02}$ & $-10.48_{-0.03}^{+0.02}$ & $-14.49_{-0.50}^{+0.48}$
\\
00033282002 & $-10.44\pm0.02$ & $-10.57\pm0.02$ & $-11.01_{-0.09}^{+0.08}$ & $-10.33_{-3.59}^{+0.16}$ & $-10.33_{-3.46}^{+0.16}$ & $-14.02_{-0.61}^{+0.46}$
\\
00095000016 & $-10.06\pm0.03$ & $-10.18_{-0.03}^{+0.02}$ & $-10.68\pm0.09$ & $-10.45_{-3.9}^{+0.12}$ & $-10.45_{-3.99}^{+0.12}$ & $>-16.94$
\\
00095653007 & $-9.91\pm0.02$ & $-10.06_{-0.02}^{+0.03}$ & $-10.45\pm0.06$ & $-10.35_{-0.53}^{+0.21}$ & $-10.35_{-0.44}^{+0.21}$ & $<-16.33$
\\
00092237005 & $-9.89^{+0.03}_{-0.02}$ & $-10.02\pm0.02$ & $-10.51^{+0.01}_{-0.09}$ & $-10.00^{+0.74}_{-0.23}$ & $-10.63^{+0.47}_{-5.45}$ & $<-11.19$
\\
00094000018 & $-10.21_{-0.04}^{+0.05}$ & $-10.39\pm0.04$ & $-10.67_{-0.11}^{+0.10}$ & $-10.54_{-0.26}^{+0.16}$ & $-10.54_{-0.26}^{+0.16}$ & $<-12.98$
\\
00095000001 & $-10.20_{-0.04}^{+0.05}$ & $-10.32\pm0.03$ & $-10.82_{-0.15}^{+0.14}$ & $-10.38_{-0.26}^{+0.10}$ & $-10.39_{-0.25}^{+0.17}$ & $<-11.42$ 
\\
00094000017 & $-10.01_{-0.02}^{+0.03}$ & $-10.12\pm0.03$ & $-10.64_{-0.07}^{+0.06}$ & $-10.35_{-0.24}^{+0.09}$ & $-10.35_{-0.24}^{+0.11}$ & $<-14.74$
\\
00093158014 & $-10.26_{-0.03}^{+0.06}$ & $-10.37\pm0.03$ & $-10.89_{-0.12}^{+0.20}$ & $-10.55_{-0.48}^{+0.41}$ & $-10.56_{-0.78}^{+0.22}$ & $<-11.21$
\\
00093158010 & $-10.18\pm0.03$ & $-10.30\pm0.03$ & $-10.79\pm0.11$ & $-10.51_{-0.12}^{+0.10}$ & $-10.51_{-0.24}^{+0.07}$ & $<-12.33$
\\
00095000015 & $-10.13\pm0.04$ & $-10.31\pm0.05$ & $-10.60_{-0.09}^{+0.08}$ & $-10.48_{-0.17}^{+0.10}$ & $-10.48_{-0.17}^{+0.10}$ & $<-13.76$
\\
00095000003 & $-10.09\pm0.03$ & $-10.22_{-0.03}^{+0.02}$ & $-10.69\pm0.09$ & $-10.38_{-0.10}^{+0.12}$ & $-10.38_{-0.10}^{+0.11}$ & $<-11.89$ 
\\
00094000013 & $-10.32\pm0.03$ & $-10.41\pm0.03$ & $-11.08_{-0.10}^{+0.09}$ & $-10.68^{+0.25}_{-4.32}$ & $-10.68^{+0.25}_{-4.32}$ & $<-14.72$
\\
00094000015 &  $-10.01\pm0.03$ & $-10.14\pm0.03$ & $-10.61^{+0.07}_{-0.08}$ &  $-10.41^{+0.11}_{-0.21}$ &  $-10.41^{+0.10}_{-0.21}$ & $<-12.32$
\\
00096113001 & $-10.29\pm0.04$ & $-10.41\pm0.03$ & $-10.91^{+0.09}_{-0.21}$ & $-10.69^{+0.14}_{-0.24}$ & $-10.69^{+0.13}_{-0.26}$ & $<-11.59$ 
\\
00095653010 & $-10.24_{-0.02}^{+0.04}$ & $-10.36_{-0.05}^{+0.04}$ & $-10.84_{-0.13}^{+0.11}$ & $-10.52_{-0.16}^{+0.25}$ & $-10.52_{-0.16}^{+0.17}$ &$<-13.34$
\\
00096050002 & $-10.37_{-0.06}^{+0.05}$ & $-10.63_{-0.07}^{+0.06}$ & $-10.73_{-0.19}^{+0.10}$ & $-10.76_{-0.20}^{+0.04}$ & $-10.76_{-0.20}^{+0.04}$ & $<-13.38$
\\
00093158006 & $-10.09\pm0.03$ & $-10.21\pm0.03$ & $-10.72_{-0.07}^{+0.09}$ & $-10.52_{-0.12}^{+0.07}$ & $-10.52_{-0.02}^{+0.07}$ & $<-15.84$
\\
00096113002 & $-10.41_{-0.28}^{+0.02}$ & $-10.50_{-0.04}^{+0.03}$ & $-11.06_{-0.08}^{+0.10}$ & $-10.71_{-0.09}^{+0.19}$ & $-10.71_{-0.05}^{+0.19}$ & $<-15.72$
\\
00095653014 & $-10.04_{-0.03}^{+0.04}$ & $-10.13_{-0.02}^{+0.03}$ & $-10.64_{-0.08}^{+0.13}$ & $-10.27_{-0.15}^{+0.03}$ & $-10.28_{-0.15}^{+0.05}$ & $<-16.99$
\\
00095000003 & $-10.09_{-0.03}^{+0.04}$ & $-10.22\pm0.03$ & $-10.69_{-0.10}^{+0.09}$ & $-10.38_{-0.11}^{+0.13}$ & $-10.38_{-0.11}^{+0.12}$ & $<-14.45$
\\
00095000004 & $-9.98_{-0.04}^{+0.05}$ & $-10.11\pm0.03$ & $-10.59_{-0.11}^{+0.17}$ & $-10.43_{-2.87}^{+0.59}$ & $-10.43_{-2.87}^{+0.59}$ & $<-16.75$
\\
00095000005 & $-9.86_{-0.02}^{+0.03}$ & $-9.96\pm0.03$ & $-10.53\pm0.06$ & $<-10.63$ & $<-10.63$ & $<-20.81$
\\
00095000007 & $-10.41\pm0.04$ & $-10.51\pm0.04$ & $-11.07_{-0.17}^{+0.08}$ & $-10.80_{-3.37}^{+0.12}$ & $<-10.68$ & $<-22.12$
\\
00094000016 & $-9.89_{-0.04}^{+0.02}$ & $-10.02\pm0.02$ & $-10.47\pm0.06$ & $-10.40_{-0.34}^{+0.22}$ & $-10.40_{-0.34}^{+0.21}$ & $<-16.46$
\\
00094000017 & $-10.01\pm0.03$ & $-10.12\pm0.03$ & $-10.64\pm0.07$ & $-10.35_{-0.25}^{+0.09}$ & $-10.35_{-0.25}^{+0.11}$ & $<-16.15$
\\
00093158011 & $-10.04\pm0.03$ & $-10.16\pm0.03$ & $-10.67\pm0.07$ & $-10.29_{-0.11}^{+0.07}$ & $-10.29_{-0.11}^{+0.07}$ & $<-14.35$
\\
00093158012 & $-9.98\pm0.04$ & $-10.11_{-0.04}^{+0.02}$ & $-10.54_{-0.11}^{+0.09}$ & $-10.49_{-0.17}^{+1.4}$ & $-10.49_{-1.13}^{+0.17}$ & $<-18.95$
\\
00093158013 & $-10.12_{-0.07}^{+0.04}$ & $-10.29\pm0.03$ & $-10.62_{-0.23}^{+0.34}$ & $-10.57_{-0.45}^{+0.04}$ & $-10.57_{-0.46}^{+0.02}$ & $<-20.75$
\\
00093158008 & $-9.84_{-0.01}^{+0.02}$ & $-9.99\pm0.03$ & $-10.39\pm0.05$ & $-10.27_{-0.31}^{+0.07}$ & $-10.27_{-0.31}^{+0.15}$ & $<-20.56$
\\
00092237001 & $-10.13\pm0.03$ & $-10.26\pm0.03$ & $-10.71\pm0.07$ & $-10.42_{-0.09}^{+0.06}$ & $-10.42_{-0.09}^{+0.06}$ & $<-15.56$
\\
00092237002 & $-10.05_{-0.01}^{+0.02}$ & $-10.22\pm0.03$ & $-10.54\pm0.06$ & $-10.42_{-0.04}^{+0.08}$ & $-10.42_{-0.03}^{+0.07}$ & $<-17.73$
\\
\end{longtable}
}

\begin{table}
\centering
\caption{Comparison of the warm-corona and relativistic reflection models for the highest- and lowest-flux observations. Parameters marked with (f) were fixed during fitting. 
}
\label{tab:model_comparison}

\begin{tabular}{lcc}
\toprule
Parameter & Warm corona model & Reflection model \\
& \texttt{tbabs*Simpl*NthComp} & \texttt{tbabs*relxillCp} \\
\midrule

\multicolumn{3}{c}{ObsID: 96113002} \\
\midrule

$\Gamma_{Simpl}$ & $2.47_{-0.43}^{+0.31}$ & - \\
$f_{\rm sc}$ & $0.32_{-0.24}^{+0.29}$ & - \\
$\Gamma_{nthComp}$&$<2.36$&-\\
$Norm_{NthComp}(\times10^{-2})$&$0.30\pm0.02$&-\\
$kT_e$ (keV) & $0.13_{-0.04}^{+0.11}$ & - \\

$\Gamma_{relxillcp}$ & - & $>2.08$ \\ 
Incl (deg) & - & 30 (f) \\
$a_*$ & - & 0.99 (f) \\
$A_{Fe}$ (solar) & - & 1 (f) \\
$\log N$ ($\mathrm{cm^{-3}}$) & - & 20.0 \\
$R_{\rm frac}$ & - & $0.12$ \\
$N_{\rm relxillCp}(\times10^{-5})$ & - & $<9.18$ \\
$R_{in}$ ($R_{ISCO}$) &-& $-1.00$(f) \\
$R_{out}$ ($R_{g}$) & - & 1000(f)\\
$R_{br}$ ($R_{g}$) &-& 15.00(f)\\
$Index1$ &-& 3.00(f)\\
$Index2$ &-& 3.00(f)\\
$kT_e$ (keV) & - & 15 (f) \\
$z$ &-& 0.024(f)\\
$\log \xi$ ($\mathrm{erg\,cm\,s^{-1}}$) &-& $2.17^{+1.18}_{-1.09}$\\
$\chi^2/\mathrm{dof}$ & 38.80/40 & 41.57/40 \\

\midrule
\multicolumn{3}{c}{ObsID: 92237011} \\
\midrule

$\Gamma_{Simpl}$ & $2.31_{-0.30}^{+0.32}$ &  \\
$f_{\rm sc}$ & $0.31_{-0.24}^{+0.09}$ & - \\
$\Gamma_{nthComp}$& $1.27_{-1.00}^{+1.49}$& -\\
$Norm_{NthComp}(\times10^{-2})$& $1.22\pm0.01$& -\\
$kT_e$ (keV) & $0.13_{-0.03}^{+0.13}$ & - \\

$\Gamma_{relxillcp}$ & - & $2.17^{+0.54}_{-0.51}$ \\
Incl (deg) & - & 30 (f) \\
$a_*$ & - & 0.99 (f) \\
$A_{Fe}$ (solar) & - & 1 (f) \\
$\log N$ ($\mathrm{cm^{-3}}$) & - & 19.99 \\
$R_{\rm frac}$ & - &  0.69 \\
$N_{\rm relxillCp}(\times10^{-4})$ & - & 1.46 \\
$R_{in}$ ($R_{ISCO}$) &-&$-1.00$(f) \\
$R_{out}$ ($R_{g}$) &-&1000(f)\\
$R_{br}$ ($R_{g}$) &-&15.00(f)\\
$Index1$ &-&3.00(f)\\
$Index2$ &-&3.00(f)\\
$kT_e$ (keV) & - & 15 (f) \\
$z$ &-&0.024(f)\\
$\log \xi$ ($\mathrm{erg\,cm\,s^{-1}}$) & - & $2.39^{+0.72}_{-1.56}$ \\
$\chi^2/\mathrm{dof}$ & 20.97/20 & 20.92/20 \\

\bottomrule
\end{tabular}
\end{table}




\end{appendices}


\clearpage
\bibliography{sn-bibliography}

@ARTICLE{2024RAA....24f5025A,
       author = {{Akhila}, K. and {Misra}, Ranjeev and {Ezhikode}, Savithri H. and {Jeena}, K.},
        title = "{Long Term X-Ray Spectral Variations of the Seyfert-1 Galaxy Mrk 279}",
      journal = {Research in Astronomy and Astrophysics},
         year = 2024,
        month = jun,
       volume = {24},
       number = {6},
          eid = {065025},
        pages = {065025},
          doi = {10.1088/1674-4527/ad4962},
archivePrefix = {arXiv},
       eprint = {2405.08478},
 primaryClass = {astro-ph.HE},
       adsurl = {https://ui.adsabs.harvard.edu/abs/2024RAA....24f5025A}
}

@ARTICLE{2025JHEAp..45..418A,
       author = {{Akhila}, K. and {Misra}, Ranjeev and {Sarma}, Rathin and {Ezhikode}, Savithri H. and {Jeena}, K.},
        title = "{Modelling the energy dependent X-ray variability of Mrk 335}",
      journal = {Journal of High Energy Astrophysics},
         year = 2025,
        month = mar,
       volume = {45},
        pages = {418-427},
          doi = {10.1016/j.jheap.2025.01.014},
archivePrefix = {arXiv},
       eprint = {2501.13458},
 primaryClass = {astro-ph.HE},
       adsurl = {https://ui.adsabs.harvard.edu/abs/2025JHEAp..45..418A}
}

@ARTICLE{1993ARA&A..31..473A,
       author = {{Antonucci}, Robert},
        title = "{Unified models for active galactic nuclei and quasars.}",
      journal = {\araa},
         year = 1993,
        month = jan,
       volume = {31},
        pages = {473-521},
          doi = {10.1146/annurev.aa.31.090193.002353},
       adsurl = {https://ui.adsabs.harvard.edu/abs/1993ARA&A..31..473A}
}

@INPROCEEDINGS{1996ASPC..101...17A,
       author = {{Arnaud}, K.~A.},
        title = "{XSPEC: The First Ten Years}",
    booktitle = {Astronomical Data Analysis Software and Systems V},
         year = 1996,
       editor = {{Jacoby}, George H. and {Barnes}, Jeannette},
       series = {Astronomical Society of the Pacific Conference Series},
       volume = {101},
        month = jan,
        pages = {17},
       adsurl = {https://ui.adsabs.harvard.edu/abs/1996ASPC..101...17A}
}

@ARTICLE{2020MNRAS.492.3041B,
       author = {{Barua}, Samuzal and {Jithesh}, V. and {Misra}, Ranjeev and {Dewangan}, Gulab C. and {Sarma}, Rathin and {Pathak}, Amit},
        title = "{NuSTAR observation of Ark 564 reveals the variation of coronal temperature with flux}",
      journal = {\mnras},
         year = 2020,
        month = feb,
       volume = {492},
       number = {2},
        pages = {3041-3046},
          doi = {10.1093/mnras/staa067},
archivePrefix = {arXiv},
       eprint = {2001.02570},
 primaryClass = {astro-ph.HE},
       adsurl = {https://ui.adsabs.harvard.edu/abs/2020MNRAS.492.3041B}
}

@ARTICLE{2009A&A...495..421B,
       author = {{Bianchi}, S. and {Guainazzi}, M. and {Matt}, G. and {Fonseca Bonilla}, N. and {Ponti}, G.},
        title = "{CAIXA: a catalogue of AGN in the XMM-Newton archive. I. Spectral analysis}",
      journal = {\aap},
         year = 2009,
        month = feb,
       volume = {495},
       number = {2},
        pages = {421-430},
          doi = {10.1051/0004-6361:200810620},
archivePrefix = {arXiv},
       eprint = {0811.1126},
 primaryClass = {astro-ph},
       adsurl = {https://ui.adsabs.harvard.edu/abs/2009A&A...495..421B}
}

@ARTICLE{1996A&A...305...53B,
       author = {{Boller}, T. and {Brandt}, W.~N. and {Fink}, H.},
        title = "{Soft X-ray properties of narrow-line Seyfert 1 galaxies.}",
      journal = {\aap},
         year = 1996,
        month = jan,
       volume = {305},
        pages = {53},
          doi = {10.48550/arXiv.astro-ph/9504093},
archivePrefix = {arXiv},
       eprint = {astro-ph/9504093},
 primaryClass = {astro-ph},
       adsurl = {https://ui.adsabs.harvard.edu/abs/1996A&A...305...53B}
}

@ARTICLE{2005MNRAS.356..789B,
       author = {{Botte}, V. and {Ciroi}, S. and {di Mille}, F. and {Rafanelli}, P. and {Romano}, A.},
        title = "{Stellar velocity dispersion in narrow-line Seyfert 1 galaxies}",
      journal = {\mnras},
         year = 2005,
        month = jan,
       volume = {356},
       number = {2},
        pages = {789-793},
          doi = {10.1111/j.1365-2966.2004.08499.x},
       adsurl = {https://ui.adsabs.harvard.edu/abs/2005MNRAS.356..789B}
}

@ARTICLE{1994MNRAS.271..958B,
       author = {{Brandt}, W.~N. and {Fabian}, A.~C. and {Nandra}, K. and {Reynolds}, C.~S. and {Brinkmann}, W.},
        title = "{ROSAT PSPC observations of the Seyfert 1 galaxies Ark 564, NGC 985, KAZ 163, MRK 79 and RX J2256.6+0525.}",
      journal = {\mnras},
         year = 1994,
        month = dec,
       volume = {271},
        pages = {958-966},
          doi = {10.1093/mnras/271.4.958},
archivePrefix = {arXiv},
       eprint = {astro-ph/9411063},
 primaryClass = {astro-ph},
       adsurl = {https://ui.adsabs.harvard.edu/abs/1994MNRAS.271..958B}
}

@ARTICLE{2007A&A...465..107B,
       author = {{Brinkmann}, W. and {Papadakis}, I.~E. and {Raeth}, C.},
        title = "{Spectral variability analysis of an XMM-Newton observation of Ark 564}",
      journal = {\aap},
         year = 2007,
        month = apr,
       volume = {465},
       number = {1},
        pages = {107-118},
          doi = {10.1051/0004-6361:20066420},
archivePrefix = {arXiv},
       eprint = {astro-ph/0701023},
 primaryClass = {astro-ph},
       adsurl = {https://ui.adsabs.harvard.edu/abs/2007A&A...465..107B}
}

@ARTICLE{2002ApJ...568..610E,
       author = {{Edelson}, Rick and {Turner}, T.~J. and {Pounds}, Ken and {Vaughan}, Simon and {Markowitz}, Alex and {Marshall}, Herman and {Dobbie}, Paul and {Warwick}, Robert},
        title = "{X-Ray Spectral Variability and Rapid Variability of the Soft X-Ray Spectrum Seyfert 1 Galaxies Arakelian 564 and Ton S180}",
      journal = {\apj},
         year = 2002,
        month = apr,
       volume = {568},
       number = {2},
        pages = {610-626},
          doi = {10.1086/323779},
archivePrefix = {arXiv},
       eprint = {astro-ph/0108387},
 primaryClass = {astro-ph},
       adsurl = {https://ui.adsabs.harvard.edu/abs/2002ApJ...568..610E}
}

@ARTICLE{2002A&A...391..875G,
       author = {{Gliozzi}, M. and {Brinkmann}, W. and {R{\"a}th}, C. and {Papadakis}, I.~E. and {Negoro}, H. and {Scheingraber}, H.},
        title = "{On the nature of X-ray variability in Ark 564}",
      journal = {\aap},
         year = 2002,
        month = sep,
       volume = {391},
        pages = {875-886},
          doi = {10.1051/0004-6361:20020886},
archivePrefix = {arXiv},
       eprint = {astro-ph/0206190},
 primaryClass = {astro-ph},
       adsurl = {https://ui.adsabs.harvard.edu/abs/2002A&A...391..875G}
}

@ARTICLE{2015MNRAS.448.1541S,
       author = {{Sarma}, R. and {Tripathi}, S. and {Misra}, R. and {Dewangan}, G. and {Pathak}, A. and {Sarma}, J.~K.},
        title = "{Relationship between X-ray spectral index and X-ray Eddington ratio for Mrk 335 and Ark 564}",
      journal = {\mnras},
         year = 2015,
        month = apr,
       volume = {448},
       number = {2},
        pages = {1541-1550},
          doi = {10.1093/mnras/stv005},
archivePrefix = {arXiv},
       eprint = {1501.00908},
 primaryClass = {astro-ph.GA},
       adsurl = {https://ui.adsabs.harvard.edu/abs/2015MNRAS.448.1541S}
}

@ARTICLE{2007MNRAS.375.1479S,
       author = {{Smith}, R. and {Vaughan}, S.},
        title = "{X-ray and optical variability of Seyfert 1 galaxies as observed with XMM-Newton}",
      journal = {\mnras},
         year = 2007,
        month = mar,
       volume = {375},
       number = {4},
        pages = {1479-1487},
          doi = {10.1111/j.1365-2966.2006.11413.x},
archivePrefix = {arXiv},
       eprint = {astro-ph/0701206},
 primaryClass = {astro-ph},
       adsurl = {https://ui.adsabs.harvard.edu/abs/2007MNRAS.375.1479S}
}

@ARTICLE{2024A&A...689A.116L,
       author = {{Lyu}, M. and {Fei}, Z.~Y. and {Zhang}, G.~B. and {Yang}, X.~J.},
        title = "{Investigation into the origin of the soft excess in Ark 564 using principal component analysis}",
      journal = {\aap},
         year = 2024,
        month = sep,
       volume = {689},
          eid = {A116},
        pages = {A116},
          doi = {10.1051/0004-6361/202348783},
archivePrefix = {arXiv},
       eprint = {2406.18073},
 primaryClass = {astro-ph.HE},
       adsurl = {https://ui.adsabs.harvard.edu/abs/2024A&A...689A.116L}
}

@ARTICLE{2022ApJ...939..109L,
       author = {{Lewin}, Collin and {Kara}, Erin and {Wilkins}, Dan and {Mastroserio}, Guglielmo and {Garc{\'\i}a}, Javier A. and {Zhang}, Rachel C. and {Alston}, William N. and {Connors}, Riley and {Dauser}, Thomas and {Fabian}, Andrew and {Ingram}, Adam and {Jiang}, Jiachen and {Lohfink}, Anne and {Lucchini}, Matteo and {Reynolds}, Christopher S. and {Tombesi}, Francesco and {Klis}, Michiel van der and {Wang}, Jingyi},
        title = "{X-Ray Reverberation Mapping of Ark 564 Using Gaussian Process Regression}",
      journal = {\apj},
         year = 2022,
        month = nov,
       volume = {939},
       number = {2},
          eid = {109},
        pages = {109},
          doi = {10.3847/1538-4357/ac978f},
archivePrefix = {arXiv},
       eprint = {2210.01810},
 primaryClass = {astro-ph.HE},
       adsurl = {https://ui.adsabs.harvard.edu/abs/2022ApJ...939..109L}
}

@ARTICLE{2004ApJ...611.1005G,
       author = {{Gehrels}, N. and {Chincarini}, G. and {Giommi}, P. and {Mason}, K.~O. and {Nousek}, J.~A. and {Wells}, A.~A. and {White}, N.~E. and {Barthelmy}, S.~D. and {Burrows}, D.~N. and {Cominsky}, L.~R. and {Hurley}, K.~C. and {Marshall}, F.~E. and {M{\'e}sz{\'a}ros}, P. and {Roming}, P.~W.~A. and {Angelini}, L. and {Barbier}, L.~M. and {Belloni}, T. and {Campana}, S. and {Caraveo}, P.~A. and {Chester}, M.~M. and {Citterio}, O. and {Cline}, T.~L. and {Cropper}, M.~S. and {Cummings}, J.~R. and {Dean}, A.~J. and {Feigelson}, E.~D. and {Fenimore}, E.~E. and {Frail}, D.~A. and {Fruchter}, A.~S. and {Garmire}, G.~P. and {Gendreau}, K. and {Ghisellini}, G. and {Greiner}, J. and {Hill}, J.~E. and {Hunsberger}, S.~D. and {Krimm}, H.~A. and {Kulkarni}, S.~R. and {Kumar}, P. and {Lebrun}, F. and {Lloyd-Ronning}, N.~M. and {Markwardt}, C.~B. and {Mattson}, B.~J. and {Mushotzky}, R.~F. and {Norris}, J.~P. and {Osborne}, J. and {Paczynski}, B. and {Palmer}, D.~M. and {Park}, H.-S. and {Parsons}, A.~M. and {Paul}, J. and {Rees}, M.~J. and {Reynolds}, C.~S. and {Rhoads}, J.~E. and {Sasseen}, T.~P. and {Schaefer}, B.~E. and {Short}, A.~T. and {Smale}, A.~P. and {Smith}, I.~A. and {Stella}, L. and {Tagliaferri}, G. and {Takahashi}, T. and {Tashiro}, M. and {Townsley}, L.~K. and {Tueller}, J. and {Turner}, M.~J.~L. and {Vietri}, M. and {Voges}, W. and {Ward}, M.~J. and {Willingale}, R. and {Zerbi}, F.~M. and {Zhang}, W.~W.},
        title = "{The Swift Gamma-Ray Burst Mission}",
      journal = {\apj},
         year = 2004,
        month = aug,
       volume = {611},
       number = {2},
        pages = {1005-1020},
          doi = {10.1086/422091},
archivePrefix = {arXiv},
       eprint = {astro-ph/0405233},
 primaryClass = {astro-ph},
       adsurl = {https://ui.adsabs.harvard.edu/abs/2004ApJ...611.1005G}
}

@ARTICLE{2005SSRv..120..165B,
       author = {{Burrows}, David N. and {Hill}, J.~E. and {Nousek}, J.~A. and {Kennea}, J.~A. and {Wells}, A. and {Osborne}, J.~P. and {Abbey}, A.~F. and {Beardmore}, A. and {Mukerjee}, K. and {Short}, A.~D.~T. and {Chincarini}, G. and {Campana}, S. and {Citterio}, O. and {Moretti}, A. and {Pagani}, C. and {Tagliaferri}, G. and {Giommi}, P. and {Capalbi}, M. and {Tamburelli}, F. and {Angelini}, L. and {Cusumano}, G. and {Br{\"a}uninger}, H.~W. and {Burkert}, W. and {Hartner}, G.~D.},
        title = "{The Swift X-Ray Telescope}",
      journal = {\ssr},
         year = 2005,
        month = oct,
       volume = {120},
       number = {3-4},
        pages = {165-195},
          doi = {10.1007/s11214-005-5097-2},
archivePrefix = {arXiv},
       eprint = {astro-ph/0508071},
 primaryClass = {astro-ph},
       adsurl = {https://ui.adsabs.harvard.edu/abs/2005SSRv..120..165B}
}

@ARTICLE{1996MNRAS.283..193Z,
       author = {{Zdziarski}, A.~A. and {Johnson}, W.~N. and {Magdziarz}, P.},
        title = "{Broad-band {\ensuremath{\gamma}}-ray and X-ray spectra of NGC 4151 and their implications for physical processes and geometry.}",
      journal = {\mnras},
         year = 1996,
        month = nov,
       volume = {283},
       number = {1},
        pages = {193-206},
          doi = {10.1093/mnras/283.1.193},
archivePrefix = {arXiv},
       eprint = {astro-ph/9607015},
 primaryClass = {astro-ph},
       adsurl = {https://ui.adsabs.harvard.edu/abs/1996MNRAS.283..193Z}
}

@ARTICLE{2009PASP..121.1279S,
       author = {{Steiner}, James F. and {Narayan}, Ramesh and {McClintock}, Jeffrey E. and {Ebisawa}, Ken},
        title = "{A Simple Comptonization Model}",
      journal = {\pasp},
         year = 2009,
        month = nov,
       volume = {121},
       number = {885},
        pages = {1279},
          doi = {10.1086/648535},
archivePrefix = {arXiv},
       eprint = {0810.1758},
 primaryClass = {astro-ph},
       adsurl = {https://ui.adsabs.harvard.edu/abs/2009PASP..121.1279S}
}

@ARTICLE{1993ApJ...414L..85L,
       author = {{Lawrence}, A. and {Papadakis}, I.},
        title = "{X-Ray Variability of Active Galactic Nuclei: A Universal Power Spectrum with Luminosity-dependent Amplitude}",
      journal = {\apjl},
         year = 1993,
        month = sep,
       volume = {414},
        pages = {L85},
          doi = {10.1086/187002},
       adsurl = {https://ui.adsabs.harvard.edu/abs/1993ApJ...414L..85L}
}

@ARTICLE{1981MNRAS.194..987M,
       author = {{Marshall}, N. and {Warwick}, R.~S. and {Pounds}, K.~A.},
        title = "{The variability of X-ray emission from active galaxies.}",
      journal = {\mnras},
         year = 1981,
        month = mar,
       volume = {194},
        pages = {987-1002},
          doi = {10.1093/mnras/194.4.987},
       adsurl = {https://ui.adsabs.harvard.edu/abs/1981MNRAS.194..987M}
}

@ARTICLE{2017MNRAS.468.3489K,
       author = {{Kara}, E. and {Garc{\'\i}a}, J.~A. and {Lohfink}, A. and {Fabian}, A.~C. and {Reynolds}, C.~S. and {Tombesi}, F. and {Wilkins}, D.~R.},
        title = "{The high-Eddington NLS1 Ark 564 has the coolest corona}",
      journal = {\mnras},
         year = 2017,
        month = jul,
       volume = {468},
       number = {3},
        pages = {3489-3498},
          doi = {10.1093/mnras/stx792},
archivePrefix = {arXiv},
       eprint = {1703.09815},
 primaryClass = {astro-ph.HE},
       adsurl = {https://ui.adsabs.harvard.edu/abs/2017MNRAS.468.3489K}
}

@ARTICLE{2014MNRAS.439.2307F,
       author = {{Fabian}, A.~C. and {Parker}, M.~L. and {Wilkins}, D.~R. and {Miller}, J.~M. and {Kara}, E. and {Reynolds}, C.~S. and {Dauser}, T.},
        title = "{On the determination of the spin and disc truncation of accreting black holes using X-ray reflection}",
      journal = {\mnras},
         year = 2014,
        month = apr,
       volume = {439},
       number = {3},
        pages = {2307-2313},
          doi = {10.1093/mnras/stu045},
archivePrefix = {arXiv},
       eprint = {1401.1615},
 primaryClass = {astro-ph.HE},
       adsurl = {https://ui.adsabs.harvard.edu/abs/2014MNRAS.439.2307F}
}

@ARTICLE{2007ApJ...671.1284D,
       author = {{Dewangan}, G.~C. and {Griffiths}, R.~E. and {Dasgupta}, Surajit and {Rao}, A.~R.},
        title = "{An Investigation of the Origin of Soft X-Ray Excess Emission from Ark 564 and Mrk 1044}",
      journal = {\apj},
         year = 2007,
        month = dec,
       volume = {671},
       number = {2},
        pages = {1284-1296},
          doi = {10.1086/523683},
archivePrefix = {arXiv},
       eprint = {0709.1962},
 primaryClass = {astro-ph},
       adsurl = {https://ui.adsabs.harvard.edu/abs/2007ApJ...671.1284D}
}

@ARTICLE{2018A&A...611A..59P,
       author = {{Petrucci}, P.-O. and {Ursini}, F. and {De Rosa}, A. and {Bianchi}, S. and {Cappi}, M. and {Matt}, G. and {Dadina}, M. and {Malzac}, J.},
        title = "{Testing warm Comptonization models for the origin of the soft X-ray excess in AGNs}",
      journal = {\aap},
         year = 2018,
        month = mar,
       volume = {611},
          eid = {A59},
        pages = {A59},
          doi = {10.1051/0004-6361/201731580},
archivePrefix = {arXiv},
       eprint = {1710.04940},
 primaryClass = {astro-ph.HE},
       adsurl = {https://ui.adsabs.harvard.edu/abs/2018A&A...611A..59P}
}

@ARTICLE{2020MNRAS.498.3888J,
       author = {{Jiang}, Jiachen and {Gallo}, Luigi C. and {Fabian}, Andrew C. and {Parker}, Michael L. and {Reynolds}, Christopher S.},
        title = "{A disc reflection model for ultra-soft narrow-line Seyfert 1 galaxies}",
      journal = {\mnras},
         year = 2020,
        month = nov,
       volume = {498},
       number = {3},
        pages = {3888-3901},
          doi = {10.1093/mnras/staa2625},
       adsurl = {https://ui.adsabs.harvard.edu/abs/2020MNRAS.498.3888J}
}

@ARTICLE{2006MNRAS.365.1067C,
       author = {{Crummy}, J. and {Fabian}, A.~C. and {Gallo}, L. and {Ross}, R.~R.},
        title = "{An explanation for the soft X-ray excess in active galactic nuclei}",
      journal = {\mnras},
         year = 2006,
        month = feb,
       volume = {365},
       number = {4},
        pages = {1067-1081},
          doi = {10.1111/j.1365-2966.2005.09844.x},
archivePrefix = {arXiv},
       eprint = {astro-ph/0511457},
 primaryClass = {astro-ph},
       adsurl = {https://ui.adsabs.harvard.edu/abs/2006MNRAS.365.1067C}
}

@ARTICLE{2020MNRAS.495.3373E,
       author = {{Ezhikode}, Savithri H. and {Dewangan}, Gulab C. and {Misra}, Ranjeev and {Philip}, Ninan Sajeeth},
        title = "{Correlation between relativistic reflection fraction and photon index in NuSTAR sample of Seyfert 1 AGN}",
      journal = {\mnras},
         year = 2020,
        month = jan,
       volume = {495},
       number = {3},
        pages = {3373-3386},
          doi = {10.1093/mnras/staa1288},
archivePrefix = {arXiv},
       eprint = {2005.03307},
 primaryClass = {astro-ph.HE},
       adsurl = {https://ui.adsabs.harvard.edu/abs/2020MNRAS.495.3373E}
}

@ARTICLE{2023AN....34430042E,
       author = {{Ezhikode}, Savithri H. and {Shyam Prakash V.}, P. and {Dewangan}, Gulab C. and {Mathew}, Blesson},
        title = "{Soft excess in AGN with relativistic X{\ensuremath{-}}ray reflection}",
      journal = {Astronomische Nachrichten},
         year = 2023,
        month = may,
       volume = {344},
       number = {4},
          eid = {e20230042},
        pages = {e20230042},
          doi = {10.1002/asna.20230042},
       adsurl = {https://ui.adsabs.harvard.edu/abs/2023AN....34430042E}
}

@ARTICLE{1993A&A...274..105W,
       author = {{Walter}, R. and {Fink}, H.~H.},
        title = "{The ultraviolet to soft X-ray bump of Seyfert 1 type active galactic nuclei.}",
      journal = {\aap},
         year = 1993,
        month = jul,
       volume = {274},
        pages = {105},
       adsurl = {https://ui.adsabs.harvard.edu/abs/1993A&A...274..105W}
}

@ARTICLE{2016RAA....16..108E,
       author = {{Ezhikode}, Savithri H. and {Dewangan}, Gulab C. and {Misra}, Ranjeev and {Tripathi}, Shruti and {Sajeeth Philip}, Ninan and {Kembhavi}, Ajit K.},
        title = "{UV and X-ray variability of the narrow-line Seyfert 1 galaxy Ark 564}",
      journal = {Research in Astronomy and Astrophysics},
         year = 2016,
        month = jul,
       volume = {16},
       number = {7},
          eid = {108},
        pages = {108},
          doi = {10.1088/1674-4527/16/7/108},
archivePrefix = {arXiv},
       eprint = {1603.02404},
 primaryClass = {astro-ph.HE},
       adsurl = {https://ui.adsabs.harvard.edu/abs/2016RAA....16..108E}
}

@ARTICLE{2010ApJ...718..695G,
       author = {{Garc{\'\i}a}, J. and {Kallman}, T.~R.},
        title = "{X-ray Reflected Spectra from Accretion Disk Models. I. Constant Density Atmospheres}",
      journal = {\apj},
         year = 2010,
        month = aug,
       volume = {718},
       number = {2},
        pages = {695-706},
          doi = {10.1088/0004-637X/718/2/695},
archivePrefix = {arXiv},
       eprint = {1006.0485},
 primaryClass = {astro-ph.HE},
       adsurl = {https://ui.adsabs.harvard.edu/abs/2010ApJ...718..695G}
}

@ARTICLE{2014ApJ...782...76G,
       author = {{Garc{\'\i}a}, J. and {Dauser}, T. and {Lohfink}, A. and {Kallman}, T.~R. and {Steiner}, J.~F. and {McClintock}, J.~E. and {Brenneman}, L. and {Wilms}, J. and {Eikmann}, W. and {Reynolds}, C.~S. and {Tombesi}, F.},
        title = "{Improved Reflection Models of Black Hole Accretion Disks: Treating the Angular Distribution of X-Rays}",
      journal = {\apj},
         year = 2014,
        month = feb,
       volume = {782},
       number = {2},
          eid = {76},
        pages = {76},
          doi = {10.1088/0004-637X/782/2/76},
archivePrefix = {arXiv},
       eprint = {1312.3231},
 primaryClass = {astro-ph.HE},
       adsurl = {https://ui.adsabs.harvard.edu/abs/2014ApJ...782...76G}
}

@ARTICLE{2014MNRAS.444L.100D,
       author = {{Dauser}, T. and {Garcia}, J. and {Parker}, M.~L. and {Fabian}, A.~C. and {Wilms}, J.},
        title = "{The role of the reflection fraction in constraining black hole spin.}",
      journal = {\mnras},
         year = 2014,
        month = oct,
       volume = {444},
        pages = {L100-L104},
          doi = {10.1093/mnrasl/slu125},
archivePrefix = {arXiv},
       eprint = {1408.2347},
 primaryClass = {astro-ph.HE},
       adsurl = {https://ui.adsabs.harvard.edu/abs/2014MNRAS.444L.100D}
}

@ARTICLE{2004MNRAS.347..854V,
       author = {{Vignali}, Cristian and {Brandt}, W.~N. and {Boller}, Th. and {Fabian}, A.~C. and {Vaughan}, Simon},
        title = "{Arakelian 564: an XMM-Newton view}",
      journal = {\mnras},
         year = 2004,
        month = jan,
       volume = {347},
       number = {3},
        pages = {854-860},
          doi = {10.1111/j.1365-2966.2004.07269.x},
archivePrefix = {arXiv},
       eprint = {astro-ph/0310278},
 primaryClass = {astro-ph},
       adsurl = {https://ui.adsabs.harvard.edu/abs/2004MNRAS.347..854V}
}

@ARTICLE{2006Natur.444..730M,
       author = {{McHardy}, I.~M. and {Koerding}, E. and {Knigge}, C. and {Uttley}, P. and {Fender}, R.~P.},
        title = "{Active galactic nuclei as scaled-up Galactic black holes}",
      journal = {\nat},
         year = 2006,
        month = dec,
       volume = {444},
       number = {7120},
        pages = {730-732},
          doi = {10.1038/nature05389},
archivePrefix = {arXiv},
       eprint = {astro-ph/0612273},
 primaryClass = {astro-ph},
       adsurl = {https://ui.adsabs.harvard.edu/abs/2006Natur.444..730M}
}

@ARTICLE{2011A&A...534A..39M,
       author = {{Mehdipour}, M. and {Branduardi-Raymont}, G. and {Kaastra}, J.~S. and {Petrucci}, P.~O. and {Kriss}, G.~A. and {Ponti}, G. and {Blustin}, A.~J. and {Paltani}, S. and {Cappi}, M. and {Detmers}, R.~G. and {Steenbrugge}, K.~C.},
        title = "{Multiwavelength campaign on Mrk 509. IV. Optical-UV-X-ray variability and the nature of the soft X-ray excess}",
      journal = {\aap},
         year = 2011,
        month = oct,
       volume = {534},
          eid = {A39},
        pages = {A39},
          doi = {10.1051/0004-6361/201116875},
archivePrefix = {arXiv},
       eprint = {1107.0659},
 primaryClass = {astro-ph.CO},
       adsurl = {https://ui.adsabs.harvard.edu/abs/2011A&A...534A..39M}
}

@ARTICLE{2003MNRAS.345.1271V,
       author = {{Vaughan}, S. and {Edelson}, R. and {Warwick}, R.~S. and {Uttley}, P.},
        title = "{On characterizing the variability properties of X-ray light curves from active galaxies}",
      journal = {\mnras},
         year = 2003,
        month = nov,
       volume = {345},
       number = {4},
        pages = {1271-1284},
          doi = {10.1046/j.1365-2966.2003.07042.x},
archivePrefix = {arXiv},
       eprint = {astro-ph/0307420},
 primaryClass = {astro-ph},
       adsurl = {https://ui.adsabs.harvard.edu/abs/2003MNRAS.345.1271V}
}

@ARTICLE{1985ApJ...297..166O,
       author = {{Osterbrock}, D.~E. and {Pogge}, R.~W.},
        title = "{The spectra of narrow-line Seyfert 1 galaxies.}",
      journal = {\apj},
         year = 1985,
        month = oct,
       volume = {297},
        pages = {166-176},
          doi = {10.1086/163513},
       adsurl = {https://ui.adsabs.harvard.edu/abs/1985ApJ...297..166O}
}

@ARTICLE{2015ARA&A..53..365N,
       author = {{Netzer}, Hagai},
        title = "{Revisiting the Unified Model of Active Galactic Nuclei}",
      journal = {\araa},
         year = 2015,
        month = aug,
       volume = {53},
        pages = {365-408},
          doi = {10.1146/annurev-astro-082214-122302},
archivePrefix = {arXiv},
       eprint = {1505.00811},
 primaryClass = {astro-ph.GA},
       adsurl = {https://ui.adsabs.harvard.edu/abs/2015ARA&A..53..365N}
}

@ARTICLE{1999ApJS..125..297L,
       author = {{Leighly}, Karen M.},
        title = "{A Comprehensive Spectral and Variability Study of Narrow-Line Seyfert 1 Galaxies Observed by ASCA. I. Observations and Time Series Analysis}",
      journal = {\apjs},
         year = 1999,
        month = dec,
       volume = {125},
       number = {2},
        pages = {297-316},
          doi = {10.1086/313277},
archivePrefix = {arXiv},
       eprint = {astro-ph/9907294},
 primaryClass = {astro-ph},
       adsurl = {https://ui.adsabs.harvard.edu/abs/1999ApJS..125..297L}
}

@ARTICLE{1989ApJ...342..224G,
       author = {{Goodrich}, Robert W.},
        title = "{Spectropolarimetry of ``Narrow-Line'' Seyfert 1 Galaxies}",
      journal = {\apj},
         year = 1989,
        month = jul,
       volume = {342},
        pages = {224},
          doi = {10.1086/167586},
       adsurl = {https://ui.adsabs.harvard.edu/abs/1989ApJ...342..224G}
}

@ARTICLE{2000MNRAS.314L..17M,
       author = {{Mathur}, Smita},
        title = "{Narrow-line Seyfert 1 galaxies and the evolution of galaxies and active galaxies}",
      journal = {\mnras},
         year = 2000,
        month = jun,
       volume = {314},
       number = {4},
        pages = {L17-L20},
          doi = {10.1046/j.1365-8711.2000.03530.x},
archivePrefix = {arXiv},
       eprint = {astro-ph/0003111},
 primaryClass = {astro-ph},
       adsurl = {https://ui.adsabs.harvard.edu/abs/2000MNRAS.314L..17M}
}

@ARTICLE{1995MNRAS.277L...5P,
       author = {{Pounds}, K.~A. and {Done}, C. and {Osborne}, J.~P.},
        title = "{RE 1034+39: a high-state Seyfert galaxy?}",
      journal = {\mnras},
         year = 1995,
        month = nov,
       volume = {277},
       number = {1},
        pages = {L5-L10},
          doi = {10.1093/mnras/277.1.L5},
       adsurl = {https://ui.adsabs.harvard.edu/abs/1995MNRAS.277L...5P}
}

@ARTICLE{1993ApJ...413..507H,
       author = {{Haardt}, Francesco and {Maraschi}, Laura},
        title = "{X-Ray Spectra from Two-Phase Accretion Disks}",
      journal = {\apj},
         year = 1993,
        month = aug,
       volume = {413},
        pages = {507},
          doi = {10.1086/173020},
       adsurl = {https://ui.adsabs.harvard.edu/abs/1993ApJ...413..507H}
}

@ARTICLE{1991ApJ...380L..51H,
       author = {{Haardt}, F. and {Maraschi}, L.},
        title = "{A Two-Phase Model for the X-Ray Emission from Seyfert Galaxies}",
      journal = {\apjl},
         year = 1991,
        month = oct,
       volume = {380},
        pages = {L51},
          doi = {10.1086/186171},
       adsurl = {https://ui.adsabs.harvard.edu/abs/1991ApJ...380L..51H}
}

@ARTICLE{1997ARA&A..35..445U,
       author = {{Ulrich}, Marie-Helene and {Maraschi}, Laura and {Urry}, C. Megan},
        title = "{Variability of Active Galactic Nuclei}",
      journal = {\araa},
         year = 1997,
        month = jan,
       volume = {35},
        pages = {445-502},
          doi = {10.1146/annurev.astro.35.1.445},
       adsurl = {https://ui.adsabs.harvard.edu/abs/1997ARA&A..35..445U}
}

@ARTICLE{1974ApJ...192..581K,
       author = {{Khachikian}, E.~Y. and {Weedman}, D.~W.},
        title = "{An atlas of Seyfert galaxies.}",
      journal = {\apj},
         year = 1974,
        month = sep,
       volume = {192},
        pages = {581-589},
          doi = {10.1086/153093},
       adsurl = {https://ui.adsabs.harvard.edu/abs/1974ApJ...192..581K}
}

@BOOK{2006agna.book.....O,
       author = {{Osterbrock}, Donald E. and {Ferland}, Gary J.},
        title = "{Astrophysics of gaseous nebulae and active galactic nuclei}",
         year = 2006,
       adsurl = {https://ui.adsabs.harvard.edu/abs/2006agna.book.....O}
}

@BOOK{1997iagn.book.....P,
       author = {{Peterson}, Bradley M.},
        title = "{An Introduction to Active Galactic Nuclei}",
         year = 1997,
       adsurl = {https://ui.adsabs.harvard.edu/abs/1997iagn.book.....P}
}

@ARTICLE{2004ApJ...617..939M,
       author = {{Markowitz}, A. and {Edelson}, R.},
        title = "{An Expanded Rossi X-Ray Timing Explorer Survey of X-Ray Variability in Seyfert 1 Galaxies}",
      journal = {\apj},
         year = 2004,
        month = dec,
       volume = {617},
       number = {2},
        pages = {939-965},
          doi = {10.1086/425559},
archivePrefix = {arXiv},
       eprint = {astro-ph/0409045},
 primaryClass = {astro-ph},
       adsurl = {https://ui.adsabs.harvard.edu/abs/2004ApJ...617..939M}
}

@ARTICLE{2002MNRAS.335L...1F,
       author = {{Fabian}, A.~C. and {Vaughan}, S. and {Nandra}, K. and {Iwasawa}, K. and {Ballantyne}, D.~R. and {Lee}, J.~C. and {De Rosa}, A. and {Turner}, A. and {Young}, A.~J.},
        title = "{A long hard look at MCG-6-30-15 with XMM-Newton}",
      journal = {\mnras},
         year = 2002,
        month = sep,
       volume = {335},
       number = {1},
        pages = {L1-L5},
          doi = {10.1046/j.1365-8711.2002.05740.x},
archivePrefix = {arXiv},
       eprint = {astro-ph/0206095},
 primaryClass = {astro-ph},
       adsurl = {https://ui.adsabs.harvard.edu/abs/2002MNRAS.335L...1F}
}

@ARTICLE{2015MNRAS.449..129W,
       author = {{Wilkins}, D.~R. and {Gallo}, L.~C.},
        title = "{Driving extreme variability: the evolving corona and evidence for jet launching in Markarian 335}",
      journal = {\mnras},
         year = 2015,
        month = may,
       volume = {449},
       number = {1},
        pages = {129-146},
          doi = {10.1093/mnras/stv162},
archivePrefix = {arXiv},
       eprint = {1501.05302},
 primaryClass = {astro-ph.HE},
       adsurl = {https://ui.adsabs.harvard.edu/abs/2015MNRAS.449..129W}
}

@ARTICLE{2004MNRAS.349.1435M,
       author = {{Miniutti}, G. and {Fabian}, A.~C.},
        title = "{A light bending model for the X-ray temporal and spectral properties of accreting black holes}",
      journal = {\mnras},
         year = 2004,
        month = apr,
       volume = {349},
       number = {4},
        pages = {1435-1448},
          doi = {10.1111/j.1365-2966.2004.07611.x},
archivePrefix = {arXiv},
       eprint = {astro-ph/0309064},
 primaryClass = {astro-ph},
       adsurl = {https://ui.adsabs.harvard.edu/abs/2004MNRAS.349.1435M}
}

@ARTICLE{2001ApJ...561..131T,
       author = {{Turner}, T.~J. and {Romano}, P. and {George}, I.~M. and {Edelson}, R. and {Collier}, S.~J. and {Mathur}, S. and {Peterson}, B.~M.},
        title = "{Multiwavelength Monitoring of the Narrow-Line Seyfert 1 Galaxy Arakelian 564. I. ASCA Observations and the Variability of the X-Ray Spectral Components}",
      journal = {\apj},
         year = 2001,
        month = nov,
       volume = {561},
       number = {1},
        pages = {131-145},
          doi = {10.1086/323232},
archivePrefix = {arXiv},
       eprint = {astro-ph/0105283},
 primaryClass = {astro-ph},
       adsurl = {https://ui.adsabs.harvard.edu/abs/2001ApJ...561..131T}
}

@ARTICLE{1999ApJ...524..667T,
       author = {{Turner}, T.~J. and {George}, I.~M. and {Nandra}, K. and {Turcan}, D.},
        title = "{On X-Ray Variability in Seyfert Galaxies}",
      journal = {\apj},
         year = 1999,
        month = oct,
       volume = {524},
       number = {2},
        pages = {667-673},
          doi = {10.1086/307834},
archivePrefix = {arXiv},
       eprint = {astro-ph/9906050},
 primaryClass = {astro-ph},
       adsurl = {https://ui.adsabs.harvard.edu/abs/1999ApJ...524..667T}
}

@ARTICLE{2004MNRAS.349L...7G,
       author = {{Gierli{\'n}ski}, Marek and {Done}, Chris},
        title = "{Is the soft excess in active galactic nuclei real?}",
      journal = {\mnras},
         year = 2004,
        month = mar,
       volume = {349},
       number = {1},
        pages = {L7-L11},
          doi = {10.1111/j.1365-2966.2004.07687.x},
archivePrefix = {arXiv},
       eprint = {astro-ph/0312271},
 primaryClass = {astro-ph},
       adsurl = {https://ui.adsabs.harvard.edu/abs/2004MNRAS.349L...7G}
}

@ARTICLE{2012MNRAS.420.1848D,
       author = {{Done}, Chris and {Davis}, S.~W. and {Jin}, C. and {Blaes}, O. and {Ward}, M.},
        title = "{Intrinsic disc emission and the soft X-ray excess in active galactic nuclei}",
      journal = {\mnras},
         year = 2012,
        month = mar,
       volume = {420},
       number = {3},
        pages = {1848-1860},
          doi = {10.1111/j.1365-2966.2011.19779.x},
archivePrefix = {arXiv},
       eprint = {1107.5429},
 primaryClass = {astro-ph.HE},
       adsurl = {https://ui.adsabs.harvard.edu/abs/2012MNRAS.420.1848D}
}

@ARTICLE{2013A&A...549A..73P,
       author = {{Petrucci}, P.-O. and {Paltani}, S. and {Malzac}, J. and {Kaastra}, J.~S. and {Cappi}, M. and {Ponti}, G. and {De Marco}, B. and {Kriss}, G.~A. and {Steenbrugge}, K.~C. and {Bianchi}, S. and {Branduardi-Raymont}, G. and {Mehdipour}, M. and {Costantini}, E. and {Dadina}, M. and {Lubi{\'n}ski}, P.},
        title = "{Multiwavelength campaign on Mrk 509. XII. Broad band spectral analysis}",
      journal = {\aap},
         year = 2013,
        month = jan,
       volume = {549},
          eid = {A73},
        pages = {A73},
          doi = {10.1051/0004-6361/201219956},
archivePrefix = {arXiv},
       eprint = {1209.6438},
 primaryClass = {astro-ph.HE},
       adsurl = {https://ui.adsabs.harvard.edu/abs/2013A&A...549A..73P}
}

@ARTICLE{2020A&A...634A..85P,
       author = {{Petrucci}, P.-O. and {Gronkiewicz}, D. and {Rozanska}, A. and {Belmont}, R. and {Bianchi}, S. and {Czerny}, B. and {Matt}, G. and {Malzac}, J. and {Middei}, R. and {De Rosa}, A. and {Ursini}, F. and {Cappi}, M.},
        title = "{Radiation spectra of warm and optically thick coronae in AGNs}",
      journal = {\aap},
         year = 2020,
        month = feb,
       volume = {634},
          eid = {A85},
        pages = {A85},
          doi = {10.1051/0004-6361/201937011},
archivePrefix = {arXiv},
       eprint = {2001.02026},
 primaryClass = {astro-ph.HE},
       adsurl = {https://ui.adsabs.harvard.edu/abs/2020A&A...634A..85P}
}

@ARTICLE{2020A&A...640A..99M,
       author = {{Middei}, R. and {Petrucci}, P.-O. and {Bianchi}, S. and {Ursini}, F. and {Cappi}, M. and {Clavel}, M. and {De Rosa}, A. and {Marinucci}, A. and {Matt}, G. and {Tortosa}, A.},
        title = "{The soft excess of the NLS1 galaxy Mrk 359 studied with an XMM-Newton-NuSTAR monitoring campaign}",
      journal = {\aap},
         year = 2020,
        month = aug,
       volume = {640},
          eid = {A99},
        pages = {A99},
          doi = {10.1051/0004-6361/202038112},
archivePrefix = {arXiv},
       eprint = {2006.09005},
 primaryClass = {astro-ph.HE},
       adsurl = {https://ui.adsabs.harvard.edu/abs/2020A&A...640A..99M}
}

@ARTICLE{2005MNRAS.358..211R,
       author = {{Ross}, R.~R. and {Fabian}, A.~C.},
        title = "{A comprehensive range of X-ray ionized-reflection models}",
      journal = {\mnras},
         year = 2005,
        month = mar,
       volume = {358},
       number = {1},
        pages = {211-216},
          doi = {10.1111/j.1365-2966.2005.08797.x},
archivePrefix = {arXiv},
       eprint = {astro-ph/0501116},
 primaryClass = {astro-ph},
       adsurl = {https://ui.adsabs.harvard.edu/abs/2005MNRAS.358..211R}
}

@ARTICLE{2013MNRAS.434.1129K,
       author = {{Kara}, E. and {Fabian}, A.~C. and {Cackett}, E.~M. and {Uttley}, P. and {Wilkins}, D.~R. and {Zoghbi}, A.},
        title = "{Discovery of high-frequency iron K lags in Ark 564 and Mrk 335}",
      journal = {\mnras},
         year = 2013,
        month = sep,
       volume = {434},
       number = {2},
        pages = {1129-1137},
          doi = {10.1093/mnras/stt1055},
archivePrefix = {arXiv},
       eprint = {1306.2551},
 primaryClass = {astro-ph.HE},
       adsurl = {https://ui.adsabs.harvard.edu/abs/2013MNRAS.434.1129K}
}

@ARTICLE{2013MNRAS.428.2901W,
       author = {{Walton}, D.~J. and {Nardini}, E. and {Fabian}, A.~C. and {Gallo}, L.~C. and {Reis}, R.~C.},
        title = "{Suzaku observations of `bare' active galactic nuclei}",
      journal = {\mnras},
         year = 2013,
        month = feb,
       volume = {428},
       number = {4},
        pages = {2901-2920},
          doi = {10.1093/mnras/sts227},
archivePrefix = {arXiv},
       eprint = {1210.4593},
 primaryClass = {astro-ph.HE},
       adsurl = {https://ui.adsabs.harvard.edu/abs/2013MNRAS.428.2901W}
}

@ARTICLE{2001ApJ...547..684M,
       author = {{Markowitz}, A. and {Edelson}, R.},
        title = "{An RXTE Survey of Long-Term X-Ray Variability in Seyfert 1 Galaxies}",
      journal = {\apj},
         year = 2001,
        month = feb,
       volume = {547},
       number = {2},
        pages = {684-692},
          doi = {10.1086/318402},
archivePrefix = {arXiv},
       eprint = {astro-ph/0009422},
 primaryClass = {astro-ph},
       adsurl = {https://ui.adsabs.harvard.edu/abs/2001ApJ...547..684M}
}

@ARTICLE{2009MNRAS.399.1597S,
       author = {{Sobolewska}, M.~A. and {Papadakis}, I.~E.},
        title = "{The long-term X-ray spectral variability of AGN}",
      journal = {\mnras},
         year = 2009,
        month = nov,
       volume = {399},
       number = {3},
        pages = {1597-1610},
          doi = {10.1111/j.1365-2966.2009.15382.x},
archivePrefix = {arXiv},
       eprint = {0911.0265},
 primaryClass = {astro-ph.CO},
       adsurl = {https://ui.adsabs.harvard.edu/abs/2009MNRAS.399.1597S}
}

@ARTICLE{2015A&A...580A..77R,
       author = {{R{\'o}{\.z}a{\'n}ska}, A. and {Malzac}, J. and {Belmont}, R. and {Czerny}, B. and {Petrucci}, P.-O.},
        title = "{Warm and optically thick dissipative coronae above accretion disks}",
      journal = {\aap},
         year = 2015,
        month = aug,
       volume = {580},
          eid = {A77},
        pages = {A77},
          doi = {10.1051/0004-6361/201526288},
archivePrefix = {arXiv},
       eprint = {1504.03160},
 primaryClass = {astro-ph.GA},
       adsurl = {https://ui.adsabs.harvard.edu/abs/2015A&A...580A..77R}
}

@ARTICLE{2007MNRAS.381.1235V,
       author = {{Vasudevan}, R.~V. and {Fabian}, A.~C.},
        title = "{Piecing together the X-ray background: bolometric corrections for active galactic nuclei}",
      journal = {\mnras},
         year = 2007,
        month = nov,
       volume = {381},
       number = {3},
        pages = {1235-1251},
          doi = {10.1111/j.1365-2966.2007.12328.x},
archivePrefix = {arXiv},
       eprint = {0708.4308},
 primaryClass = {astro-ph},
       adsurl = {https://ui.adsabs.harvard.edu/abs/2007MNRAS.381.1235V}
}

@ARTICLE{2020A&A...634A..92U,
       author = {{Ursini}, F. and {Petrucci}, P.-O. and {Bianchi}, S. and {Matt}, G. and {Middei}, R. and {Marcel}, G. and {Ferreira}, J. and {Cappi}, M. and {De Marco}, B. and {De Rosa}, A. and {Malzac}, J. and {Marinucci}, A. and {Ponti}, G. and {Tortosa}, A.},
        title = "{NuSTAR/XMM-Newton monitoring of the Seyfert 1 galaxy HE 1143-1810. Testing the two-corona scenario}",
      journal = {\aap},
         year = 2020,
        month = feb,
       volume = {634},
          eid = {A92},
        pages = {A92},
          doi = {10.1051/0004-6361/201936486},
archivePrefix = {arXiv},
       eprint = {1912.08720},
 primaryClass = {astro-ph.HE},
       adsurl = {https://ui.adsabs.harvard.edu/abs/2020A&A...634A..92U}
}

@ARTICLE{2007A&A...461..931P,
       author = {{Papadakis}, I.~E. and {Brinkmann}, W. and {Page}, M.~J. and {McHardy}, I. and {Uttley}, P.},
        title = "{XMM-Newton observation of the NLS1 galaxy Ark 564. I. Spectral analysis of the time-average spectrum}",
      journal = {\aap},
         year = 2007,
        month = jan,
       volume = {461},
       number = {3},
        pages = {931-942},
          doi = {10.1051/0004-6361:20065527},
archivePrefix = {arXiv},
       eprint = {astro-ph/0610154},
 primaryClass = {astro-ph},
       adsurl = {https://ui.adsabs.harvard.edu/abs/2007A&A...461..931P}
}

@ARTICLE{2017MNRAS.472.3492E,
       author = {{Ezhikode}, Savithri H. and {Gandhi}, Poshak and {Done}, Chris and {Ward}, Martin and {Dewangan}, Gulab C. and {Misra}, Ranjeev and {Philip}, Ninan Sajeeth},
        title = "{Determining the torus covering factors for a sample of type 1 AGN in the local Universe}",
      journal = {\mnras},
         year = 2017,
        month = dec,
       volume = {472},
       number = {3},
        pages = {3492-3511},
          doi = {10.1093/mnras/stx2160},
archivePrefix = {arXiv},
       eprint = {1610.00429},
 primaryClass = {astro-ph.HE},
       adsurl = {https://ui.adsabs.harvard.edu/abs/2017MNRAS.472.3492E}
}

@ARTICLE{2004ApJ...613..682P,
       author = {{Peterson}, B.~M. and {Ferrarese}, L. and {Gilbert}, K.~M. and {Kaspi}, S. and {Malkan}, M.~A. and {Maoz}, D. and {Merritt}, D. and {Netzer}, H. and {Onken}, C.~A. and {Pogge}, R.~W. and {Vestergaard}, M. and {Wandel}, A.},
        title = "{Central Masses and Broad-Line Region Sizes of Active Galactic Nuclei. II. A Homogeneous Analysis of a Large Reverberation-Mapping Database}",
      journal = {\apj},
         year = 2004,
        month = oct,
       volume = {613},
       number = {2},
        pages = {682-699},
          doi = {10.1086/423269},
archivePrefix = {arXiv},
       eprint = {astro-ph/0407299},
 primaryClass = {astro-ph},
       adsurl = {https://ui.adsabs.harvard.edu/abs/2004ApJ...613..682P}
}

@ARTICLE{2013MNRAS.431.2441D,
       author = {{De Marco}, B. and {Ponti}, G. and {Cappi}, M. and {Dadina}, M. and {Uttley}, P. and {Cackett}, E.~M. and {Fabian}, A.~C. and {Miniutti}, G.},
        title = "{Discovery of a relation between black hole mass and soft X-ray time lags in active galactic nuclei}",
      journal = {\mnras},
         year = 2013,
        month = may,
       volume = {431},
       number = {3},
        pages = {2441-2452},
          doi = {10.1093/mnras/stt339},
archivePrefix = {arXiv},
       eprint = {1201.0196},
 primaryClass = {astro-ph.HE},
       adsurl = {https://ui.adsabs.harvard.edu/abs/2013MNRAS.431.2441D}
}

@ARTICLE{2007ApJ...661...38P,
       author = {{Papadakis}, I.~E. and {Ioannou}, Z. and {Kazanas}, D.},
        title = "{Fourier-Resolved Spectroscopy of Active Galactic Nuclei Using XMM-Newton Data. I. The 3-10 keV Band Results}",
      journal = {\apj},
         year = 2007,
        month = may,
       volume = {661},
       number = {1},
        pages = {38-51},
          doi = {10.1086/513307},
archivePrefix = {arXiv},
       eprint = {astro-ph/0701809},
 primaryClass = {astro-ph},
       adsurl = {https://ui.adsabs.harvard.edu/abs/2007ApJ...661...38P}
}

@ARTICLE{2004ApJ...602..635R,
       author = {{Romano}, P. and {Mathur}, S. and {Turner}, T.~J. and {Kraemer}, S.~B. and {Crenshaw}, D.~M. and {Peterson}, B.~M. and {Pogge}, R.~W. and {Brandt}, W.~N. and {George}, I.~M. and {Horne}, K. and {Kriss}, G.~A. and {Netzer}, H. and {Shemmer}, O. and {Wamsteker}, W.},
        title = "{The Spectral Energy Distribution and Emission-Line Properties of the Narrow-Line Seyfert 1 Galaxy Arakelian 564}",
      journal = {\apj},
         year = 2004,
        month = feb,
       volume = {602},
       number = {2},
        pages = {635-647},
          doi = {10.1086/381235},
archivePrefix = {arXiv},
       eprint = {astro-ph/0311206},
 primaryClass = {astro-ph},
       adsurl = {https://ui.adsabs.harvard.edu/abs/2004ApJ...602..635R}
}

@ARTICLE{2015MNRAS.446..633G,
       author = {{Gallo}, L.~C. and {Wilkins}, D.~R. and {Bonson}, K. and {Chiang}, C.-Y. and {Grupe}, D. and {Parker}, M.~L. and {Zoghbi}, A. and {Fabian}, A.~C. and {Komossa}, S. and {Longinotti}, A.~L.},
        title = "{Suzaku observations of Mrk 335: confronting partial covering and relativistic reflection}",
      journal = {\mnras},
         year = 2015,
        month = jan,
       volume = {446},
       number = {1},
        pages = {633-650},
          doi = {10.1093/mnras/stu2108},
archivePrefix = {arXiv},
       eprint = {1410.2330},
 primaryClass = {astro-ph.HE},
       adsurl = {https://ui.adsabs.harvard.edu/abs/2015MNRAS.446..633G}
}

@ARTICLE{2014MNRAS.443.1723P,
       author = {{Parker}, M.~L. and {Wilkins}, D.~R. and {Fabian}, A.~C. and {Grupe}, D. and {Dauser}, T. and {Matt}, G. and {Harrison}, F.~A. and {Brenneman}, L. and {Boggs}, S.~E. and {Christensen}, F.~E. and {Craig}, W.~W. and {Gallo}, L.~C. and {Hailey}, C.~J. and {Kara}, E. and {Komossa}, S. and {Marinucci}, A. and {Miller}, J.~M. and {Risaliti}, G. and {Stern}, D. and {Walton}, D.~J. and {Zhang}, W.~W.},
        title = "{The NuSTAR spectrum of Mrk 335: extreme relativistic effects within two gravitational radii of the event horizon?}",
      journal = {\mnras},
         year = 2014,
        month = sep,
       volume = {443},
       number = {2},
        pages = {1723-1732},
          doi = {10.1093/mnras/stu1246},
archivePrefix = {arXiv},
       eprint = {1407.8223},
 primaryClass = {astro-ph.HE},
       adsurl = {https://ui.adsabs.harvard.edu/abs/2014MNRAS.443.1723P}
}

@ARTICLE{2019MNRAS.490..683P,
       author = {{Parker}, M.~L. and {Longinotti}, A.~L. and {Schartel}, N. and {Grupe}, D. and {Komossa}, S. and {Kriss}, G. and {Fabian}, A.~C. and {Gallo}, L. and {Harrison}, F.~A. and {Jiang}, J. and {Kara}, E. and {Krongold}, Y. and {Matzeu}, G.~A. and {Pinto}, C. and {Santos-Lle{\'o}}, M.},
        title = "{The nuclear environment of the NLS1 Mrk 335: Obscuration of the X-ray line emission by a variable outflow}",
      journal = {\mnras},
         year = 2019,
        month = nov,
       volume = {490},
       number = {1},
        pages = {683-697},
          doi = {10.1093/mnras/stz2566},
archivePrefix = {arXiv},
       eprint = {1909.04924},
 primaryClass = {astro-ph.HE},
       adsurl = {https://ui.adsabs.harvard.edu/abs/2019MNRAS.490..683P}
}

@ARTICLE{2021JApA...42...51E,
       author = {{Ezhikode}, Savithri H. and {Dewangan}, Gulab C. and {Misra}, Ranjeev},
        title = "{AstroSat view of the NLS1 galaxy Mrk 335}",
      journal = {Journal of Astrophysics and Astronomy},
         year = 2021,
        month = oct,
       volume = {42},
       number = {2},
          eid = {51},
        pages = {51},
          doi = {10.1007/s12036-021-09704-8},
archivePrefix = {arXiv},
       eprint = {2102.00805},
 primaryClass = {astro-ph.HE},
       adsurl = {https://ui.adsabs.harvard.edu/abs/2021JApA...42...51E}
}

@ARTICLE{1999ApJS..121..287H,
       author = {{Huchra}, John P. and {Vogeley}, Michael S. and {Geller}, Margaret J.},
        title = "{The CFA Redshift Survey: Data for the South Galactic CAP}",
      journal = {\apjs},
         year = 1999,
        month = apr,
       volume = {121},
       number = {2},
        pages = {287-368},
          doi = {10.1086/313194},
       adsurl = {https://ui.adsabs.harvard.edu/abs/1999ApJS..121..287H}
}

@ARTICLE{2016A&A...594A.116H,
       author = {{HI4PI Collaboration} and {Ben Bekhti}, N. and {Fl{\"o}er}, L. and {Keller}, R. and {Kerp}, J. and {Lenz}, D. and {Winkel}, B. and {Bailin}, J. and {Calabretta}, M.~R. and {Dedes}, L. and {Ford}, H.~A. and {Gibson}, B.~K. and {Haud}, U. and {Janowiecki}, S. and {Kalberla}, P.~M.~W. and {Lockman}, F.~J. and {McClure-Griffiths}, N.~M. and {Murphy}, T. and {Nakanishi}, H. and {Pisano}, D.~J. and {Staveley-Smith}, L.},
        title = "{HI4PI: A full-sky H I survey based on EBHIS and GASS}",
      journal = {\aap},
         year = 2016,
        month = oct,
       volume = {594},
          eid = {A116},
        pages = {A116},
          doi = {10.1051/0004-6361/201629178},
archivePrefix = {arXiv},
       eprint = {1610.06175},
 primaryClass = {astro-ph.GA},
       adsurl = {https://ui.adsabs.harvard.edu/abs/2016A&A...594A.116H}
}

@ARTICLE{1973A&A....24..337S,
       author = {{Shakura}, N.~I. and {Sunyaev}, R.~A.},
        title = "{Black holes in binary systems. Observational appearance.}",
      journal = {\aap},
         year = 1973,
        month = jan,
       volume = {24},
        pages = {337-355},
       adsurl = {https://ui.adsabs.harvard.edu/abs/1973A&A....24..337S}
}

@ARTICLE{2012MNRAS.420.1825J,
       author = {{Jin}, Chichuan and {Ward}, Martin and {Done}, Chris and {Gelbord}, Jonathan},
        title = "{A combined optical and X-ray study of unobscured type 1 active galactic nuclei - I. Optical spectra and spectral energy distribution modelling}",
      journal = {\mnras},
         year = 2012,
        month = mar,
       volume = {420},
       number = {3},
        pages = {1825-1847},
          doi = {10.1111/j.1365-2966.2011.19805.x},
archivePrefix = {arXiv},
       eprint = {1109.2069},
 primaryClass = {astro-ph.HE},
       adsurl = {https://ui.adsabs.harvard.edu/abs/2012MNRAS.420.1825J}
}

@ARTICLE{1999MNRAS.309..561Z,
       author = {{{\.Z}ycki}, Piotr T. and {Done}, Chris and {Smith}, David A.},
        title = "{The 1989 May outburst of the soft X-ray transient GS 2023+338 (V404 Cyg)}",
      journal = {\mnras},
         year = 1999,
        month = nov,
       volume = {309},
       number = {3},
        pages = {561-575},
          doi = {10.1046/j.1365-8711.1999.02885.x},
archivePrefix = {arXiv},
       eprint = {astro-ph/9904304},
 primaryClass = {astro-ph},
       adsurl = {https://ui.adsabs.harvard.edu/abs/1999MNRAS.309..561Z}
}

@ARTICLE{2018ApJ...864...25G,
       author = {{Garc{\'\i}a}, Javier A. and {Steiner}, James F. and {Grinberg}, Victoria and {Dauser}, Thomas and {Connors}, Riley M.~T. and {McClintock}, Jeffrey E. and {Remillard}, Ronald A. and {Wilms}, J{\"o}rn and {Harrison}, Fiona A. and {Tomsick}, John A.},
        title = "{Reflection Spectroscopy of the Black Hole Binary XTE J1752-223 in Its Long-stable Hard State}",
      journal = {\apj},
         year = 2018,
        month = sep,
       volume = {864},
       number = {1},
          eid = {25},
        pages = {25},
          doi = {10.3847/1538-4357/aad231},
archivePrefix = {arXiv},
       eprint = {1807.01949},
 primaryClass = {astro-ph.HE},
       adsurl = {https://ui.adsabs.harvard.edu/abs/2018ApJ...864...25G}
}

\end{document}